\documentclass[a4paper,10pt]{article}

\usepackage[top=1in, left=0.9in, bottom=1in, right=0.9in]{geometry}
\def\baselinestretch{1.3}

\usepackage{amsmath,amssymb,amsfonts}

\usepackage[utf8]{inputenc}
\usepackage{cite}

\usepackage{graphicx}
\usepackage{xcolor}
\usepackage{bm}

\usepackage{booktabs}

\usepackage[colorlinks=true,linkcolor=blue,citecolor=blue,urlcolor=blue]{hyperref}

\let\counterwithin\relax
\usepackage{chngcntr}

\usepackage{longtable}
\usepackage{multirow}
\usepackage{makecell}
\usepackage{diagbox}
\usepackage{adjustbox}

\usepackage{tikz}
\usepackage[normalem]{ulem}
\usepackage{verbatim}
\usepackage[dvipsnames]{xcolor}
\usepackage{xcolor}
\usepackage[vcentermath,enableskew]{youngtab}
\usepackage{ytableau}
\usepackage{young}
\usepackage{tabularx}
\usepackage{diagbox}
\usepackage[toc]{appendix}
\usepackage{longtable}
\usepackage{extarrows}
\usepackage{enumerate}
\usepackage{float}
\usepackage{subfig}
\usepackage{makecell}

\usepackage{simplewick}
\usepackage{makeidx} 

\usepackage[section]{placeins}
\newcommand{\bea}{\begin{eqnarray}}
\newcommand{\eea}{\end{eqnarray}}

\newcommand{\la}{\langle}
\newcommand{\ra}{\rangle}
\newcommand{\lb}{\lbrack}
\newcommand{\rb}{\rbrack}

\begin{document}
\begin{center}

{\Large \textbf  {Heavy-Particle Effective Theory: from On-Shell Formalism to Gravitational Compton Amplitudes}}\\[10mm]



Yong-Kang Li$^{a, b}$\footnote{liyongkang@itp.ac.cn}, Yi-Ning Wang$^{a, b}$\footnote{wangyining@itp.ac.cn}, Jiang-Hao Yu$^{a, b, c, d, e}$\footnote{jhyu@itp.ac.cn}\\[10mm]

\noindent 
$^a${\em \small CAS Key Laboratory of Theoretical Physics, Institute of Theoretical Physics, Chinese Academy of Sciences,    \\ Beijing 100190, P. R. China}  \\
$^b${\em \small School of Physical Sciences, University of Chinese Academy of Sciences,   Beijing 100049, P.R. China}   \\
$^c${\em \small Center for High Energy Physics, Peking University, Beijing 100871, China} \\
$^d${\em \small School of Fundamental Physics and Mathematical Sciences, Hangzhou Institute for Advanced Study, UCAS, Hangzhou 310024, China} \\
$^e${\em \small International Centre for Theoretical Physics Asia-Pacific, Beijing/Hangzhou, China}\\[10mm]

\date{\today}   
          
\end{center}

\begin{abstract}

We formulate on-shell heavy-particle effective theory (HPET) amplitudes by projecting the spinor-helicity variables onto the $SU(2)$ little group of a timelike reference velocity $v^{\mu}$. Retaining both large and small components of the velocity-projected spinors gives an intermediate velocity-projected amplitudes in which reparameterization invariance can be imposed before expanding in the inverse heavy-particle mass. Integrating out the small component then yields HPET amplitudes organized by heavy-particle power counting, with a direct correspondence to local operators at fixed order. We determine the three-point velocity-projected amplitudes for two equal-mass particles of arbitrary spin $S$ coupled to a soft massless boson of helicity $h$. Four-point gravitational Compton amplitudes can be constructively obtained using velocity-projected three-point amplitudes, and expanded into HPET amplitudes which is free of spurious poles up to sub-leading order. As a gravitational application, we show that the projected gravitational minimal coupling reproduces the Kerr spin-induced multipoles up to order $2S$, and find the classical limits of the tree-level elastic gravitational scattering and gravitational Compton amplitudes agreed with the literature up to contact terms undetermined by factorization.


\end{abstract}

\newpage
\setcounter{tocdepth}{3}
\setcounter{secnumdepth}{3}

\tableofcontents

\setcounter{footnote}{0}

\def\baselinestretch{1.5}
\counterwithin{equation}{section}

\newpage

\section{Introduction}
Black hole gravitational scattering is an important arena for gravitational-wave physics, from the first direct detection of a binary black hole merger~\cite{LIGOScientific:2016aoc} and the accompanying tests of general relativity~\cite{LIGOScientific:2016lio}, to the future space-based observatories~\cite{LISA:2017pwj}. In recent years, scattering-amplitude and effective field theory methods have significantly advanced the description of black hole gravitational scattering and compact binary dynamics~\cite{Bjerrum-Bohr:2002gqz,Holstein:2008sx,Damour:2016gwp,Cheung:2018wkq,Kosower:2018adc,Bern:2019crd,Maybee:2019jus,Damgaard:2019lfh,Johansson:2019dnu,Bern:2019nnu,Bern:2020buy,Haddad:2020tvs,Brandhuber:2021kpo,Brandhuber:2021eyq,Bern:2022kto,Brandhuber:2023hhy,Herderschee:2023fxh,Bern:2023ity,Bini:2024rsy,Alaverdian:2024spu,Alaverdian:2025jtw}. Many of these discussions focus on spinless~\cite{Cheung:2018wkq,Damour:2016gwp,Bern:2019nnu,Bjerrum-Bohr:2002gqz,Brandhuber:2021kpo,Brandhuber:2021eyq,Brandhuber:2023hhy} or low-spin~\cite{Holstein:2008sx,Damgaard:2019lfh} configurations, whereas astrophysical black holes generically carry spin.

For spinning black holes, several complementary descriptions are available, including amplitudes-based approach~\cite{Bern:2022kto,Bern:2023ity}, worldline effective field theory~\cite{Levi:2015msa,Vines:2017hyw,Haddad:2024ebn}, on-shell heavy particle effective theory~\cite{Aoude:2020onz,Aoude:2022trd}, heavy-mass effective field theory~\cite{Bjerrum-Bohr:2023jau}, and higher-spin gauge constructions~\cite{Cangemi:2023bpe,Cangemi:2023ysz}, etc. Among these, on-shell methods~\cite{Dixon:1996wi,Elvang:2013cua,Henn:2014yza,Cheung:2017pzi,Guevara:2017csg,Badger:2023eqz} play an important role, in particular through the massive spinor-helicity formalism of Arkani-Hamed, Huang and Huang (AHH)~\cite{Arkani-Hamed:2017jhn}, which has significantly advanced the study of spinning black hole gravitational scattering~\cite{Chung:2018kqs,Guevara:2018wpp,Chung:2019yfs,Guevara:2019fsj,Chung:2019duq,Arkani-Hamed:2019ymq,Chung:2020rrz,Guevara:2020xjx,Parra-Martinez:2020dzs,Aoude:2021oqj,Bautista:2022wjf,Aoki:2024boe,Bohnenblust:2024hkw}. Matching the spin-induced multipole moments of the Kerr black hole~\cite{Arkani-Hamed:2019ymq,Maybee:2019jus,Aoude:2020onz,Bautista:2022wjf}, constructing high-spin Compton amplitudes free of spurious poles~\cite{Chiodaroli:2021eug,Bjerrum-Bohr:2023jau,Bjerrum-Bohr:2023iey,Ema:2025qgd}, and pushing amplitude-based calculations to the second post-Minkowskian (2PM) order~\cite{Aoude:2022trd,Bohnenblust:2024hkw} still require additional treatment.

Heavy particle effective theory (HPET) applies when the mass $m$ of the heavy particle is much larger than the interaction energy scale, and organizes the physics into an expansion in $1/m$. The paradigm is heavy quark effective theory (HQET)~\cite{Isgur:1989vq,Georgi:1990um,Eichten:1989zv,Grinstein:1990mj,Eichten:1990vp,Manohar:1997qy,Manohar_Wise_2023}, in which the residual rotational symmetry organizes the heavy mass expansion; the same strategy underlies the heavy particle effective descriptions of black holes~\cite{Damgaard:2019lfh}. On the on-shell side, Refs.~\cite{Aoude:2020onz,Aoude:2022trd} introduced on-shell heavy particle variables, which retain only the large component of the massive spinors, and advanced the applications of massive spinors to black hole gravitational-wave physics. These variables provide a starting point for a further organization in which the reference-velocity little group, the small-component expansion, and the correspondence to HPET operators are treated together. In particular, it is useful to separate the implementation of reparameterization invariance from the elimination of the small component, while expressing both heavy-particle and soft-boson variables in the same $SU(2)$ little group.

In this work, we introduce such an organization by projecting the spinor variables onto a timelike reference velocity $v^{\mu}$, building on the constructive heavy particle effective theory of Ref.~\cite{Li:2025ejk}. Three successive definitions of heavy-particle spinors make the residual rotational symmetry manifest step by step. The first is the standard massive spinor-helicity variables in Ref.~\cite{Arkani-Hamed:2017jhn}, the AHH spinor
\begin{eqnarray}
  \left[\begin{array}{r}
         |\boldsymbol{p}^I\rangle  \\
         |\boldsymbol{p}^I\rbrack
    \end{array}\right],
\end{eqnarray}
defined by the decomposition $p_{\alpha\dot\alpha}=|\boldsymbol{p}^I\rangle\lbrack\boldsymbol{p}_I|$, where $I$ is the little group index of $p^{\mu}$ itself. The second, the on-shell heavy particle variables of Ref.~\cite{Aoude:2020onz}, retains only the large component of the heavy state, projecting the massive spinors with $(1+v\!\!\!/)/2$,
\begin{eqnarray}
     \left[\begin{array}{c}
         |\boldsymbol{p}_v^I\rangle  \\
         |\boldsymbol{p}_v^I\rbrack
    \end{array}\right]=\frac{1+v\!\!\!/}{2}\left[\begin{array}{c}
         |\boldsymbol{p}^I\rangle  \\
         |\boldsymbol{p}^I\rbrack
    \end{array}\right] ,
\end{eqnarray}
but they still carry the Lorentz indices $\alpha$ and $\dot\alpha$, so the residual rotational symmetry is not yet manifest. We propose the third definition which acts on the large component of the heavy state and starts from the momentum decomposition along the $v$ direction $p_{\alpha\dot\alpha} = m v_{\alpha\dot\alpha} + k_{\alpha\dot\alpha}$. Using the reference spinor $v_{\alpha\dot\alpha}=|v^L\rangle\lbrack v_L|$, one projects the massive spinors onto the $SU(2)$ little group that leaves $v^{\mu}$ invariant, and thus defines the velocity-projected spinors ($v\!\!\!/$-spinors), with the spinor indices contracted and only little group indices left,
\begin{eqnarray}
|\boldsymbol{k}^+\rfloor_{L}^{I}\equiv\frac{1}{2}\Big(\la\boldsymbol{p}^{I} v_{L}\ra+\lb v_{L}\boldsymbol{p}^{I}\rb\Big) .
\end{eqnarray}
The index $I$ continues to label the massive-particle little group, while $L$ labels the reference little group; no Lorentz spinor indices remain. Projecting the soft-boson variables in the same way expresses all contractions in a common rotation group.

The small component is represented by $|\boldsymbol{k}^-\rfloor_{L}^{I}\equiv\frac{1}{2}\big(-\la\boldsymbol{p}^{I}v_L\ra+\lb v_L\boldsymbol{p}^{I}\rb\big)$. Keeping it as an auxiliary variable defines an intermediate velocity-projected amplitude between the relativistic and HPET amplitudes. Reparameterization invariance (RPI)~\cite{Luke:1992cs,Chen:1993np,Finkemeier:1997re,Sundrum:1997ut,Heinonen:2012km,Li:2025ejk} acts on the large and small components through coupled linear transformations. It can therefore be imposed before eliminating the small component, implementing the Lorentz symmetry left non-manifest by the choice of $v^{\mu}$. The subsequent integrating out expands the amplitude in powers of residual momenta over $m$.


Following the logic of constructive on-shell amplitudes, we start from the three-particle amplitude of two heavy particles of spin $S$ and one soft massless boson of helicity $h$. Little group covariance fixes the most general reference-$SU(2)$-invariant three-particle amplitude, and imposing reparameterization invariance then constrains the Wilson coefficients and yields the Lorentz-invariant amplitude in agreement with the literature~\cite{Arkani-Hamed:2017jhn,Aoude:2020onz}. The four-particle Compton amplitude is subsequently bootstrapped from the three-particle amplitude through consistent factorization, and the resulting amplitudes are applied to the gravitational scattering of Kerr black holes.

The paper is organized as follows. Section~\ref{sec:spinor} introduces the velocity-projected spinors and their reference-$SU(2)$ transformation properties. Section~\ref{sec:threep} projects the relativistic three-particle amplitudes onto the velocity-projected amplitudes, expands the small component to obtain the HPET amplitudes, and constructs the same amplitudes bottom-up from little group covariance and reparameterization invariance. Section~\ref{sec:fourp} determines the factorizing Compton amplitudes through the next-to-leading order in $1/m$ and analyzes their heavy mass limit. Section~\ref{sec:spmultipole} matches the minimal graviton coupling to the multipole Lagrangian of the worldline effective field theory of the Kerr black hole and derives the classical elastic and gravitational Compton amplitudes at the first post-Minkowskian (1PM) order in a general frame. Appendix~\ref{app:HPET} reviews the relation between the projected spinors and heavy-particle states, Appendix~\ref{subsec:comp} compares the velocity-projected spinors with the on-shell heavy particle variables of Ref.~\cite{Aoude:2020onz}, Appendix~\ref{app:aoc} discusses the amplitude--operator correspondence, and Appendix~\ref{apprpi} provides an explicit reparameterization calculation.

\section{Velocity-projected Spinor-Helicity Formalism}\label{sec:spinor}

We introduce velocity-projected spinors ($v\!\!\!/$-spinors) that express the heavy and soft boson variables in a common rotation group. The construction separates three ingredients: the spin index of each massive particle, rotations preserving the reference velocity, and the expansion in residual momenta over the heavy mass. Reparameterization invariance relates different reference decompositions and constrains the coefficients of the projected amplitudes.

Throughout, $v^2=1$ and $p^{\mu}=mv^{\mu}+k^{\mu}$. We distinguish the massive-particle little group indices $I,J$ from the reference-velocity indices $L,L'$. For real timelike momenta the relevant little group is $SU(2)$; spinor expressions are analytically continued when complex on-shell kinematics are needed. The large and small components are denoted by $|\boldsymbol{k}^{+}\rfloor$ and $|\boldsymbol{k}^{-}\rfloor$, respectively. The signs label components, not particle helicities.

Subsection~\ref{subsec:rspinor} fixes the relativistic spinor conventions. Subsection~\ref{subsec:vpspinor} defines the projected heavy-particle variables, their contractions, and their equations of motion. Subsection~\ref{subsec:expand} gives the inverse spinor relations and projects the massless variables. Appendix~\ref{subsec:comp} provides the dictionary to the on-shell HPET variables of Ref.~\cite{Aoude:2020onz}.

\subsection{Massless and Massive Relativistic Spinors}\label{subsec:rspinor}

We briefly review the relativistic spinor-helicity formalism for particles of arbitrary mass and spin~\cite{Elvang:2013cua,Cheung:2017pzi,Arkani-Hamed:2017jhn}. The massless and massive spinors furnish representations of the Lorentz group and their respective Wigner little groups; we also summarize the associated polarization vectors needed below.

A complex relativistic on-shell momentum satisfying $p^2 = m^2$ is mapped into a $2\times 2$ complex matrix via the four-vector of Pauli matrices $\sigma^{\mu} = (1, \vec\sigma)$: $p^{\mu}\rightarrow p_{\alpha\dot\alpha}\equiv p_{\mu}\sigma^{\mu}_{\alpha\dot\alpha}$, preserving the number of degrees of freedom. Here $\alpha, \dot\alpha = 1, 2$ are two independent $SL(2,\mathbb{C})$ group indices. For complex momenta, the dotted and undotted indices are independent, so the symmetry group is $SL(2,\mathbb{C})\times SL(2,\mathbb{C})$; for real momenta, the two sets of indices are related by Hermitian conjugation, and the symmetry reduces to $SL(2,\mathbb{C})$. The determinant of the matrix yields the squared mass, $\det[p_{\alpha\dot\alpha}] = p^2 = m^2$. Consequently, massless ($m=0$) and massive ($m\neq0$) particles correspond to matrices of different ranks, leading to distinct spinor decompositions in this formalism.

For a massless particle with complex momentum $q^{\mu}$, we have $\det[q_{\alpha\dot\alpha}]=0$; this rank-1 matrix can be expressed as $q_{\alpha\dot\alpha}=q_{\alpha}\,\tilde q_{\dot\alpha}$, where $q_{\alpha}$ and $\tilde q_{\dot\alpha}$ are independent two-component spinors. The $SL(2,\mathbb{C})\times SL(2,\mathbb{C})$ anti-symmetric invariant tensors $\epsilon_{\alpha\beta}$ and $\epsilon^{\dot\alpha\dot\beta}$ (with $\epsilon^{12}=1$) are used to raise and lower indices: $q^{\alpha}=\epsilon^{\alpha\beta}q_{\beta}$, $\tilde q^{\dot\alpha}=\epsilon^{\dot\alpha\dot\beta}\tilde q_{\dot\beta}$. In the spinor-helicity formalism, the \emph{angle} and \emph{bracket} notations are introduced, $|q\rangle\equiv q_{\alpha}$, $\langle q|\equiv q^{\alpha}$, $|q]\equiv \tilde q^{\dot\alpha}$, and $[q|\equiv \tilde q_{\dot\alpha}$, so that products of massless spinors (e.g., $q$ and $\xi$) are written as
\begin{eqnarray}\label{eq:anglebracket}
\langle q\xi\rangle \equiv \epsilon_{\alpha\beta}\,q^{\alpha}\xi^{\beta},\qquad
[  q\xi ] \equiv \epsilon^{\dot\alpha\dot\beta}\,\tilde q_{\dot\alpha}\tilde\xi_{\dot\beta}.
\end{eqnarray}
The on-shell conditions for massless spinors follow directly from the decomposition: $q^{\dot\alpha\alpha}q_{\alpha}=|q\rbrack\langle q q\rangle=0$ and $q_{\alpha\dot\alpha}\tilde q^{\dot\alpha}=|q\rangle\lbrack qq\rbrack=0$.

For a massless helicity $|h|=1$ vector boson, the polarization vectors are
\begin{eqnarray}\label{eq:photonepsilon}
\varepsilon_{\alpha\dot{\alpha}}^{+}= \frac{\xi_{\alpha}\,\tilde q_{\dot{\alpha}}}{\langle\xi q\rangle},\qquad
\varepsilon_{\alpha\dot{\alpha}}^{-}= \frac{q_{\alpha}\,\tilde\xi_{\dot{\alpha}}}{[ q\xi]},
\end{eqnarray}
where $\xi,\tilde\xi$ are arbitrary reference spinors reflecting gauge redundancy. These polarization vectors are transverse to $q^{\mu}$: $q^{\dot\alpha\alpha}\varepsilon_{\alpha\dot\alpha}^{+}=0$ and $q^{\dot\alpha\alpha}\varepsilon_{\alpha\dot\alpha}^{-}=0$. The gauge transformation caused by $|\xi\rangle\rightarrow a|\xi\rangle+b|q\rangle$ only shifts the polarization vector $\varepsilon^+_{\alpha\dot\alpha}\rightarrow \varepsilon^+_{\alpha\dot\alpha}+\frac{b}{a\langle\xi q\rangle}q_{\alpha\dot\alpha}$ for some parameters $a$ and $b$.
The polarization vector of a $|h|=2$ graviton is then expressed by the symmetric tensor
\begin{eqnarray}
    \varepsilon^{++}_{\alpha\beta\dot\alpha\dot\beta}= \varepsilon^+_{\alpha\dot\alpha}\varepsilon^+_{\beta\dot\beta},\quad \varepsilon^{--}_{\alpha\beta\dot\alpha\dot\beta}= \varepsilon^-_{\alpha\dot\alpha}\varepsilon^-_{\beta\dot\beta},
\end{eqnarray}
where $\alpha,\beta$ and $\dot\alpha,\dot\beta$ are individually symmetrized.

In terms of the on-shell massless spinors, the Lorentz transformations read $q_{\beta}\rightarrow \Lambda_{\beta}^{\;\alpha}q_{\alpha}$ and $\tilde q^{\dot\beta}\rightarrow \tilde\Lambda^{\dot\beta}_{\;\dot\alpha}\tilde q^{\dot\alpha}$, where $\Lambda$ and $\tilde\Lambda$ are independent elements of $SL(2,\mathbb C)\times SL(2,\mathbb C)$.  The massless little group transformations that leave $q^{\mu}$ invariant are $q_{\alpha}\rightarrow \omega^{-1}q_{\alpha}$ and $\tilde q_{\dot\alpha}\rightarrow \omega\,\tilde q_{\dot\alpha}$, with $\omega\in GL(1,\mathbb{C})$ a nonzero complex number. With the reference spinors held fixed, the photon polarizations scale as $\varepsilon^{\pm}\to\omega^{\pm2}\varepsilon^{\pm}$, and the graviton polarization tensors as $\varepsilon^{\pm\pm}\to\omega^{\pm4}\varepsilon^{\pm\pm}$. Thus the polarization tensor of helicity $h$ has weight $\omega^{2h}$ in this convention.

For a massive particle with momentum $p^{\mu}$, we have $\det[p_{\alpha\dot\alpha}]=m^{2}\neq0$; this rank-2 matrix can be written as $p_{\alpha\dot\alpha}= \lambda_{\alpha}^{I}\,\tilde\lambda_{\dot\alpha I}$, where the little group index $I=1,2$ can be raised and lowered using $\epsilon^{IJ}$ and $\epsilon_{IJ}$. In the \textbf{BOLD} notation we have the angle and square spinors 
\begin{eqnarray}
    |\boldsymbol{p}\rangle\equiv \lambda_{\alpha}^{I}, \quad|\boldsymbol{p}]\equiv\tilde\lambda^{I\dot\alpha}.
\end{eqnarray}
The massive spinors $\lambda_{\alpha}^{I}$ and $\tilde\lambda_{\dot\alpha I}$ are independent complex spinors satisfying the two-component Dirac equations: $p_{\alpha\dot\alpha}\,\tilde\lambda^{\dot\alpha I}=m\,\lambda_{\alpha}^{I}$ and $p^{\dot\alpha\alpha}\,\lambda_{\alpha }^I=m\,\tilde\lambda_{\dot\alpha}^{I}$.
The angle and bracket contractions of Eq.~\eqref{eq:anglebracket} also apply to those of massive spinors. The massive little group transformations are given by $\lambda_{\alpha}^{I}\rightarrow \lambda_{\alpha}^{J}\,W_{J}^{\;I}$ and $\tilde\lambda_{\dot\alpha I}\rightarrow (W^{-1})_{I}^{\;J}\,\tilde\lambda_{\dot\alpha J}$, where $W_{J}^{\;I}\in SL(2,\mathbb C)$.

For the spin-1 massive particle, the polarization vector reads
\begin{eqnarray}
    \varepsilon^{I_1I_2}_{\alpha\dot\alpha}=\frac{\lambda_{\alpha}^{\{I_1}\tilde\lambda_{\dot\alpha}^{I_2\}}}{m}=\frac{|\boldsymbol{p}\rangle\lbrack\boldsymbol{p}|}{m}.
\end{eqnarray}
The arbitrary spin-$S$ representation is obtained by taking the totally symmetric tensor product of $2S$ massive spinors:
\begin{eqnarray}\label{eq:relspins}
    \text{spin-$S$}:\quad \frac{1}{m^{S}}\,\lambda^{\{I_1}_{\alpha_1}\lambda^{I_2}_{\alpha_2}\cdots\tilde\lambda^{I_{2S-1}}_{\dot\alpha_{2S-1}}\tilde\lambda^{I_{2S}\}}_{\dot\alpha_{2S}}=\,\frac{1}{m^{S}}\underbrace{|\boldsymbol{p}\rangle|\boldsymbol{p}\rangle\cdots|\boldsymbol{p}\rbrack|\boldsymbol{p}\rbrack}_{2S}\,,
\end{eqnarray}
where the massive little group indices $\{I_1\cdots I_{2S}\}$ are totally symmetrized.

\subsection{Velocity-projected Spinors for Massive Particles}\label{subsec:vpspinor}

Similar to the relativistic spinors for a massive particle, the heavy particle can also be expressed in the spinor-helicity formalism. We derive the projected spinors from the large and small components of a massive Dirac field. This identifies the on-shell variables with the two-component fields used in HPET and fixes the relation needed to integrate out the small component.

\paragraph{Heavy Particle Fields}

The Dirac field contains the left- and right-handed Weyl spinors, transforming under the $(\frac{1}{2},0)$ and $(0,\frac{1}{2})$ representations of the Lorentz group, respectively. The four-component Dirac field is related to the massive spinor through the mode expansion
\begin{eqnarray}\label{eq:dirac}
    \Psi_{\bar\alpha}(x)=\int \frac{d^3\vec p}{(2\pi)^3\sqrt{2E}}\sum_I \left\{\left[\begin{array}{c}
         \lambda_{\alpha}^I  \\
         \tilde \lambda^{I\dot\alpha}
    \end{array}\right]a_{\vec p}^{I}\,e^{-\mathrm{i}px} + \left[\begin{array}{c}
         -\lambda_{\alpha}^I  \\
         \tilde \lambda^{I\dot\alpha}
    \end{array}\right]b_{\vec p}^{I\dagger}\,e^{\mathrm{i}px}\right\}\,,
\end{eqnarray}
where the spinor index $\bar \alpha$ denotes $\bar\alpha= (\alpha, \dot\alpha) = 1,2,3,4$, and $I$ is the little group index.

For residual energies well below the heavy-particle pair-production scale, modes separated by an energy of order $2m$ can be eliminated. This can be treated by projecting the Dirac field onto the large- and small-component spinor fields $Q_v(x)$ and $B_v(x)$:
\begin{eqnarray}
\label{Qvdef}
\Psi_{\bar\alpha}(x) =\sum_{v} e^{-\mathrm{i}mv\cdot x} \left[  \underbrace{e^{\mathrm{i}mv\cdot x}\left( \frac{1+{v\!\!\!/}}{2}\right)\Psi(x)}_{\equiv ~Q_{v}(x)} + \underbrace{e^{\mathrm{i}mv\cdot x}\left( \frac{1-{v\!\!\!/}}{2}\right)\Psi(x)}_{\equiv ~{B}_{v}(x)} \right]\,,
\end{eqnarray}
where the timelike reference momentum $v^\mu$ (with $v^2=1$) is introduced for the system. 
Extracting the phase factor $e^{\mathrm{i}mvx}$ out, the momentum of the heavy particle should be decomposed into a large part and a small residue momentum
\begin{eqnarray}\label{eq:pdecomp}
    p^{\mu}=mv^{\mu}+k^{\mu},\quad k\ll m, \quad \textrm{with RPI } (v^{\mu},~k^{\mu})\longleftrightarrow (v^{\mu}-\frac{q^{\mu}}{m},~k^{\mu}+q^{\mu})\,,
\end{eqnarray}
where the reparameterization invariance (RPI) is preserved under the Lorentz boost with parameter $q^{\mu}/m$.

After momentum decomposition, the mode expansion for the $Q_v(x)$ field takes the form
\begin{eqnarray}\label{eq:heavyQv}
    Q_{v,\bar\alpha}(x)=\int \frac{d^3\vec p}{(2\pi)^3\sqrt{2E}}  \sum_I\left\{\left[\frac{1+v\!\!\!/}{2}\right]_{\bar\alpha\bar\beta}\left[\begin{array}{c}
         \lambda_{\beta}^I  \\
         \tilde \lambda^{I\dot\beta}
    \end{array}\right]\,a_{\vec p}^{I}\,e^{-\mathrm{i}kx} + \left[\frac{1+v\!\!\!/}{2}\right]_{\bar\alpha\bar\beta}\left[\begin{array}{c}
         -\lambda_{\beta}^I  \\
         \tilde \lambda^{I\dot\beta}
    \end{array}\right]b_{\vec p}^{I\dagger}\,e^{\mathrm{i}(2mv +k)x}\,\right\}.
\end{eqnarray}
The second term $e^{\mathrm{i}(2mv +k)x}$ contains fast oscillation, and is thus suppressed compared to the first term $e^{-\mathrm{i}kx}$. Similar to the momentum $p_{\alpha \dot\alpha}$, we introduce the spinor form for the reference velocity $v^{\mu}$:
\begin{eqnarray}
   v_{\alpha\dot\alpha} \equiv v_{\alpha}^{L}\tilde v_{\dot\alpha L} = | v^L \rangle [v_L|, \qquad \det[v_{\alpha\dot\alpha}]=1.
\end{eqnarray} 
Thus the four-component wave function can be written in the spinor form
\begin{eqnarray}
\left[\frac{1+v\!\!\!/}{2}\right]_{\bar\alpha\bar\beta} \left[\begin{array}{c}
         \lambda_{\beta}^I  \\
         \tilde \lambda^{I\dot\beta}
    \end{array}\right] 
    = \frac{1}{2}
\left[\begin{array}{c}
         | v^L\rangle \ \left(\la\boldsymbol{p}^{I} v_{L}\ra+\lb v_{L}\boldsymbol{p}^{I}\rb \right)   \\
         | v^L ]\ \left(\la\boldsymbol{p}^{I} v_{L}\ra+\lb v_{L}\boldsymbol{p}^{I}\rb\right)
    \end{array}\right]     
\end{eqnarray}
This gives rise to two new two-component spinors
\begin{eqnarray}
    |\boldsymbol{p}_v\rangle = | v^L\rangle \ \frac{\left(\la\boldsymbol{p}^{I} v_{L}\ra+\lb v_{L}\boldsymbol{p}^{I}\rb \right)}{2}, \quad |\boldsymbol{p}_v ]= | v^L\rbrack \ \frac{\left(\la\boldsymbol{p}^{I} v_{L}\ra+\lb v_{L}\boldsymbol{p}^{I}\rb \right)}{2}, 
\end{eqnarray}
which is exactly the spinor form in Ref.~\cite{Aoude:2020onz}.
Since the two new spinors contain the same term, let us redefine the new spinor without the $| v^L\rangle$ or $| v^L]$ basis. This velocity-projected spinor ($v\!\!\!/$-spinor) contains two little group indices, and is defined with a new type of bracket
\begin{eqnarray}  |\boldsymbol{k}^+\rfloor_{L}^{I}&\equiv&\frac{1}{2}\Big(\la\boldsymbol{p}^{I} v_{L}\ra+\lb v_{L}\boldsymbol{p}^{I}\rb\Big)\,.
\end{eqnarray}
Compared with Ref.~\cite{Aoude:2020onz}, only a single spinor is introduced, instead of two, and furthermore, different spinor indices indicate different symmetry structures that can be utilized.

Let us investigate the physical meaning of this $v\!\!\!/$-spinor and its corresponding field.  Defining the four-component row vector $\bar \zeta_{L}^{\bar\alpha}=\frac{1}{2}\big[-\langle v_{L}|,\, \lbrack v_{L}|\,\big]$, projecting the $Q_v$ field, we obtain 
\begin{eqnarray}
    N_{v,L}(x) \equiv \bar \zeta_{L}^{\bar \alpha}\, Q_{v,\bar \alpha}(x) = \int \frac{d^3\vec k}{(2\pi)^3\sqrt{2E}}  \sum_I\left( |\boldsymbol{k}^+\rfloor_{L}^{I} \,a_{v,k}^{I}\,e^{-\mathrm{i}kx} +|\boldsymbol{k}^-\rfloor_L^I \,b_{v,k}^{I\dagger}\,e^{\mathrm{i}(2mv+k)x} \right)\,,
\end{eqnarray}
where the anti-particle wave function is $|\boldsymbol{k}^-\rfloor_L^I\equiv\frac{1}{2}\left(-\langle\boldsymbol{p}^Iv_L\rangle+\lbrack v_L\boldsymbol{p}^I\rbrack\right)$.
In the frame where the particle is at rest, the reference momentum takes the form $v_0^{\mu} = (1,0,0,0)$, and the large component spinor field $Q_v(x)$ is reduced to the non-relativistic field $N(x)$. In the NR theory, the spacetime symmetry is Euclidean, and thus the little group index $L$ is reduced to the $\ell$ index 
\begin{eqnarray}
    N_{v_0, \ell}(x) = \int \frac{d^3\vec k}{(2\pi)^3\sqrt{2E}}\sum_{I}\sqrt{m+E}\,\delta_{\ell}^{I}\,a_{v_0,k}^{I}\,e^{-\mathrm{i} kx}\,.
\end{eqnarray}
Therefore, $N_{v,L}(x)$ represents the heavy particle field with general $v$, after boosting the NR field to the $v$ frame. More discussion about the HPET wave function and field can be found in Appendix~\ref{app:HPET}.

\paragraph{$v\!\!\!/$-spinors for Spin-1/2 Heavy Particles}

After introducing the velocity-projected spinor, we are ready to present its properties in the spinor-helicity formalism.
The large-component $v\!\!\!/$-spinor of the heavy particle reads
\begin{eqnarray}\label{eq:spinhalfheavyspinor}  |\boldsymbol{k}^+\rfloor_{L}^{I}&\equiv&\frac{1}{2}\Big(\la\boldsymbol{p}^{I} v_{L}\ra+\lb v_{L}\boldsymbol{p}^{I}\rb\Big)\,,
\end{eqnarray}
while the small-component $v\!\!\!/$-spinor takes a simple form
\begin{eqnarray}\label{eq:spinhalfheavyspinor2}
    |\boldsymbol{k}^-\rfloor_{L}^{I}&\equiv& \frac{1}{2}\Big(-\la\boldsymbol{p}^{I} v_{L}\ra+\lb v_{L}\boldsymbol{p}^{I}\rb\Big)\,.
\end{eqnarray}
The $SU(2)$ anti-symmetric invariant tensor $\epsilon^{L_1L_2}$ raises the $L$ index in $|\boldsymbol{k}^{\pm}\rfloor_L^I$:
\begin{eqnarray}
    \lceil\boldsymbol{k}^{\pm}|^{IL}\equiv\epsilon^{LL'}|\boldsymbol{k}^{\pm}\rfloor^I_{L'}\,,
\end{eqnarray}
Here we refer to $\lceil\boldsymbol{k}^{\pm}|$ and $|\boldsymbol{k}^{\pm}\rfloor$ as the \textbf{ceil} and \textbf{floor} spinors, respectively.

If one would like to keep the small component, both $|\boldsymbol{k}^\pm\rfloor_{L}^{I}$ can be used in the spinor formalism. 
The large component scales as $|\boldsymbol{k}^+\rfloor\sim\sqrt{m}$, while the small component scales as $|\boldsymbol{k}^-\rfloor\sim k/\sqrt{m}$. 
When the small auxiliary component is kept, the RPI form (defined as RPI-I) of the $v\!\!\!/$-spinors is much simpler:
\begin{eqnarray}
    |\boldsymbol{k}^+\rfloor &\longrightarrow&|\boldsymbol{k}^+\rfloor- \delta v|\boldsymbol{k}^-\rfloor,\nonumber\\
        |\boldsymbol{k}^-\rfloor&\longrightarrow& |\boldsymbol{k}^-\rfloor- \delta v|\boldsymbol{k}^+\rfloor.
\end{eqnarray}
Simple RPI would simplify the calculation considerably. 

Both $v\!\!\!/$-spinors transform under the Lorentz transformation $\Lambda$ as
\begin{eqnarray}\label{eq:Wigneru}
 |\boldsymbol{k}^{\pm}\rfloor_L^I\,\rightarrow\, |\boldsymbol{k}^{\pm}\rfloor_L^J\,W(\Lambda p,p_0)_{J}^{~I},\quad 
 |\boldsymbol{k}^{\pm}\rfloor_L^I\,\rightarrow\, |\boldsymbol{k}^{\pm}\rfloor_{L'}^IW^{-1}(\Lambda p, v)^{L'}_{~L}\,,
\end{eqnarray}
where the Wigner rotations for the momenta $p^{\mu}_0$ and $v^{\mu}$ are $W(\Lambda p,p_0)=L(\Lambda p,p_0)^{-1} \Lambda\, L(p,p_0)$, and $W(\Lambda p,v)=L(\Lambda p,v)^{-1} \Lambda\, L(p,v)$, respectively.

In a heavy particle amplitude with ceil/floor spinors, the $L$ index is always contracted. Using the $SU(2)$ anti-symmetric invariant tensor $\epsilon^{L_1L_2}$, the \textbf{ceil--floor} contraction reads 
\begin{eqnarray}
    \lceil \boldsymbol{1}^+\boldsymbol{2}^+\rfloor\equiv\lceil\boldsymbol{1}^+|^{L_1}|\boldsymbol{2}^+\rfloor_{L_1}=\epsilon^{L_1L_2}|\boldsymbol{1}^+\rfloor_{L_2}|\boldsymbol{2}^+\rfloor_{L_1}\,,
\end{eqnarray}
which is the building block of heavy particle amplitudes.
It satisfies the antisymmetry relation
$\lceil \boldsymbol{1}^+\boldsymbol{2}^+\rfloor=- \lceil \boldsymbol{2}^+\boldsymbol{1}^+\rfloor$
and the Schouten identity
$\lceil\boldsymbol{1}^{+}\boldsymbol{4}^{+}\rfloor\lceil\boldsymbol{2}^{+}\boldsymbol{3}^{+}\rfloor=-\lceil\boldsymbol{1}^{+}\boldsymbol{2}^{+}\rfloor\lceil\boldsymbol{3}^{+}\boldsymbol{4}^{+}\rfloor+\lceil\boldsymbol{1}^{+}\boldsymbol{3}^{+}\rfloor\lceil\boldsymbol{2}^{+}\boldsymbol{4}^{+}\rfloor$.
These relations also hold for the auxiliary component $|\boldsymbol{k}^-\rfloor$.

On the other hand, the $I$ indices, representing the spin information of the particle, should not be contracted since scattering amplitudes transform covariantly under the little group. Under the Lorentz transformation, the ceil--floor contraction should transform by the little group rotation
\begin{eqnarray}
    \lceil \boldsymbol{1}^+\boldsymbol{2}^+\rfloor^{I_1I_2}\longrightarrow \left(W_1\right)^{I_1}_{~J_1}\left(W_2\right)^{I_2}_{~J_2}\lceil \boldsymbol{1}^+\boldsymbol{2}^+\rfloor^{J_1J_2}\,,
\end{eqnarray}
where the spin index defines the tensor notation for the spin space.

\paragraph{$v\!\!\!/$-spinors for Arbitrary Spin Heavy Particles}

From the tensor notation, higher-spin particles can be similarly defined. 
For the spin-1 heavy particle, the spin-1 $v\!\!\!/$-spinor can be written as 
\begin{eqnarray}
  \boldsymbol{ \varepsilon}(\boldsymbol{k}^+)^{\{I_1,I_2\}}_{\{L_1,L_2\}}\equiv  \frac{1}{2m}\left(|\boldsymbol{k}^+\rfloor^{I_1}_{L_1}|\boldsymbol{k}^+\rfloor^{I_2}_{L_2}+|\boldsymbol{k}^+\rfloor^{I_1}_{L_2}|\boldsymbol{k}^+\rfloor^{I_2}_{L_1}\right),
\end{eqnarray} 
where $I_1,I_2, L_1,L_2=1,2$, and the notation $\{I_1,\cdots,I_n\}$ means that the indices inside are totally symmetrized.
For general spin $S>0$, the $v\!\!\!/$-spinor is obtained from the symmetric product of $2S$ spin-$1/2$ variables, 
\begin{eqnarray}\label{eq:spinsk+}
    \boldsymbol{\varepsilon}(\boldsymbol{k}^+)^{\{I_1\cdots I_{2S}\}}_{\{L_1\cdots L_{2S}\}}:\quad\frac{1}{m^S}\,|\boldsymbol{k}^+\rfloor_{\{L_1}^{\{I_1}\,|\boldsymbol{k}^+\rfloor_{L_2}^{I_2}\cdots|\boldsymbol{k}^+\rfloor_{L_{2S-1}}^{I_{2S-1}}\, |\boldsymbol{k}^+\rfloor_{L_{2S}\}}^{I_{2S}\}}\,,
\end{eqnarray} with the little group indices $I$ and $L$ totally symmetric. These symmetrizations ensure that the spinors transform under the irreducible representations of the $SU(2)$ little groups. For brevity, we suppress the heavy-particle little group index $I$ in the $\textbf{BOLD}$ notation below.

\paragraph{Residual Momentum and Equations of Motion}

The momenta $p^{\mu}$ and $k^{\mu}$ can also be projected by the $v$-spinor. The projected momentum tensor is defined as
\begin{eqnarray}
   p_{LL'}= \la v_{L}|p_{\alpha\dot\alpha}| v_{L'}]
    = \la v_{L}\boldsymbol{p}^I\ra [\boldsymbol{p}_I v_{L'}]\,.
\end{eqnarray}
In the tensor analysis, a general tensor can always be decomposed into symmetric and anti-symmetric parts $p^{\mu}=p_{\perp}^{\mu}+(v\cdot p)v^{\mu}$, and thus similarly for $p_{LL'}$  
\begin{eqnarray}
    (p_{\perp})_{LL'}=\la v_{L}|p_{\perp}| v_{L'}]=\frac{1}{2}\left(p_{LL'} + p_{L'L}\right), \quad 
    v\cdot p\, \epsilon_{LL'}=\frac{1}{2}\left(p_{LL'} - p_{L'L}\right)\,.
\end{eqnarray}
Rewriting the momentum $p^{\mu} = mv^{\mu} + k^{\mu}$, we can also decompose the residual momentum $k^{\mu}$ as $k^{\mu}=k_{\perp}^{\mu}+(v\cdot k)v^{\mu}$: 
\begin{eqnarray} \label{eq:ktij}
(k_{\perp})_{LL'}\equiv \la v_{L}|k_{\perp}| v_{L'}] = (p_{\perp})_{LL'}, \quad 
   (v\cdot k)\epsilon_{LL'} =  (v\cdot p -m)\epsilon_{LL'}
\end{eqnarray}
They can be converted into the ceil and floor spinors by
\begin{eqnarray}   |\boldsymbol{k}^+\rfloor^I_{L}\lceil\boldsymbol{k}^+|_{IL'}=\left(m + \frac{v\cdot k}{2}\right)\epsilon_{LL'}, \quad 
|\boldsymbol{k}^-\rfloor^I_{L}\lceil\boldsymbol{k}^-|_{IL'}=-\frac{v\cdot k}{2}\epsilon_{LL'}, \quad 
|\boldsymbol{k}^+\rfloor^I_{L}\lceil\boldsymbol{k}^-|_{IL'}= \frac{(k_{\perp})_{LL'}}{2}\,.
\end{eqnarray}

Projecting the massive Dirac equations of Sec.~\ref{subsec:rspinor} onto the velocity spinors gives the equations of motion for the heavy particle,
\begin{equation}\label{eq:EOMfermion}
   \begin{array}{rcl}
      p|\boldsymbol{p}\rangle  &=&m |\boldsymbol{p}], \\[3pt]
      p|\boldsymbol{p}]  & =&m|\boldsymbol{p}\rangle,
   \end{array} \quad\Longrightarrow\quad\begin{array}{rcl}
         v\cdot k|\boldsymbol{k}^{+}\rfloor&=&k_{\perp}|\boldsymbol{k}^{-}\rfloor,\\[3pt]
     (2m+v\cdot k)|\boldsymbol{k}^{-}\rfloor&=&k_{\perp}|\boldsymbol{k}^{+}\rfloor.
    \end{array}
\end{equation}
Here $k_{\perp}|\boldsymbol{k}^{+}\rfloor \equiv (k_{\perp})_{L}^{~L'}|\boldsymbol{k}^{+}\rfloor_{L'}=-(k_{\perp})_{LL'}|\boldsymbol{k}^{+}\rfloor^{L'}$. The second equation of motion determines the small component in terms of the large one,
\begin{eqnarray}\label{eq:integrateoutk-spinor}
    |\boldsymbol{k}^-\rfloor=\frac{1}{2m+v\cdot k}k_{\perp}|\boldsymbol{k}^+\rfloor \quad\Longleftrightarrow\quad \tilde N(x)=-\frac{1}{2m+\mathrm{i}\partial_t}\mathrm{i}\vec\sigma\cdot \vec\nabla N(x).
\end{eqnarray}
The operator on the right-hand side is nonlocal and admits a local-operator expansion in powers of $v\cdot k/(2m)$, $|\boldsymbol{k}^-\rfloor=\frac{1}{2m}\sum_{n=0}^{\infty}\left(-\frac{v\cdot k}{2m}\right)^{n}k_{\perp}|\boldsymbol{k}^+\rfloor$. After integrating out the small component, the equation of motion for the large component $|\boldsymbol{k}^+\rfloor$ becomes
\begin{eqnarray}
v\cdot k|\boldsymbol{k}^+\rfloor=k_{\perp}\frac{1}{2m}\left(\sum_{n=0}^{\infty}(-1)^n(\frac{v\cdot k}{2m})^n\right)k_{\perp}|\boldsymbol{k}^+\rfloor,\nonumber
\end{eqnarray}
whose leading-order contribution reduces to the Schr\"odinger equation $v\cdot k|\boldsymbol{k}^+\rfloor=\frac{1}{2m}k_{\perp}k_{\perp}|\boldsymbol{k}^+\rfloor$, corresponding to $\mathrm{i}\partial_t N(x)=-\frac{1}{2m}\nabla^2 N(x)$.

\paragraph{Advantages of $v\!\!\!/$-spinors}
Before proceeding, we explain why the 
$v\!\!\!/$-spinors are advantageous. Unlike the conventional spinors, 
$v\!\!\!/$-spinors carry only little group indices, have simpler large and small components, and simplify reparameterization invariance. These features can be summarized as follows:
\begin{itemize}

    \item The $v\!\!\!/$-spinor contains the little group indices only and is consistent with the non-relativistic spinor definition, while the other is obtained by acting on the $v^{\mu}$ and thus contains the $SL(2,\mathbb{C})$ indices with angle and square brackets
\begin{eqnarray}
    \left[\begin{array}{c}
         |\boldsymbol{p}_v^I\rangle  \\
         |\boldsymbol{p}_v^I\rbrack 
    \end{array}\right]=\left[\begin{array}{c}
        |v^L\rangle\\
        | v^L\rbrack
    \end{array}\right]\, |\boldsymbol{k}^+\rfloor^I_L.
\end{eqnarray}
Thus, although both describe the large component of the heavy state, the $v\!\!\!/$-spinor is simpler with its manifest Euclidean symmetry.  Amplitudes in this formulation carry the same symmetry and power counting as an HPET operator, so that they map directly onto the corresponding operators.

\item Given the large component, the small component is usually obtained via the integrate-out version of the HPET equation of motion 
\begin{eqnarray}
    |\boldsymbol{k}^-\rfloor=\frac{1}{2m+v\cdot k}k_{\perp}|\boldsymbol{k}^+\rfloor, \qquad
    |\boldsymbol{p}_-^I\rangle=\frac{1}{2m+v\cdot k}\big(p|\boldsymbol{p}_v^I\rbrack-v\cdot p|\boldsymbol{p}_v^I\rangle\big). 
\end{eqnarray}
On the other hand, using the $v\!\!\!/$-spinor there is a simpler form for the small component, which is orthogonal to the large component
\begin{eqnarray}
    |\boldsymbol{k}^-\rfloor_{L}^{I}\equiv \frac{1}{2}\Big(-\la\boldsymbol{p}^{I} v_{L}\ra+\lb v_{L}\boldsymbol{p}^{I}\rb\Big).
\end{eqnarray}
Thus, if one would like to keep the small component, both $|\boldsymbol{k}^\pm\rfloor_{L}^{I}$ can be used in the spinor formalism. 
The large component scales as $|\boldsymbol{k}^+\rfloor\sim\sqrt{m}$, while the small component scales as $|\boldsymbol{k}^-\rfloor\sim k/\sqrt{m}$. Contracting all Lorentz indices with the reference spinors, including those of the soft bosons, makes the expansion in $k/m$ explicit for every degree of freedom.

\item Finally, there should be the reparameterization invariance for the large component, defined as RPI-II: 
\begin{eqnarray}\label{eq:audorpi}
|\mathbf{p}_v\rangle &= \left(1 - \frac{(k-2m\delta v)^2}{4m^2}\right)^{-1} \left[ \left(1 - \frac{k^2}{4m^2} + \frac{k\!\!\!/\delta v\!\!\!/}{2m}\right) |\mathbf{p}_w\rangle - \delta v\!\!\!/ |\mathbf{p}_w] \right], \nonumber\\
|\mathbf{p}_v] &= \left(1 - \frac{(k-2m\delta v)^2}{4m^2}\right)^{-1} \left[ \left(1 - \frac{k^2}{4m^2} + \frac{k\!\!\!/\delta v\!\!\!/}{2m}\right) |\mathbf{p}_w] - \delta v\!\!\!/ |\mathbf{p}_w\rangle \right].
\end{eqnarray}
We find that, if the small auxiliary component is kept, the reparameterization of the $v\!\!\!/$-spinors is much simpler, defined as RPI-I:
\begin{eqnarray}\label{eq:rpiI}
    |\boldsymbol{k}^+\rfloor &\longrightarrow&|\boldsymbol{k}^+\rfloor- \delta v|\boldsymbol{k}^-\rfloor,\nonumber\\
        |\boldsymbol{k}^-\rfloor&\longrightarrow& |\boldsymbol{k}^-\rfloor- \delta v|\boldsymbol{k}^+\rfloor.
\end{eqnarray}
This form is used in the bottom-up construction of Section~\ref{sec:threep}.  The choice of $v^{\mu}$
makes boosts non-manifest rather than removing Lorentz symmetry: reparameterization invariance relates
different decompositions of the same physical momentum, and imposing it restores the underlying Lorentz
symmetry of the rotationally invariant amplitudes.

\end{itemize}
The detailed comparison with the literature can be found in Appendix~\ref{subsec:comp}.

\subsection{Projection of Relativistic Spinors}\label{subsec:expand}
\paragraph{Projection of the AHH Spinors}
In terms of the heavy particle $v\!\!\!/$-spinors $|\boldsymbol{k}^+\rfloor$ and $|\boldsymbol{k}^-\rfloor$ that we introduced in the previous subsection, we can directly project the relativistic AHH spinors:
\begin{eqnarray}\label{eq:expandAHH}
|\boldsymbol{p}\rangle=|v_L\rangle\langle v^L\boldsymbol{p}\rangle=|v^L\rangle\left(|\boldsymbol{k}^+\rfloor_{L}-|\boldsymbol{k}^-\rfloor_{L}\right) ,\quad
|\boldsymbol{p}\rbrack=|v^L\rbrack\lbrack v_L\boldsymbol{p}\rbrack=|v^L\rbrack\left(|\boldsymbol{k}^+\rfloor_{L}+|\boldsymbol{k}^-\rfloor_{L}\right).
\end{eqnarray}
The spin-1 polarization vector can be projected as
\begin{eqnarray}
    \frac{|\boldsymbol{p}\rangle\lbrack\boldsymbol{p}|}{m}=\frac{|v^{L_1}\rangle\lbrack v^{L_2}|}{m}\left(|\boldsymbol{k}^+\rfloor_{L_1}-|\boldsymbol{k}^-\rfloor_{L_1}\right)\left(|\boldsymbol{k}^+\rfloor_{L_2}+|\boldsymbol{k}^-\rfloor_{L_2}\right).
\end{eqnarray}
The arbitrary spin-$S$ massive polarization vector in Eq.~\eqref{eq:relspins} can be projected as
\begin{eqnarray}
 \text{spin-}S:\quad\frac{1}{m^S}\,\underbrace{|\boldsymbol{p}\rangle  |\boldsymbol{p}\rangle \cdots|\boldsymbol{p}\rbrack  |\boldsymbol{p}\rbrack}_{2S}= \frac{1}{m^S}\,|v^{L_1}\rangle\cdots|v^{L_{2S}}\rbrack\left(|\boldsymbol{k}^+\rfloor_{L_1}-|\boldsymbol{k}^-\rfloor_{L_1}\right)\cdots\left(|\boldsymbol{k}^+\rfloor_{L_{2S}}+|\boldsymbol{k}^-\rfloor_{L_{2S}}\right),
\end{eqnarray}
where the leading contribution reduces to the heavy particle state $\boldsymbol{\varepsilon}(\boldsymbol{k}^+)$ in Eq.~\eqref{eq:spinsk+}.

\paragraph{Projection of the Massless Spinors}
We now extend the velocity-projection formalism to the massless spinors describing soft bosons coupled to the heavy particle.

Generally, the relativistic massless spinors $|q\rangle$ and $|q\rbrack$ can be expanded by
\begin{eqnarray}\label{eq:expandmassless}
    |q\rangle=|v_L\rangle\langle v^Lq\rangle= |v^L\rangle\, |q\rfloor_L 
 ,\quad |q\rbrack=|v^L\rbrack\lbrack v_Lq\rbrack=|v^L\rbrack \,|\tilde q\rfloor_L,
\end{eqnarray}
where we contract the massless spinors with $|v_L\rangle$ and $|v_L\rbrack$, and define the \textbf{floor} $v\!\!\!/$-spinors for massless particles,
\begin{eqnarray}
    |q\rfloor_{ L}  \equiv \langle q  v_{ L} \rangle,\quad
   |\tilde q\rfloor_L \equiv[v_{L} \tilde{q}],\label{eq:qfloordef}
\end{eqnarray}
as well as the \textbf{ceil} spinors $\lceil q|^L\equiv\epsilon^{LL'}|q\rfloor_{L'}$ and $\lceil \tilde q|^L\equiv\epsilon^{LL'}|\tilde q\rfloor_{L'}$.
These $v\!\!\!/$-spinors are built solely from the gauge-invariant spinors, so they are independent of the choice of the polarization reference spinors $\xi$.

For the $|h|=1$ photon, the polarization vectors $\varepsilon_{\alpha\dot\alpha}^{\pm}$ in Eq.~\eqref{eq:photonepsilon} are projected by
\begin{eqnarray}
    \varepsilon^{+}_{LL'}=\langle v_L|^{\alpha}\varepsilon^{+}_{\alpha\dot\alpha}|v_{L'}\rbrack^{\dot\alpha}=\frac{\langle v_L\xi\rangle\lbrack qv_{L'}\rbrack}{\langle\xi q\rangle},\quad \varepsilon^{-}_{LL'}=\langle v_L|\varepsilon^{-}|v_{L'}\rbrack=\frac{\langle v_L q\rangle\lbrack\xi v_{L'}\rbrack}{\lbrack q\xi\rbrack}.
\end{eqnarray}
Similar to the momentum $p_{LL'}$, the symmetric parts of these vectors $\varepsilon^{\pm}_{LL'}$ are the non-trivial irreducible representations under the $SU(2)$ group of $v^{\mu}$:
\begin{eqnarray}
    \varepsilon^+_{\{LL'\}}=\frac{\varepsilon^+_{LL'}+\varepsilon^+_{L'L}}{2}\equiv \frac{|\xi\rfloor_L\,|\tilde q\rfloor_{L'}+|\xi\rfloor_{L'}\,|\tilde q\rfloor_{L}}{2\lceil q\xi\rfloor},\quad
    \varepsilon^-_{\{LL'\}}\equiv \frac{|\xi\rfloor_L\,| q\rfloor_{L'}+|\xi\rfloor_{L'}\,| q\rfloor_{L}}{2\lceil \tilde q\xi\rfloor}.
\end{eqnarray}

Specifically, we can fix the gauge by aligning the reference spinors with the heavy-particle velocity $v^{\mu}$. We construct $\xi$ and $\tilde\xi$ from the velocity spinors $v_{\alpha}^{L}, \tilde v_{\dot\alpha,L}$ and the photon spinors $q_{\alpha}$, $\tilde q_{\dot\alpha}$, $ \la\xi|= -\lb \tilde q  v^{L}\rb\la  v_{L}| ,\quad |\tilde\xi\rbrack=-|  v^{L}]\la  v_{L}q\ra$. With this gauge choice, the photon polarization vectors have simple forms:
\begin{eqnarray}\label{eq:polarizationdef}
     \varepsilon^+_{\{LL'\}} = \frac{|\tilde q\rfloor_L|\tilde q\rfloor_{L'}}{2\,v\cdot q}  ,\quad
    \varepsilon^-_{LL'} =   \frac{ |q\rfloor_{ L}  |q\rfloor_{ L'}}{2\,v\cdot q} \,  ,
\end{eqnarray}
where the normalization factor $\lceil q \tilde{q} \rfloor \equiv \langle q  v^{L} \rangle [v_{L} \tilde{q}] = 2 v \cdot q$ is a rotational scalar.  Note that in the bottom-up construction of $SU(2)$-invariant amplitudes, this scalar factor can often be absorbed into Wilson coefficients that are determined by matching.  This gauge choice requires $v\cdot q\neq0$; at exceptional complex kinematics one must use another reference spinor. 

For the $|h|=2$ graviton, the polarization vectors can be similarly projected as
\begin{eqnarray}
    \varepsilon^{++}_{\{L_1L_2L_3L_4\}}=\varepsilon^+_{\{L_1L_2}\varepsilon^+_{L_3L_4\}},\quad \varepsilon^{--}_{\{L_1L_2L_3L_4\}}=\varepsilon^-_{\{L_1L_2}\varepsilon^-_{L_3L_4\}}.
\end{eqnarray}
Here the four indices $L_1, L_2, L_3$ and $L_4$ are totally symmetrized.

Furthermore, the momentum of the massless leg admits a decomposition,
\begin{eqnarray}\label{eq:qperp}
    \langle  v_{L} | q | v_{L'} ] = (v\cdot q)\epsilon_{LL'} + (q_{\perp})_{LL'},\quad (q_{\perp})_{LL'}= \langle  v_{L} | q_{\perp} | v_{L'} ]=\frac{1}{2}\Big(|q\rfloor_L|\tilde q\rfloor_{L'}+|q\rfloor_{L'}|\tilde q\rfloor_{L}\Big),
\end{eqnarray}
where the first term $v\cdot q=v_{\mu}q^{\mu}$ represents the time component in the $v^{\mu}$ frame and the second term corresponds to the spatial triplet component in the $v^{\mu}$ frame.
The on-shell conditions $q|q\rangle=q|q]=0$, projected onto the velocity spinors with the decomposition in Eq.~\eqref{eq:qperp}, yield the equations of motion
\begin{eqnarray}\label{eq:EOMphoton}
    \begin{array}{rll}
         (v\cdot q)|\tilde q\rfloor &=&q_{\perp}|\tilde q\rfloor,  \\
         -(v\cdot q)|q\rfloor&=&q_{\perp}|q\rfloor.
    \end{array}
\end{eqnarray}
Here $q_{\perp}|q\rfloor\equiv(q_{\perp})_{L}^{~L'}|q\rfloor_{L'}=- (q_{\perp})_{LL'}|q\rfloor^{L'}$. This completes the treatment of the heavy and soft sectors within the $SU(2)$ $v\!\!\!/$-spinor framework.

\section{Three-point Amplitudes}\label{sec:threep}

\begin{figure}
    \centering
    \includegraphics[width=12cm, height=8cm]{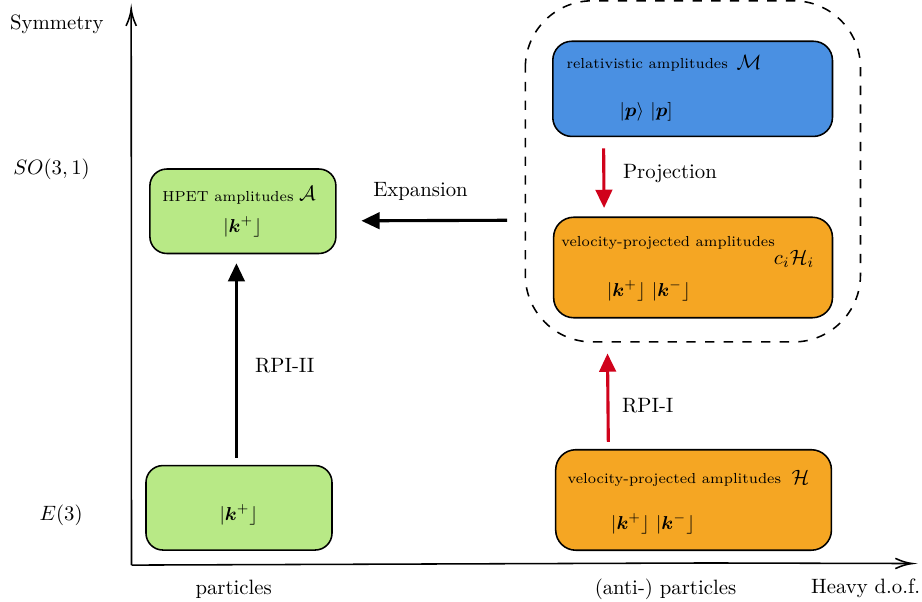}
    \caption{Workflow for deriving the HPET amplitudes. Top-down: relativistic amplitudes are projected onto velocity-projected amplitudes and expanded in $1/m$ to obtain the HPET amplitudes. Bottom-up: the same HPET amplitudes are obtained by imposing RPI-I on the velocity-projected amplitudes and expansion.}
    \label{fig:workflow3amps}
\end{figure}

In this section, we use the $v\!\!\!/$-spinors of Section~\ref{sec:spinor} to construct the three-particle amplitudes involving two heavy particles of spin $S$ and one soft massless boson of helicity $h$. Before proceeding, we first distinguish three kinds of amplitudes, then introduce two operations relating them, and finally present two separate routes to the HPET amplitudes. The workflow is exhibited in Fig.~\ref{fig:workflow3amps}.

Following the above spinor definitions, we distinguish three kinds of amplitudes:
\begin{enumerate}
   \item \emph{Relativistic Amplitudes} $\mathcal{M}^{(S,S,h)}$: the Lorentz-invariant amplitudes composed of relativistic spinors such as $|\boldsymbol{p}\rangle$ and $|q\rangle$.
    \item \emph{Velocity-projected Amplitudes}  $\mathcal{H}^{(S,S,h)}$: the ceil-floor contractions built from   $|\boldsymbol{k}^{\pm}\rfloor$ and $|q\rfloor, |\tilde q\rfloor$.   In particular, unlike the traditional amplitudes, it contains the small component $|\boldsymbol{k}^-\rfloor$.
    \item \emph{HPET amplitudes} $\mathcal{A}^{(S,S,h)}$: the integrated-out amplitudes obtained after expanding $|\boldsymbol{k}^-\rfloor$ in terms of $|\boldsymbol{k}^+\rfloor$ with $1/m$ suppression.
\end{enumerate}
These three kinds of amplitudes are related by two operations:
\begin{itemize}
     \item \emph{Projection}: directly project the relativistic amplitudes onto the velocity-projected amplitudes,
     \begin{eqnarray} \mathcal{M}^{(S,S,h)}\quad=\quad\sum_{i}c_i\  \mathcal{H}^{(S,S,h)}_i ,
    \end{eqnarray}
     by using the reference spinor $v_{\alpha\dot\alpha}=|v^L\rangle\lbrack v_L|$.
     On the other hand, the \emph{inverse operation of projection} is imposing RPI-I on the velocity-projected amplitudes $\mathcal{H}^{(S,S,h)}$, which gives
    \begin{eqnarray}   \mathcal{M}^{(S,S,h)}\quad=\quad\sum_{i}\underbrace{c_i(v,k,q)}_{\text{determined by RPI-I}}\times\underbrace{\mathcal{H}^{(S,S,h)}_i}_{\text{determined by little group covariance}}.
    \end{eqnarray}
    Here RPI-I refers to the reparameterization involving both $|\boldsymbol{k}^+\rfloor$ and $|\boldsymbol{k}^-\rfloor$, as in Eq.~\eqref{eq:rpiI}.

    \item \emph{Expansion}: directly expand the velocity-projected amplitudes in terms of the HPET amplitudes,
    \begin{eqnarray}  \mathcal{H}^{(S,S,h)}\quad=\quad\sum_{n}\frac{1}{m^n}\mathcal{A}^{(S,S,h)}_{(n)}\quad=\quad \mathcal{A}^{(S,S,h)}_{(0)}+\frac{1}{m}\mathcal{A}^{(S,S,h)}_{(1)}+\cdots \,,
    \end{eqnarray}
    where the expansion is taken by replacing $|\boldsymbol{k}^-\rfloor=\frac{1}{2m+v\cdot k}k_{\perp}|\boldsymbol{k}^+\rfloor$.
    On the other hand, the inverse operation of expansion is imposing RPI-II involving only the variation of $|\boldsymbol{k}^+\rfloor$, as in Eq.~\eqref{eq:audorpi}, on the HPET amplitudes. However, we do not use it in this work because of its complexity.
\end{itemize}
After defining these operations, there are two separate approaches to obtaining the HPET amplitudes used in this work:
\begin{itemize}
    \item top-down approach:
\begin{eqnarray}
\mathcal{M}^{(S,S,h)}\xrightarrow{\text{projection}}\sum_ic_i\mathcal{H}_i^{(S,S,h)}\xrightarrow{\text{expansion}}\sum_n\frac{1}{m^n}\mathcal{A}_{(n)}^{(S,S,h)}.
\end{eqnarray}
In Subsection~\ref{sec:expandrel} we project the most general relativistic three-particle amplitudes of Ref.~\cite{Arkani-Hamed:2017jhn} onto the velocity-projected amplitudes. In Subsection~\ref{sec:integrateoutamp} we expand the small component using its equation of motion and obtain the HPET amplitudes with explicit $1/m$ power counting.
\item bottom-up approach:
\begin{eqnarray}
    \mathcal{H}^{(S,S,h)}\xrightarrow{\text{RPI-I}} \sum_ic_i\mathcal{H}^{(S,S,h)}_i\xrightarrow{\text{expansion}}\sum_n\frac{1}{m^n}\mathcal{A}_{(n)}^{(S,S,h)}.
\end{eqnarray}
In Subsection~\ref{sec:bottomup}, the same HPET amplitudes are obtained from the $SU(2)$ little group and RPI-I. We find that the HPET amplitudes $\mathcal{A}_{(n)}^{(S,S,h)}$ correspond to HPET operators, giving the amplitude-operator correspondence.
\end{itemize}

The HPET amplitudes, including the 3-point and higher-point amplitudes, can be systematically expanded on the HPET operator basis $\mathcal{O}$. 
Appendix~\ref{app:aoc} organizes the HPET amplitude-operator correspondence through the $SU(2)$ semi-standard Young tableaux of Ref.~\cite{Li:2026eym}.

\subsection{Projection of  Relativistic Amplitudes}\label{sec:expandrel}

In this subsection we start from the most general relativistic three-particle amplitudes of Ref.~\cite{Arkani-Hamed:2017jhn}, and then we expand them in terms of the $v\!\!\!/$-spinors of Section~\ref{sec:spinor}.

As an illustration, consider the three-particle amplitude for the spin-1/2 fermions and a photon. For each helicity of the photon, the two independent relativistic amplitudes in Ref.~\cite{Arkani-Hamed:2017jhn} read
\begin{eqnarray}\label{eq:1/2minrel3pt}
       \mathcal{M}^{(\frac{1}{2},\frac{1}{2},+1)}&=&g_0\,x\la\textbf{12}\ra+g_1~\frac{1}{m}\lb\textbf{1}3\rb\lb\textbf{2}3\rb,
\nonumber\\
    \mathcal{M}^{(\frac{1}{2},\frac{1}{2},-1)}&=&g_0\,\frac{1}{x}\lb\textbf{12}\rb+g_1~\frac{1}{m}\la\textbf{1}3\ra\la\textbf{2}3\ra.
\end{eqnarray}
The gauge-invariant $x$ factor is introduced, which carries the helicity weight of the soft massless leg and is defined by its action on the massless spinors:
\begin{eqnarray}\label{eq:defx}
    x|3\ra=\frac{p_2}{m}|3\rb,\quad x^{-1}|3\rb=\frac{p_2}{m}|3\ra.
\end{eqnarray}
The first term is the minimal coupling of the heavy particle to the photon, while the second term, which carries two spinor factors on the photon leg, is the non-minimal coupling through the photon field strength. From the definition of the $x$ factor in Eq.~\eqref{eq:defx}, the massive spinor contractions satisfy the redundancy relations $\la\textbf{1}3\ra=-\frac{1}{x}\lb\textbf{1}3\rb$, $\la\textbf{2}3\ra=-\frac{1}{x}\lb\textbf{2}3\rb$ and $\lb\textbf{12}\rb=-\la\textbf{12}\ra-\frac{1}{mx}\lb\textbf{1}3\rb\lb\textbf{2}3\rb$, which translate between the angle and square bracket bases. In terms of the Dirac field $\Psi$, the photon vector potential $A^{\mu}$ and the field strength $F^{\mu\nu}$, the $g_0$ terms correspond to the configuration space operator $\bar \Psi A\!\!\!/\Psi$, while the $g_1$ terms correspond to the effective operator $\frac{1}{m}\bar \Psi \sigma_{\mu\nu}F^{\mu\nu}\Psi$.

The massive and massless spinors in $\mathcal{M}^{(\frac{1}{2},\frac{1}{2},\pm1)}$ are expanded in terms of the $v\!\!\!/$-spinors by Eqs.~\eqref{eq:expandAHH} and \eqref{eq:expandmassless}.
Each relativistic contraction in Eq.~\eqref{eq:1/2minrel3pt} separates into four terms, one for each sign combination of the large and small components of the two heavy legs, and the amplitudes are written in terms of the $SU(2)$ ceil-floor contraction:
\begin{equation}\label{eq:minimal3ptx12p1}
    \begin{cases}
    x\la\textbf{12}\ra  &= x\left(\lceil\boldsymbol{1}^+\boldsymbol{2}^+\rfloor-\lceil\boldsymbol{1}^+\boldsymbol{2}^-\rfloor-\lceil\boldsymbol{1}^-\boldsymbol{2}^+\rfloor+\lceil\boldsymbol{1}^-\boldsymbol{2}^-\rfloor\right),  \\
    \\
    \frac{1}{x}\lb\textbf{12}\rb &=\frac{1}{x}\left(-\lceil\boldsymbol{1}^+\boldsymbol{2}^+\rfloor-\lceil\boldsymbol{1}^+\boldsymbol{2}^-\rfloor-\lceil\boldsymbol{1}^-\boldsymbol{2}^+\rfloor-\lceil\boldsymbol{1}^-\boldsymbol{2}^-\rfloor\right), \\
    \\
\frac{1}{m}\lb\textbf{1}3\rb\lb\textbf{2}3\rb&=\frac{1}{m}\lceil\boldsymbol{1}^+\tilde3\rfloor\lceil\boldsymbol{2}^+\tilde3\rfloor +
\frac{1}{m}\lceil\boldsymbol{1}^+\tilde3\rfloor\lceil\boldsymbol{2}^-\tilde3\rfloor+
\frac{1}{m}\lceil\boldsymbol{1}^-\tilde3\rfloor\lceil\boldsymbol{2}^+\tilde3\rfloor+
\frac{1}{m}\lceil\boldsymbol{1}^-\tilde3\rfloor\lceil\boldsymbol{2}^-\tilde3\rfloor,\\
\\
              \frac{1}{m}\la\textbf{1}3\ra\la\textbf{2}3\ra&=\frac{1}{m}\lceil\boldsymbol{1}^+3\rfloor\lceil\boldsymbol{2}^+3\rfloor -
                \frac{1}{m}\lceil\boldsymbol{1}^+3\rfloor\lceil\boldsymbol{2}^-3\rfloor-
                \frac{1}{m}\lceil\boldsymbol{1}^-3\rfloor\lceil\boldsymbol{2}^+3\rfloor+
                \frac{1}{m}\lceil\boldsymbol{1}^-3\rfloor\lceil\boldsymbol{2}^-3\rfloor.
\end{cases}
\end{equation}
These expansions in fact convert the left- and right-handed spinors into two-component Pauli spinors in the non-relativistic case.
The correspondence to the bilinear fields can be seen as
\begin{eqnarray}
    \frac{1}{2}\Big(\la\textbf{12}\ra-\lbrack\boldsymbol{12}\rbrack\Big)=\lceil\boldsymbol{1}^+\boldsymbol{2}^+\rfloor+\lceil\boldsymbol{1}^-\boldsymbol{2}^-\rfloor& \Longleftrightarrow&  \bar\Psi(x)\Psi(x)= N^{\dagger}(x)N(x)-\tilde N^{\dagger}(x)\tilde N(x),\nonumber\\
    -\frac{1}{2}\Big(\la\textbf{12}\ra+\lbrack\boldsymbol{12}\rbrack\Big)=\lceil\boldsymbol{1}^+\boldsymbol{2}^-\rfloor+\lceil\boldsymbol{1}^-\boldsymbol{2}^+\rfloor&\Longleftrightarrow& \bar\Psi(x)\gamma_5\Psi(x)= N^{\dagger}(x)\tilde N(x)-\tilde N^{\dagger}(x)N(x).
\end{eqnarray}
For example, the large component contraction satisfies
$\lceil\boldsymbol{1}^+\boldsymbol{2}^+\rfloor\sim N^{\dagger}_1(x)N_2(x)$,
whereas the small-component contraction satisfies
$\lceil\boldsymbol{1}^-\boldsymbol{2}^-\rfloor\sim-\tilde N^{\dagger}_1(x)\tilde N_2(x)$.
The relative minus sign arises from Hermitian conjugation of the small component. 
Besides, the contractions $\la\textbf{1}3\ra$ and $\lb\textbf{1}3\rb$ can be related to the electric and magnetic fields, $\vec{B}\sim |q\rfloor|q\rfloor+|\tilde q\rfloor|\tilde q\rfloor$ and $\vec E\sim \mathrm{i}\left(|q\rfloor|q\rfloor-|\tilde q\rfloor|\tilde q\rfloor\right)$. For example, a dimension-five HPET operator is
$\lceil\boldsymbol{1}^+3\rfloor\lceil3\boldsymbol{2}^+\rfloor+\lceil\boldsymbol{1}^+\tilde3\rfloor\lceil\tilde3\boldsymbol{2}^+\rfloor \Longleftrightarrow N_1^{\dagger}(x)\vec{\sigma}\cdot\vec{B}_3(x)N_2(x)$.

The arbitrary-spin amplitude is the symmetric product of $2S$ copies of the spin-$1/2$ structures in Eq.~\eqref{eq:1/2minrel3pt}, so for spin $S$ and helicity $h$ the most general relativistic three-particle amplitude in Ref.~\cite{Arkani-Hamed:2017jhn} reads
\begin{equation}\label{eq:r3ptgeneralangle}
    \mathcal{M}^{(S,S,h)}=\sum_{i=0}^{2S}g_ix^{h-2S}\left(x\la\textbf{12}\ra\right)^{2S-i}\left(\frac{1}{m}\lb\textbf{1}3\rb\lb\textbf{2}3\rb\right)^i.
\end{equation}
There are $2S+1$ independent structures in total, with the $g_i$ being undetermined coupling constants. The $i$-th term carries $i$ insertions of the non-minimal structure $\frac{1}{m}\lb\textbf{1}3\rb\lb\textbf{2}3\rb$ and describes the $i$-th non-minimal coupling of the heavy particle. Equivalently, by the redundancy relations above, the same amplitude is expressed in the square bracket form, $\mathcal{M}^{(S,S,h)}=\sum_{i=0}^{2S}g_ix^{h+2S}\left(\frac{1}{x}\lb\textbf{12}\rb\right)^{2S-i}\left(\frac{1}{m}\la\textbf{1}3\ra\la\textbf{2}3\ra\right)^i$, where the helicity weight is distributed differently between the factors. The velocity projection of both forms follows by replacing each relativistic contraction with its expansion in Eq.~\eqref{eq:minimal3ptx12p1}. After expanding Eq.~\eqref{eq:r3ptgeneralangle}, we find that
\begin{eqnarray}\label{eq:projectedssh}
\mathcal{M}^{(S,S,h)}&=&\sum_{i=0}^{2S}g_i~x^{h-2S}\Big(x\lceil\boldsymbol{1}^+\boldsymbol{2}^+\rfloor-x\lceil\boldsymbol{1}^+\boldsymbol{2}^-\rfloor-x\lceil\boldsymbol{1}^-\boldsymbol{2}^+\rfloor+x\lceil\boldsymbol{1}^-\boldsymbol{2}^-\rfloor\Big)^{2S-i}\nonumber
\\
&&\times\left( \frac{1}{m}\lceil\boldsymbol{1}^+\tilde3\rfloor\lceil\boldsymbol{2}^+\tilde3\rfloor +
                \frac{1}{m}\lceil\boldsymbol{1}^+\tilde3\rfloor\lceil\boldsymbol{2}^-\tilde3\rfloor+
                \frac{1}{m}\lceil\boldsymbol{1}^-\tilde3\rfloor\lceil\boldsymbol{2}^+\tilde3\rfloor+
                \frac{1}{m}\lceil\boldsymbol{1}^-\tilde3\rfloor\lceil\boldsymbol{2}^-\tilde3\rfloor\right)^i.
\end{eqnarray}    
The resulting amplitude involves the auxiliary small components $|\boldsymbol{k}^-\rfloor$ explicitly, and removing them through their equations of motion leads to the HPET amplitudes constructed in Subsection~\ref{sec:integrateoutamp}.

\subsection{Expansion of Velocity-projected Amplitudes}\label{sec:integrateoutamp}

Having projected the relativistic amplitudes as in Eq.~\eqref{eq:projectedssh}, we complete the top-down route by removing the auxiliary small component. The small component $|\boldsymbol{k}^-\rfloor$ is suppressed by $1/m$ relative to the large component $|\boldsymbol{k}^+\rfloor$, giving one power of $k/m$. Using the equations of motion in Eq.~\eqref{eq:integrateoutk-spinor}, the small component is expressed in terms of the large one and expanded in powers of $v\cdot k/(2m)\ll1$:
\begin{eqnarray}\label{eq:integratekminus}
     |\boldsymbol{k}^-\rfloor=\frac{1}{2m}\left(\sum_{n=0}(-1)^n(\frac{v\cdot k}{2m})^n\right)k_{\perp}|\boldsymbol{k}^+\rfloor\equiv \frac{1}{2m_k}k_{\perp}|\boldsymbol{k}^+\rfloor,\quad\text{where}\quad m_k\equiv m+\frac{v\cdot k}{2}.
\end{eqnarray}
Here we define the mass $m_k$ for simplicity.
By integrating out the $|\boldsymbol{k}^-\rfloor$ in $\mathcal{M}^{(S,S,h)}$, we obtain the HPET amplitudes $\mathcal{A}^{(S,S,h)}$.

For instance, for the minimal coupling in the heavy mass limit $m\rightarrow\infty$, only $|\boldsymbol{k}^+\rfloor$ is kept, and the leading order result is
\begin{eqnarray}\label{eq:minleading3pt}
    \mathcal{A}_{(0),\text{min}}^{(S,S,h)}&=&g_0\,x^h\lceil\boldsymbol{1}^+\boldsymbol{2}^+\rfloor^{2S}.
\end{eqnarray}
For the minus helicity case there is an extra sign $(-1)^{2S}$ due to the definition of the amplitude $\mathcal{M}^{(\frac{1}{2},\frac{1}{2},-1)}$. In this limit the soft boson couples to the heavy particle only through the $x$ factor. For finite mass, the sub-leading corrections are obtained from the first order of Eq.~\eqref{eq:integratekminus}:
\begin{eqnarray}\label{eq:heavyp3pt}
\mathcal{A}_{(1),\text{min}}^{(S,S,h)}&=&\frac{1}{2}\binom{2S}{1}g_0\,x^h\lceil\boldsymbol{1}^+\boldsymbol{2}^+\rfloor^{2S-1}\Big(-\lceil\textbf{1}^+|k_{2\perp}|\textbf{2}^+\rfloor+\lceil\textbf{1}^+|k_{1\perp}|\textbf{2}^+\rfloor\Big).
\end{eqnarray}
Here one of the $2S$ symmetric factors of $\lceil\boldsymbol{1}^+\boldsymbol{2}^+\rfloor^{2S}$ is replaced by the first-order correction acting on either heavy leg, and the binomial coefficient counts the number of choices of the factor. 

For the general three-particle interactions including the non-minimal couplings, the higher-order terms of the heavy-particle expansion suppressed by $1/m$ follow by applying Eq.~\eqref{eq:integratekminus} to the three-particle amplitude in Eq.~\eqref{eq:projectedssh}:
\begin{eqnarray}\label{eq:ohetexpansion}
    \mathcal{M}^{(S,S,h)}\xrightarrow{\text{integrate out}    |\boldsymbol{k}^-\rfloor}\mathcal{A}^{(S,S,h)}=\mathcal{A}^{(S,S,h)}_{(0)}+\frac{1}{m}\mathcal{A}^{(S,S,h)}_{(1)}+\cdots+\frac{1}{m^n}\mathcal{A}^{(S,S,h)}_{(n)}+\cdots\, .
\end{eqnarray}
These HPET amplitudes exhibit the explicit power counting of the heavy-particle effective theory, and each term maps onto a heavy-particle effective operator of the corresponding order.
Explicitly, the HPET amplitude for general spin $S$ and positive helicity $h$ is
\begin{eqnarray}\label{eq:alloderintegrateput3pt}
    \mathcal{A}^{(S,S,h)}
    &=& \sum_{i=0}^{2S}g_i~x^{h-2S}\left( x\lceil\boldsymbol{1}^+\boldsymbol{2}^+\rfloor-\frac{x\lceil\boldsymbol{1}^+|k_{2\perp}|\boldsymbol{2}^+\rfloor}{2m_{k_2}}+\frac{x\lceil\boldsymbol{1}^+|k_{1\perp}|\boldsymbol{2}^+\rfloor}{2m_{k_1}}-\frac{x\lceil\boldsymbol{1}^+|k_{1\perp}k_{2\perp}|\boldsymbol{2}^+\rfloor}{4m_{k_1}m_{k_2}}\right)^{2S-i}\nonumber\\
    &&\times\left(\frac{\lceil\boldsymbol{1}^+\tilde 3\rfloor\lceil\boldsymbol{2}^+\tilde 3\rfloor}{m}-\frac{\lceil\boldsymbol{1}^+\tilde 3\rfloor\lceil\boldsymbol{2}^+|k_{2\perp}|\tilde 3\rfloor}{2m\,m_{k_2}}-\frac{\lceil\boldsymbol{1}^+|k_{1\perp}|\tilde 3\rfloor\lceil\boldsymbol{2}^+\tilde 3\rfloor}{2m\,m_{k_1}}+\frac{\lceil\boldsymbol{1}^+|k_{1\perp}|\tilde 3\rfloor\lceil\boldsymbol{2}^+|k_{2\perp}|\tilde 3\rfloor}{4m\,m_{k_1}m_{k_2}}\right)^i.\nonumber\\
\end{eqnarray}
The terms $x\lceil\boldsymbol{1}^+\boldsymbol{2}^+\rfloor$ and $\lceil\boldsymbol{1}^+\tilde 3\rfloor\lceil\boldsymbol{2}^+\tilde 3\rfloor$ are the leading projections of the minimal and non-minimal structures of Eq.~\eqref{eq:1/2minrel3pt}, and the remaining terms are their $1/m$ corrections generated by the expansion of the small components. The negative-helicity basis is obtained in the same way. This completes the top-down construction; the complementary bottom-up construction of the same amplitudes is presented in Subsection~\ref{sec:bottomup}.

As an illustration, consider the case of $S=1/2$ fermions and a $|h|=1$ photon of Eq.~\eqref{eq:alloderintegrateput3pt}. Up to $1/m$, the parity and time-reversal invariant combination gives the operator
\begin{eqnarray}
     \mathcal{A}^{(\frac{1}{2},\frac{1}{2},+1)}+ \mathcal{A}^{(\frac{1}{2},\frac{1}{2},-1)} &\Longleftrightarrow& eN^{\dagger}(x) A^0(x)N(x)+\frac{1}{2m}\mathrm{i}eN^{\dagger}(x)\vec A(x)\cdot \vec D N(x)+\text{h.c.}\nonumber\\
     &&+\frac{c_F}{2m}eN^{\dagger}(x)\vec\sigma\cdot\vec B(x)N(x)+\mathcal{O}(\frac{k^2}{m^2}).
\end{eqnarray}
Here $e$ is the coupling constant. For the two-component heavy-particle fermion field $N$, the $x$ factor is proportional to the polarization vector $\varepsilon$ contracted with the reference velocity, $x^{\pm1}\sim \varepsilon^{\pm}\cdot\frac{p}{m}\sim \varepsilon^{\pm}\cdot v$, so the first $g_0$ terms correspond to the configuration space operator $N^{\dagger}v\cdot AN=N^{\dagger} A^0N$, while the second $g_0$ terms correspond to the operator $N^{\dagger}A\cdot D_{\perp}N$. Together with the kinetic terms, they combine into the heavy-particle operators $N^{\dagger}\mathrm{i}v\cdot D N$ and $\frac{1}{2m}N^{\dagger}D_{\perp}^2 N$, where $D^{\mu}=(v\cdot D, D_{\perp})=(D_t,-\vec D)$. The $g_1$ terms correspond to the magnetic moment operator $\frac{1}{m}N^{\dagger}\vec \sigma\cdot \vec B N$.
The higher-order results can be obtained by the same procedure. 

\subsection{Direct Construction and RPI Condition of Velocity-projected Amplitudes}\label{sec:bottomup}

Different from the top-down amplitude, we can directly construct the 3-point velocity-projected amplitudes $\mathcal{H}$ using $v\!\!\!/$-spinors by symmetry and kinematics. These amplitudes are little group covariant in the indices $I$ and invariant in the indices $L$. Imposing the RPI-I condition restores the Lorentz symmetry. Finally, expanding the small component in the velocity-projected amplitudes $\mathcal{H}$ yields the HPET amplitudes $\mathcal{A}$.

\paragraph{Determining the Velocity-projected Amplitude by Symmetry}

The form of 3-point $v\!\!\!/$-projected amplitudes can be determined by symmetry and kinematics. Let us illustrate the construction procedure for amplitudes involving two heavy particles with equal mass $m$ and spin $S$, and one massless particle with helicity $h$.

The building blocks of the amplitudes can only be  
\begin{eqnarray}
    |\mathbf{1}^{\pm}\rfloor^{2S}, |\mathbf{2}^{\pm}\rfloor^{2S}, |q\rfloor^{|2h|}, |\tilde{q}\rfloor^{|2h|},\,k_{\perp1}, \,k_{\perp2},\, q_{\perp}. 
\end{eqnarray}
Here the multiplets $|\mathbf{1}^{\pm}\rfloor$ and $|\mathbf{2}^{\pm}\rfloor$ denote the heavy particles with both large and small components, while $|q\rfloor$ (or $|\tilde{q}\rfloor$) corresponds to the massless soft particle. Note that the time components $v\cdot k$ and $v\cdot q$ are treated as $SU(2)$ scalars, which can always be absorbed into the Wilson coefficients.

The on-shell amplitude can be factorized into external spinors and the stripped term $\mathcal{H}^{\{L_1\cdots L_{2S}\},\{L'_1\cdots L'_{2S}\},h}$:
\begin{equation}\label{eq:H1ssh}   \mathcal{H}^{(S,S,h)}=\left(|\mathbf{1}^{\pm}\rfloor^{2S}\right)_{\{L_1\cdots L_{2S}\}}\left(|\mathbf{2}^{\pm}\rfloor^{2S}\right)_{\{L'_1\cdots L'_{2S}\}} \mathcal{H}^{\{L_1\cdots L_{2S}\},\{L'_1\cdots L'_{2S}\},h}.
\end{equation}
The little group covariance for the massive legs 1 and 2 is manifest, as the $\textbf{BOLD}$ spinors carry the indices $I$. The stripped term $\mathcal{H}^{\{L_1\cdots L_{2S}\},\{L'_1\cdots L'_{2S}\},h}$ contains the reference little group indices and the massless helicity weight. These determine its allowed tensor structures, while independent scalar coefficients remain to be constrained or matched. 

The structure of the stripped amplitude $\mathcal{H}^{\{L_1\cdots L_{2S}\},\{L'_1\cdots L'_{2S}\},h}$ is constrained by the kinematics and the massless little group symmetry:
\begin{enumerate}
    \item  Momentum conservation, $k_{1\perp}=k_{2\perp}+q_{\perp}$, eliminates $k_{1\perp}$ from the amplitudes. 

\item Equations of motion in Eq.~\eqref{eq:EOMfermion} for the heavy particles restrict the appearance of the three-momenta in the multiplet $|\boldsymbol{k}^{\pm}\rfloor$, since contracting $(k_{2\perp})_{LL'}$ or $(q_{\perp})_{LL'}$ directly with the massive legs is redundant, reducing to scalar structures multiplied by $\lceil\mathbf{1}^{\pm}\mathbf{2}^{\pm}\rfloor$.

\item The independent tensor structures for the stripped term are limited to the symmetric polarization vector $|q\rfloor |q\rfloor$ (or $|\tilde{q}\rfloor |\tilde{q}\rfloor$) and the anti-symmetric invariant tensor $\epsilon^{LL'}$.

\item An additional term $\lceil q|k_{2\perp}|q\rfloor$ carrying the helicity $h$ index arises when the polarization vector contracts with the momentum $k_{2\perp}$. 
\end{enumerate}
Therefore, the amplitude should consist of 
\begin{eqnarray}\label{eq:H2IJh}
    \mathcal{H}^{\{L_1\cdots L_{2S}\},\{L'_1\cdots L'_{2S}\},h} \propto \left\{ \left(|q\rfloor|q\rfloor\right)^{LL'}, \, \epsilon^{LL'}, \, \lceil q|k_{2\perp}|q\rfloor \right\}; \quad \text{or} \quad |q\rfloor\leftrightarrow |\tilde{q}\rfloor.
\end{eqnarray}

Depending on the relation between the spin $S$ and the helicity magnitude $|h|$, the amplitudes can be determined in the following two cases:
\begin{enumerate}
    \item \textbf{$2S \leq |h|$.} The stripped term consists of $2S$ symmetric tensors $|q\rfloor|q\rfloor^{LL'}$ (or $|\tilde{q}\rfloor|\tilde{q}\rfloor^{LL'}$),  $|h|-2S$ factors of $\lceil q|k_{2\perp}|q\rfloor$ (or $\lceil \tilde{q}|k_{2\perp}|\tilde{q}\rfloor$) from helicity compensation:  
    \begin{equation}\label{eq:b3pt1}
   \boxed{\mathcal{H}^{(S,S,h)}= \begin{cases}
         \lceil\mathbf{1}^{\pm}q\rfloor^{2S}\lceil\mathbf{2}^{\pm}q\rfloor^{2S}\lceil q|k_{2\perp}|q\rfloor^{-h-2S}, & h< 0, \\
         \lceil\mathbf{1}^{\pm}\tilde{q}\rfloor^{2S}\lceil\mathbf{2}^{\pm}\tilde{q}\rfloor^{2S}\lceil \tilde{q}|k_{2\perp}|\tilde{q}\rfloor^{h-2S}, & h\geq 0.
    \end{cases}}
\end{equation}

\item \textbf{$2S >|h|$.}  The stripped term consists of $|h|$ symmetric tensors $|q\rfloor|q\rfloor^{LL'}$ (or $|\tilde{q}\rfloor|\tilde{q}\rfloor^{LL'}$) and $2S-|h|$ anti-symmetric tensors $\epsilon^{LL'}$, which effectively contract the remaining indices of the massive legs: 
\begin{equation}\label{eq:b3pt2}
\boxed{\widetilde{\mathcal{H}}^{(S,S,h)}=\begin{cases}
        \lceil\mathbf{1}^{\pm}q\rfloor^{-h}\lceil\mathbf{2}^{\pm}q\rfloor^{-h}\lceil\mathbf{1}^{\pm}\mathbf{2}^{\pm}\rfloor^{2S+h}, & h< 0, \\
        \lceil\mathbf{1}^{\pm}\tilde{q}\rfloor^{h}\lceil\mathbf{2}^{\pm}\tilde{q}\rfloor^{h}\lceil\mathbf{1}^{\pm}\mathbf{2}^{\pm}\rfloor^{2S-h}, & h\geq 0.
    \end{cases}}
\end{equation}

\end{enumerate}
We denote these amplitudes $\mathcal{H}^{(S,S,h)}$ and $\widetilde{\mathcal{H}}^{(S,S,h)}$ as the velocity-projected amplitudes, in which the small component $|\boldsymbol{k}^{-}\rfloor$ of the heavy particle remains. They are the most general three-particle amplitudes consistent with the $SU(2)$ little group symmetry alone, and the Wilson coefficients are left completely general at this stage.

\paragraph{Imposing the RPI-I Condition} The underlying Lorentz symmetry of the projected amplitudes can be restored by the reparameterization of $v^{\mu}$.
The three-particle relativistic amplitude $\mathcal{M}$ is obtained by multiplying the velocity-projected amplitude  $\mathcal{H}$ by the Wilson coefficients $c(v,k,q)$ built from rotational scalars such as $v\cdot k$, $v\cdot q$, $\text{tr}[k_{\perp}k_{\perp}]$ and mass $m$. Imposing invariance of $\mathcal{M}$ under the reparameterizations, $\delta\mathcal{M}=\delta c\,\mathcal{H}+c\,\delta\mathcal{H}=0$, then determines the coefficients $c$.

To find the RPI combinations of the amplitudes, we first introduce the infinitesimal variation $\delta v$, such that the projection spinors reparameterize as
\begin{eqnarray}
 |v^{\prime}_{L}\rangle=| v_{L}\rangle+\delta v|v_{L}],\quad |v^{\prime}_{L}]=|v_{L}]+\delta v| v_{L}\rangle.
\end{eqnarray}
This transformation can also be written as $v'_{\alpha\dot\beta}=v_{\alpha\dot\beta}+2\delta v_{\alpha\dot\beta}$. It is sufficient working to first order in $\delta v$, with $v\cdot\delta v=0$. The factor of two defines the parameter convention used in this subsection. This reparameterization is a linear transformation of the velocity with the infinitesimal parameter $\delta v$, satisfying the on-shell condition $\det[v'_{\alpha\dot\beta}]=\det[v_{\alpha\dot\beta}]=1$. Based on the variation defined above, the momenta and the on-shell $v\!\!\!/$-spinors transform as:
\begin{equation}\label{eq:rpiall}
    \left\{\begin{array}{lccccl}
        \text{momenta:}
        &v&\longrightarrow&v&+&2\delta v,\\
         &k&\longrightarrow&k&-&2m\delta v,
        \\
        &q&\longrightarrow&q,&\\
         &k_{\perp}&\longrightarrow& k_{\perp}&-&2\left(m+v\cdot k\right)\delta v,
        \\
         \text{spinors:} &|\boldsymbol{k}^+\rfloor& \longrightarrow&|\boldsymbol{k}^+\rfloor&-& \delta v|\boldsymbol{k}^-\rfloor,\\
       & |\boldsymbol{k}^-\rfloor&\longrightarrow& |\boldsymbol{k}^-\rfloor&-& \delta v|\boldsymbol{k}^+\rfloor,\\
         &|q\rfloor&\longrightarrow&|q\rfloor&+&\delta v|q\rfloor,\\
        &|\tilde q\rfloor&\longrightarrow&|\tilde q\rfloor&-&\delta v|\tilde q\rfloor.\\
    \end{array}\right.
\end{equation}
The consistency of this transformation can be verified by explicit calculation. For instance, the scalar product $v\cdot q\equiv\frac{1}{2}v_{\alpha\dot\alpha}q^{\dot\alpha\alpha}=\frac{1}{2}\la3 v^{L}\ra\lb v_{L}3\rb=\frac{1}{2}\lceil3\tilde 3\rfloor$ transforms as $v\cdot q\longrightarrow v\cdot q+2\delta v\cdot q$, where $2\delta v\cdot q\equiv(\delta v)_{\alpha\dot\alpha}q^{\dot\alpha\alpha}=\la3|\delta v|3\rb=-\lceil3|\delta v|\tilde3\rfloor$, in agreement with the transformation of the $v\!\!\!/$-spinors.

Secondly, for a boson of helicity $h$ with $2S\leq|h|$ coupled to spin-$1/2$ fermions, the velocity-projected amplitudes  are derived in Eq.~\eqref{eq:b3pt1}. When the gauge boson is the photon, the amplitudes are given by $\lceil\boldsymbol{1}^{\pm} q\rfloor\lceil\boldsymbol{2}^{\pm} q\rfloor$ and $\lceil\boldsymbol{1}^{\pm}\tilde q\rfloor\lceil\boldsymbol{2}^{\pm}\tilde q\rfloor$. When the soft boson is the graviton, the scalar factor $\lceil q|k_{2\perp}|q\rfloor$ (or $\lceil\tilde q|k_{2\perp}|\tilde q\rfloor$) remains in the velocity-projected amplitude. 
The $x$ factor admits an explicit local expression. Writing $x\equiv\frac{\la\xi|p_2|3\rb}{m\la\xi3\ra}$ with a gauge reference spinor $\xi$ and fixing the gauge $\xi^{\alpha}=\tilde\lambda_{3\dot\alpha}v^{\dot\alpha\alpha}$, we have $\frac{\lceil\tilde3|k_{2\perp}|\tilde3\rfloor}{2mv\cdot q}=x$ and $\frac{\lceil3|k_{2\perp}|3\rfloor}{2mv\cdot q}=x^{-1}$. Therefore, the amplitudes with a graviton follow from the photon amplitudes by multiplying by the appropriate powers of $x$, and the same procedure extends to a general helicity-$h$ boson.\footnote{
However, the Weinberg-Witten theorem~\cite{Weinberg:1980kq} constrains the interactions for $|h|>1/2$ massless particles coupled to a Lorentz-covariant conserved current, and $|h|>1$ massless particles coupled to a conserved Lorentz-covariant energy-momentum tensor, so that only the photon ($|h|=1$) and graviton ($|h|=2$) couplings are physically realized; the amplitudes with higher helicities are formal.}

Finally, with the detailed calculations given in Appendix~\ref{apprpi}, for general spin the reparameterization-invariant amplitudes are obtained by the symmetric tensor product of the spin-$1/2$ structures,
\begin{eqnarray}\label{eq:spins3pt1}
\mathcal{M}^{(S,S,+h)}&=&\sum_{i=0}^{2S}g_i~x^{h-2S}\Big(x\left(\lceil\boldsymbol{1}^+\boldsymbol{2}^+\rfloor-\lceil\boldsymbol{1}^+\boldsymbol{2}^-\rfloor-\lceil\boldsymbol{1}^-\boldsymbol{2}^+\rfloor+\lceil\boldsymbol{1}^-\boldsymbol{2}^-\rfloor\right)\Big)^{2S-i}\nonumber
\\
&&\times\left( \frac{1}{m}\lceil\boldsymbol{1}^+\tilde3\rfloor\lceil\boldsymbol{2}^+\tilde3\rfloor +
                \frac{1}{m}\lceil\boldsymbol{1}^+\tilde3\rfloor\lceil\boldsymbol{2}^-\tilde3\rfloor+
                \frac{1}{m}\lceil\boldsymbol{1}^-\tilde3\rfloor\lceil\boldsymbol{2}^+\tilde3\rfloor+
                \frac{1}{m}\lceil\boldsymbol{1}^-\tilde3\rfloor\lceil\boldsymbol{2}^-\tilde3\rfloor\right)^i,
\end{eqnarray}
\begin{eqnarray}\label{eq:spins3pt2}
    \mathcal{M}^{(S,S,-h)}&=&\sum_{i=0}^{2S}g_i~x^{2S-h}\Big(
x^{-1}\left(-\lceil\boldsymbol{1}^+\boldsymbol{2}^+\rfloor-\lceil\boldsymbol{1}^+\boldsymbol{2}^-\rfloor-\lceil\boldsymbol{1}^-\boldsymbol{2}^+\rfloor-\lceil\boldsymbol{1}^-\boldsymbol{2}^-\rfloor\right)\Big)^{2S-i}\nonumber \\
         &\times&\left(\frac{1}{m}\lceil\boldsymbol{1}^+3\rfloor\lceil\boldsymbol{2}^+3\rfloor -
                \frac{1}{m}\lceil\boldsymbol{1}^+3\rfloor\lceil\boldsymbol{2}^-3\rfloor-
                \frac{1}{m}\lceil\boldsymbol{1}^-3\rfloor\lceil\boldsymbol{2}^+3\rfloor+
                \frac{1}{m}\lceil\boldsymbol{1}^-3\rfloor\lceil\boldsymbol{2}^-3\rfloor\right)^i.
\end{eqnarray}
For notational simplicity we use the same symbols $g_i$ for all amplitudes, for example for $h=\pm1$ and $h=\pm2$, but they should be understood as process-dependent parameters that differ for different helicities or different types of massless bosons. Using the relations between the relativistic spinors and $v\!\!\!/$-spinors in Eqs.~\eqref{eq:expandAHH} and \eqref{eq:expandmassless}, one checks that the relativistic amplitudes  coincide with the most general relativistic three-particle amplitudes of Ref.~\cite{Arkani-Hamed:2017jhn} in Eq.~\eqref{eq:r3ptgeneralangle} for spin $S$ and helicity $h$.

Although we start from the velocity-projected amplitude  of Eq.~\eqref{eq:b3pt1} for $2S\leq|h|$, either Eq.~\eqref{eq:spins3pt1} or Eq.~\eqref{eq:spins3pt2} applies to the amplitude with general spin and helicity, including the case $2S>|h|$. After extracting the factor $x$ with the appropriate power, the remaining structure is the velocity-projected amplitude  of Eq.~\eqref{eq:b3pt1} alone. In this way, the velocity-projected amplitude  with $2S>|h|$ is obtained by $\widetilde{\mathcal{H}}^{(S,S,h)}\Longleftrightarrow \mathcal{H}^{(S,S,h-n)}\times x^n$; for instance, $\widetilde{\mathcal{H}}^{(1,1,-1)}=\lceil\mathbf{1}^{\pm}3\rfloor\lceil\mathbf{2}^{\pm}3\rfloor\lceil\mathbf{1}^{\pm}\mathbf{2}^{\pm}\rfloor\sim\lceil\mathbf{1}^{\pm}3\rfloor^2\lceil\mathbf{2}^{\pm}3\rfloor^2\times x=\mathcal{H}^{(1,1,-2)}\times x$. The cases with $2S>|h|$ are thus reduced to the cases with $2S\leq |h|$ with proper powers of $x$ carrying the helicity weights. Consequently, both relativistic amplitudes  in Eqs.~\eqref{eq:spins3pt1} and \eqref{eq:spins3pt2} hold for general spin and helicity. Besides, they can be translated into each other, just as Eq.~\eqref{eq:r3ptgeneralangle} and its square bracket counterpart can in the relativistic case.

\paragraph{Obtaining HPET amplitudes}
Following the coset construction in Ref.~\cite{Li:2025ejk}, the most general way to integrate out the small component $|\boldsymbol{k}^-\rfloor$ is
\begin{eqnarray}\label{eq:generalintegrateout}
    |\boldsymbol{k}^-\rfloor=\, F\Big(\frac{1}{m},k_{\perp},|q\rfloor, |\tilde q\rfloor\Big)|\boldsymbol{k}^+\rfloor,
\end{eqnarray}
where the general operator function $F\Big(\frac{1}{m},k_{\perp},|q\rfloor, |\tilde q\rfloor\Big)$ encodes the interactions of the small component. For example, this function can be $\frac{k_{\perp}}{2m}+c_1\frac{v\cdot q}{m^2}\,|q\rfloor\lceil q|+\cdots+\mathcal{O}(1/m^3)$, and the $c_1$ term is interaction-dependent. This function can be generally constructed, and its coefficients are constrained by the consistency condition which follows from Lorentz covariance~\cite{Li:2025ejk}.
In the coordinate space of the constructive HPET, this formalism extends the tree-level free-field integrated-out result in Eq.~\eqref{eq:integrateoutk-spinor} to the general result, 
\begin{eqnarray}
    \tilde N(x)=\left(-\frac{i\vec\sigma\cdot\vec\nabla}{2m}+\frac{i\vec\sigma\cdot\vec\nabla i\partial_t}{4m^2}+...\right)N(x)\longrightarrow\left(-\frac{i\vec{\sigma}\cdot\vec{D}}{2m}+c^{(2)}_A\frac{\{D_t,\vec{\sigma}\cdot{\vec{D}}\}}{8m^2}+c^{(2)}_B\frac{ig\vec{\sigma}\cdot\vec{E}}{8m^2}\right)N(x).
\end{eqnarray}
Equation~\eqref{eq:generalintegrateout} provides a possible extension to interaction-dependent small-component relations. Establishing the independent interactions and their coefficient constraints requires a separate analysis; we do not carry out that extension here.

For the amplitudes calculated in this work, we use the free-field relation for the on-shell $v\!\!\!/$-spinor, 
\begin{eqnarray}
     |\boldsymbol{k}^-\rfloor=\, F\Big(\frac{1}{m},k_{\perp},|q\rfloor, |\tilde q\rfloor\Big)|\boldsymbol{k}^+\rfloor\longrightarrow \frac{1}{2m+v\cdot k}k_{\perp}|\boldsymbol{k}^+\rfloor,
\end{eqnarray}
namely the equations of motion of Eq.~\eqref{eq:integratekminus} used in Subsection~\ref{sec:integrateoutamp}. The HPET amplitude $\mathcal{A}^{(S,S,h)}$ derived from the relativistic amplitude  in Eq.~\eqref{eq:spins3pt1} is exactly the same as Eq.~\eqref{eq:alloderintegrateput3pt}.

In particular, for a scalar heavy particle with $S=0$ interacting with a boson of helicity $\pm h$, the three-particle velocity-projected amplitudes  follow from Eq.~\eqref{eq:b3pt1}:
\begin{eqnarray}
    \mathcal{H}^{(0,0,+h)}=\lceil \tilde 3|k_{2\perp}|\tilde 3\rfloor^h,\quad\mathcal{H}^{(0,0,-h)}=\lceil 3|k_{2\perp}|3\rfloor^h.
\end{eqnarray}
The scalar heavy particle is obtained by integrating out the high-frequency modes of the original state and keeping the residual momentum $k=p-mv$ with $m\gg|k|$. The residual momentum of the heavy scalar in the $SU(2)$ amplitude is further projected onto the perpendicular component $(k_{\perp})_{IJ}$. There is no component like $|\boldsymbol{k}^-\rfloor$ that needs to be removed. Consequently, the HPET amplitudes equal the velocity-projected amplitudes. After imposing the reparameterization invariance, we obtain the amplitudes 
\begin{eqnarray}
    \mathcal{A}^{(0,0, h)}=\left(\frac{\lceil \tilde 3|k_{2\perp}|\tilde 3\rfloor}{2mv\cdot q}\right)^h,\quad\mathcal{A}^{(0,0, -h)}=\left(\frac{\lceil 3|k_{2\perp}|3\rfloor}{2mv\cdot q}\right)^h\Longrightarrow\, \mathcal{A}^{(0,0, h)}=x^h.
\end{eqnarray}
The above amplitude coincides with the result given by the general case $\mathcal{A}^{(S,S,h)}$ in Eq.~\eqref{eq:alloderintegrateput3pt} when $S=0$. Compared with the spin-$S$ case which contains an infinite $1/m$ expansion, $\mathcal{A}^{(0,0,h)}$ is a single term. The scalar amplitude is also compatible with the result given by Ref.~\cite{Brandhuber:2023hhy}.

To summarize, the HPET amplitude  is obtained in three steps in the bottom-up approach. The little group covariance of $v^{\mu}$ determines the velocity-projected amplitudes  $\mathcal{H}^{(S,S,h)}$; reparameterization invariance constrains the Wilson coefficients and determines the relativistic amplitudes  $\mathcal{M}^{(S,S,h)}$; expanding the small component produces the HPET amplitude  $\mathcal{A}^{(S,S,h)}$, which carries a definite $1/m$ power counting and maps directly onto the HPET operators. The relativistic amplitudes  agree with the general relativistic three-particle amplitudes of Ref.~\cite{Arkani-Hamed:2017jhn}, and the integrated-out minimal couplings provide the building blocks for the Compton amplitudes bootstrapped in Section~\ref{sec:fourp}.

\section{The Compton Scattering Amplitudes}\label{sec:fourp}

Having determined the heavy-particle three-particle amplitudes, we now consider photon and graviton Compton scattering from a spin-$S$ heavy particle, with the amplitude expanded according to the heavy-particle power counting in $1/m$. In particular, we find that our Compton amplitudes in Eq.~\eqref{eq:bottomupcompton1} and Eq.~\eqref{eq:boup4ptgravitoncomp1} are free of the spurious pole for any finite spin $S$ up to the sub-leading order $1/m$.

We use the momentum assignment displayed in the amplitudes below, where the physical incoming heavy momentum is $p_1=mv+k_1$, the outgoing heavy momentum is $p_4=mv+k_4$, and the two boson momenta $q_2,q_3$ are taken outgoing, so that $p_1=q_2+q_3+p_4$. We define
\begin{equation}
 s=(p_1-q_2)^2,\qquad u=(p_1-q_3)^2,\qquad t=(q_2+q_3)^2,
 \qquad k_s=k_1-q_2,\qquad k_u=k_1-q_3.
\end{equation}
Thus $s-m^2=2m\,v\cdot k_s+k_s^2$ and $u-m^2=2m\,v\cdot k_u+k_u^2$. Overall electromagnetic and gravitational coupling constants are suppressed in the bottom-up formulas and restored when comparing with the relativistic amplitudes.
The formulas below are written for the fixed-helicity ordering $(2^{+h},3^{-h})$. Crossing to the opposite ordering exchanges $q_2\leftrightarrow q_3$, the helicities, and the s- and u-channel residual momenta.

When an intermediate particle goes on shell, locality and unitarity require the four-particle amplitude to factorize into a product of three-particle amplitudes, which fixes the residue in the corresponding channel.
In the s-channel, the residue of the Compton amplitude is $\mathop{\rm Res}_{s=m^2}\mathcal{A}_4\equiv\lim_{s\rightarrow m^2}(s-m^2)\mathcal{A}_4=\mathcal{A}_L\mathcal{A}_R$, the product of two three-particle amplitudes, and the u-channel result follows by exchanging the boson legs $2\leftrightarrow 3$. For the photon Compton amplitude, the residue of the s-channel contains the pole in the u-channel, $\text{Res}(s)=\frac{R_{su}}{u-m^2}$, indicating the u-channel interaction.
For the graviton Compton amplitude, the residue of the s-channel is $\text{Res}(s)=\frac{R_{stu}}{(u-m^2)t}$, which includes the additional t-channel pole arising from the graviton three-particle self-interaction.

Subsection~\ref{subsec:bo4pt} constructs the Compton amplitudes from the reparameterization-invariant HPET amplitudes, Subsection~\ref{sec:bottomva} constructs the velocity-projected Compton amplitudes, while Subsection~\ref{subsec:top4pt} derives the same amplitudes top-down by expanding the relativistic results. Factorization fixes the residues at the physical poles but, by itself, does not exclude additional local contact terms.

\subsection{Bootstrapping 4-point HPET Amplitudes up to Sub-leading Order}\label{subsec:bo4pt}

In the bottom-up approach, the four-particle heavy-particle Compton amplitude is bootstrapped from the three-particle heavy-particle amplitude by factorization.  For finite spin satisfying $S\ll m$,
using Eq.~\eqref{eq:heavyp3pt}, the two integrated-out amplitudes to be glued are
\begin{eqnarray}\label{eq:alar3pt}
    \mathcal{A}_L(-\textbf{1}^{S},2^{+h},\textbf{P}^{S})&=&\left(x_{12}\right)^h\frac{\lceil\boldsymbol{1}^+\boldsymbol{P}^+\rfloor^{2S}}{m^{2S-1}}+\frac{\binom{2S}{1}}{2m}\left(x_{12}\right)^h\frac{\lceil\boldsymbol{1}^+\boldsymbol{P}^+\rfloor^{2S-1}}{m^{2S-1}}\lceil\textbf{1}^+|k_{1\perp}-k_{P\perp}|\boldsymbol{P}^+\rfloor\nonumber\\
    &&+\mathcal{O}(\frac{k^2}{m^2}),\nonumber\\
     \nonumber\\
   \mathcal{A}_R(-\textbf{P}^{S},3^{-h},\textbf{4}^{S})&=&\left(x_{34}\right)^{-h} \frac{(-\lceil\boldsymbol{P}^+\boldsymbol{4}^+\rfloor)^{2S}}{m^{2S-1}}+\frac{\binom{2S}{1}}{2m}\left(x_{34}\right)^{-h}\frac{(-\lceil\boldsymbol{P}^+\boldsymbol{4}^+\rfloor)^{2S-1}}{m^{2S-1}}\lceil\boldsymbol{P}^+|k_{P\perp}-k_{4\perp}|\textbf{4}^+\rfloor\nonumber\\
  &&+\mathcal{O}(\frac{k^2}{m^2}),
\end{eqnarray}
where the minus signs on legs $1$ and $P$ in the bracket of the left side amplitudes $\mathcal{A}$ denote that they are the incoming particles in the all-outgoing convention. The overall factor $g_0=\frac{1}{m^{2S-1}}$ comes from the normalization of the three-particle amplitude. The binomial coefficient is denoted as $\binom{2S}{n}\equiv\frac{(2S)!}{n!(2S-n)!}$. For the scalar case $S=0$, the sub-leading term vanishes. Note that the gauge-invariant $x$ factor is defined according to Eq.~\eqref{eq:defx}. Specifically, $x_{12}$ and $x_{34}$ are defined by $x_{12}\lambda_2=\frac{p_1|2\rbrack}{m}$ and $x_{34}\lambda_3=\frac{p_4|3\rbrack}{m}$. 

The s-channel residue is given by the product of the two amplitudes of Eq.~\eqref{eq:alar3pt}, glued through the intermediate state:
\begin{eqnarray}
    \lim_{s\rightarrow m^2}\text{Res}(s)=\sum_{I} \mathcal{A}_L(-\textbf{1}^{S},2^{+h},\textbf{P}^{S})\times  \mathcal{A}_R(-\textbf{P}^{S},3^{-h},\textbf{4}^{S}).
\end{eqnarray}
Here $I$ is the little group index of the leg $P$, and it is summed when gluing the two amplitudes. In the calculation, we use the identities for the boson legs $mx_{12}\lceil32\rfloor=(m+v\cdot k_1)\lceil3\tilde2\rfloor-\lceil3|k_{1\perp}|\tilde2\rfloor$ and $\frac{m}{x_{34}}\lceil\tilde2\tilde3\rfloor=-(m+v\cdot k_4)\lceil3\tilde2\rfloor+\lceil3|k_{4\perp}|\tilde2\rfloor$, so the s-channel on-shell constraint for the internal momentum reads
\begin{eqnarray}\label{eq:onshellPconstrain}
    &&\frac{1}{m}\left(|\boldsymbol{P}^{+I}\rfloor\lceil\boldsymbol{P}^{+}_I|+|\boldsymbol{P}^{+I}\rfloor\lceil\boldsymbol{P}^{+}_I|\frac{k_{P\perp}}{2m}-\frac{k_{P\perp}}{2m}|\boldsymbol{P}^{+I}\rfloor\lceil\boldsymbol{P}^{+}_I|\right)\nonumber\\
    &=&\frac{1}{\lceil3\tilde2\rfloor}\left(1+\frac{1}{2m}\frac{\lceil3|k_{1\perp}+k_{4\perp}|\tilde2\rfloor}{\lceil3\tilde2\rfloor}\right)\left(|3\rfloor\lceil\tilde2|-|\tilde2\rfloor\lceil3|+\frac{k_{1\perp}}{m}|\tilde2\rfloor\lceil3|+|\tilde2\rfloor\lceil3|\frac{k_{4\perp}}{m}\right)+\,\mathcal{O}(\frac{k^2}{m^2}).
\end{eqnarray}
The spinors scale as $|\textbf{1}^+\rfloor\sim|\textbf{4}^+\rfloor\sim|\textbf{P}^+\rfloor\sim\sqrt{m}$, and the products satisfy $\lceil\textbf{1}^+\textbf{P}^+\rfloor\sim\lceil\textbf{P}^+\textbf{4}^+\rfloor\sim m$ while $x_{12}\sim x_{34}\sim1$. Consequently, the three-particle amplitudes scale as $\mathcal{A}_L\sim\mathcal{A}_R\sim m$. In four dimensions, the energy scaling behavior of an $n$-particle amplitude is $E^{4-n}$, which is $E^{4-3}=E^1$ for the three-particle case. Since in this system the typical energy is dominated by the heavy-particle mass, $E\sim m$, the amplitudes $\mathcal{A}_L$ and $\mathcal{A}_R$ satisfy this scaling.

Using the momentum conservation $P=p_1-q_2=q_3+p_4$, and the on-shell constraint of Eq.~\eqref{eq:onshellPconstrain}, we obtain the s-channel residue
\begin{eqnarray}\label{eq:ress}
   \lim_{s\rightarrow m^2}\text{Res}(s)
    &=&\,m^2\left(\frac{x_{12}}{x_{34}}\right)^h\left[(\frac{\lceil\boldsymbol{1}^+\boldsymbol{4}^+\rfloor}{m})^{2S}+\frac{1}{2m}\binom{2S}{1}\frac{\lceil3|k_{1\perp}+k_{4\perp}|\tilde2\rfloor}{\lceil3\tilde2\rfloor}(\frac{\lceil\boldsymbol{1}^+\boldsymbol{4}^+\rfloor}{m})^{2S}\right.\nonumber\\
    &&-\,\frac{1}{2m}\binom{2S}{1}\frac{\lceil\textbf{1}^+3\rfloor\lceil\textbf{4}^+|k_{4\perp}|\tilde 2\rfloor-\lceil\textbf{4}^+3\rfloor\lceil\textbf{1}^+|k_{1\perp}|\tilde 2\rfloor}{\lceil3\tilde2\rfloor}(\frac{\lceil\boldsymbol{1}^+\boldsymbol{4}^+\rfloor}{m})^{2S-1}\nonumber\\
      &&-\,\left.\frac{1}{2m}\binom{2S}{1}\frac{\lceil\textbf{1}^+\tilde2\rfloor\lceil\textbf{4}^+|k_{4\perp}|3\rfloor-\lceil\textbf{4}^+\tilde 2\rfloor\lceil\textbf{1}^+|k_{1\perp}|3\rfloor}{\lceil3\tilde2\rfloor}(\frac{\lceil\boldsymbol{1}^+\boldsymbol{4}^+\rfloor}{m})^{2S-1}+\mathcal{O}(\frac{k^2}{m^2})\right].
\end{eqnarray}
The four-particle amplitudes follow from this residue. In the heavy mass limit $m\rightarrow\infty$, the residual momenta are suppressed, so the $\mathcal{O}(k/m)$ terms in Eq.~\eqref{eq:ress} vanish, and the only structure that remains in the spin-dependent part of the residue is free of spurious poles, $\left(-\frac{\lceil\boldsymbol{1}^+\boldsymbol{P}^{+,I}\rfloor\lceil \boldsymbol{P}^+_{I}\boldsymbol{4}^+\rfloor}{m^2}\right)^{2S}=\left(\frac{\lceil\boldsymbol{1}^+\boldsymbol{4}^+\rfloor}{m}\right)^{2S}+\mathcal{O}(\frac{k}{m})$.

\paragraph{Photon}
For the heavy-particle propagator, the on-shell condition of the heavy particle, $p=mv+k$ with $p^2-m^2=0$, is expressed by the residual momentum as $2mv\cdot k+k^2=0$. To leading order, the $x$ factors give the proper pole structure of the intermediate heavy particle,
\begin{equation}\label{eq:x12dividex34}
    \frac{x_{12}}{x_{34}}=-\frac{\lceil3\tilde2\rfloor^2}{2mv\cdot k_u+k_u^2}\left(1-\frac{1}{m}\frac{\lceil3|k_{1\perp}+k_{4\perp}|\tilde2\rfloor}{\lceil3\tilde2\rfloor}\right)+\mathcal{O}(\frac{k^2}{m^2}),
\end{equation}
where we have used $\lceil23\rfloor\lceil\tilde2\tilde3\rfloor=t=-(2mv\cdot k_u+k_u^2)$. This ratio is non-local, since in the s-channel residue it provides the u-channel pole of the amplitude. Multiplying the residue of Eq.~\eqref{eq:ress} by the $x$-factor ratio of Eq.~\eqref{eq:x12dividex34}, we find that the consistent Compton amplitude is
\begin{eqnarray}\label{eq:bottomupcompton1}
    &&\mathcal{A}_4(-\textbf{1}^S,2^{+1},3^{-1},\textbf{4}^S)\nonumber\\
    &=&-\frac{\lceil3\tilde 2\rfloor}{4\left(v\cdot k_s+\frac{k_{s}^2}{2m}\right)\left(v\cdot k_u+\frac{k_{u}^2}{2m}\right)}\left[\lceil3\tilde 2\rfloor\left(\frac{\lceil\boldsymbol{1}^+\boldsymbol{4}^+\rfloor}{m}\right)^{2S}-\frac{2-\binom{2S}{1}}{2m}\lceil3|k_{1\perp}+k_{4\perp}|\tilde2\rfloor\left(\frac{\lceil\boldsymbol{1}^+\boldsymbol{4}^+\rfloor}{m}\right)^{2S}\right.\nonumber\\
    &&-\,\frac{\binom{2S}{1}}{2m}\left(\lceil\textbf{1}^+3\rfloor\lceil\textbf{4}^+|k_{4\perp}|\tilde 2\rfloor-\lceil\textbf{4}^+3\rfloor\lceil\textbf{1}^+|k_{1\perp}|\tilde 2\rfloor\right)\left(\frac{\lceil\boldsymbol{1}^+\boldsymbol{4}^+\rfloor}{m}\right)^{2S-1}\nonumber\\
      &&-\,\left.\frac{\binom{2S}{1}}{2m}\left(\lceil\textbf{1}^+\tilde2\rfloor\lceil\textbf{4}^+|k_{4\perp}|3\rfloor-\lceil\textbf{4}^+\tilde 2\rfloor\lceil\textbf{1}^+|k_{1\perp}|3\rfloor\right)\left(\frac{\lceil\boldsymbol{1}^+\boldsymbol{4}^+\rfloor}{m}\right)^{2S-1}\right]+\mathcal{O}(\frac{k^2}{m^2}).
\end{eqnarray}
The inverse powers of the soft bracket $\lceil3\tilde2\rfloor$ that appear separately in these two building blocks cancel in the product, and the only denominators left in Eq.~\eqref{eq:bottomupcompton1} are the physical heavy-particle propagators. Both the leading term and the $1/m$ corrections are therefore free of spurious poles, for any finite spin $S$.
The leading part of this amplitude scales as $m^0$, which is the leading-order term of the four-particle amplitude with scale $E^{4-4}\sim E^0\sim m^0$. The subsequent terms are suppressed by $k/m$.
The amplitude should satisfy the spin-statistics property for the heavy particles, and indeed we find that this amplitude gives $(-1)^{2S}$ after exchanging legs $1\leftrightarrow 4$.

\paragraph{Graviton}
In the s-channel of the graviton Compton amplitude, the $x$-factor ratio is again non-local, accounting for the u-channel pole propagated by the heavy particle together with the t-channel pole of the graviton three-particle self-interaction,
\begin{equation}
    \frac{x_{12}^2}{x_{34}^2}=-\frac{\lceil3\tilde2\rfloor^4}{(2mv\cdot k_u+k_u^2)t}\left(1-\frac{2}{m}\frac{\lceil3|k_{1\perp}+k_{4\perp}|\tilde2\rfloor}{\lceil3\tilde2\rfloor}\right)+\mathcal{O}(\frac{k^2}{m^2}).
\end{equation}
Multiplying this ratio by the spin-$S$ sector of the residue in Eq.~\eqref{eq:ress}, we find the consistent residue $R_{stu}$. The graviton Compton amplitude is then obtained with the correct t-channel pole of the graviton three-particle interaction,
\begin{eqnarray}\label{eq:boup4ptgravitoncomp1}
    &&\mathcal{A}_4(-\textbf{1}^S,2^{+2},3^{-2},\textbf{4}^S)\nonumber\\
    &=&\frac{\lceil3\tilde 2\rfloor^3}{-4\left(v\cdot k_s+\frac{k_{s}^2}{2m}\right)\left(v\cdot k_u+\frac{k_{u}^2}{2m}\right)t}\left[\lceil3\tilde 2\rfloor\left(\frac{\lceil\boldsymbol{1}^+\boldsymbol{4}^+\rfloor}{m}\right)^{2S}-\frac{4-\binom{2S}{1}}{2m}\lceil3|k_{1\perp}+k_{4\perp}|\tilde2\rfloor\left(\frac{\lceil\boldsymbol{1}^+\boldsymbol{4}^+\rfloor}{m}\right)^{2S}\right.\nonumber\\
    &&-\,\frac{\binom{2S}{1}}{2m}\left(\lceil\textbf{1}^+3\rfloor\lceil\textbf{4}^+|k_{4\perp}|\tilde 2\rfloor-\lceil\textbf{4}^+3\rfloor\lceil\textbf{1}^+|k_{1\perp}|\tilde 2\rfloor\right)\left(\frac{\lceil\boldsymbol{1}^+\boldsymbol{4}^+\rfloor}{m}\right)^{2S-1}\nonumber\\
      &&-\,\left.\frac{\binom{2S}{1}}{2m}\left(\lceil\textbf{1}^+\tilde2\rfloor\lceil\textbf{4}^+|k_{4\perp}|3\rfloor-\lceil\textbf{4}^+\tilde 2\rfloor\lceil\textbf{1}^+|k_{1\perp}|3\rfloor\right)\left(\frac{\lceil\boldsymbol{1}^+\boldsymbol{4}^+\rfloor}{m}\right)^{2S-1}\right]+\mathcal{O}(\frac{k^2}{m^2}).
\end{eqnarray}
which can be verified to factorize properly in both the u- and the t-channel. As in the photon case, the amplitude is assembled directly from the building blocks, so the leading term and the displayed $1/m$ corrections are free of spurious poles, for any spin $S$. The dimensionless gravitational coupling is left undetermined in the bottom-up construction.

For the scalar case $S=0$, the amplitude can be written in the manifest double-copy form, with the field strength in momentum space $F^{\mu\nu}\rightarrow\lambda_{\alpha}\lambda_{\beta}\epsilon_{\dot\alpha\dot\beta}+\tilde\lambda_{\dot\alpha}\tilde\lambda_{\dot\beta}\epsilon_{\alpha\beta}$. Using the approximation $\lceil3\tilde2\rfloor=\frac{1}{m}\lceil3|p_1|\tilde2\rfloor$ in the heavy mass limit, Eq.~\eqref{eq:boup4ptgravitoncomp1} gives the scalar graviton Compton amplitude
\begin{equation}\label{eq:clascalar4pt}
    \mathcal{A}_4(-\textbf{1}^0,2^{+2},3^{-2},\textbf{4}^0)=\frac{1}{\left(2p_1\cdot q_2\right)\left(2p_1\cdot q_3\right)t}\left(p_1\cdot F_2\cdot F_3\cdot p_1\right)^2,
\end{equation}
which is compatible with the result in Refs.~\cite{Bjerrum-Bohr:2013bxa,Brandhuber:2023hhy}.

In the heavy mass limit $m\rightarrow\infty$, the spin effects decouple and the amplitudes reduce to $\frac{\lceil3\tilde 2\rfloor^2\left(\lceil\boldsymbol{1}^+\boldsymbol{4}^+\rfloor/m\right)^{2S}}{4(v\cdot k_s)(v\cdot k_u)}$ and $\frac{\lceil3\tilde 2\rfloor^4\left(\lceil\boldsymbol{1}^+\boldsymbol{4}^+\rfloor/m\right)^{2S}}{-4\left(v\cdot k_s\right)\left(v\cdot k_u\right)t}$
for the photon and the graviton, respectively. This is the analog of the heavy-quark spin symmetry~\cite{Isgur:1989vq,Isgur:1990yhj,Neubert:1993mb,Manohar_Wise_2023}, where the spin interactions are suppressed at $\mathcal{O}(\frac{k}{m})$. These limits hold for any finite spin $S$, including $S>1$ for the photon and $S>2$ for the graviton, and both the leading terms and the $1/m$ corrections displayed in Eqs.~\eqref{eq:bottomupcompton1} and \eqref{eq:boup4ptgravitoncomp1} contain no spin-dependent spurious poles. The velocity projection of both the heavy and soft external variables makes this cancellation manifest, since the pole amplitudes are expressed only in terms of reference-$SU(2)$ indices and the heavy-particle propagators. A systematic classification of sub-leading contact terms and possible spurious structures is beyond the scope of the present calculation.


\subsection{Bootstrapping 4-point Velocity-projected Amplitudes}\label{sec:bottomva}

Following the same procedure of subsection~\ref{subsec:bo4pt}, we generalize the bootstrapped 4-point Compton amplitudes from the consistency factorization of the velocity-projected 3-point minimal couplings.

The velocity-projected minimal couplings  are derived from Eqs.~\eqref{eq:spins3pt1} and ~\eqref{eq:spins3pt2} with $g_{i\geq 0}=0$,
\begin{eqnarray}
    \mathcal{A}_L(-\textbf{1}^{S},2^{+h},\textbf{P}^{S})&=&\,m(x_{12})^{h}\Big(\frac{\lceil\boldsymbol{1}^+\boldsymbol{P}^+\rfloor-\lceil\boldsymbol{1}^+\boldsymbol{P}^-\rfloor-\lceil\boldsymbol{1}^-\boldsymbol{P}^+\rfloor+\lceil\boldsymbol{1}^-\boldsymbol{P}^-\rfloor}{m}\Big)^{2S},\nonumber\\
    \nonumber\\
    \mathcal{A}_R(-\textbf{P}^{S},3^{-h},\textbf{4}^{S})&=&m(x_{34})^{-h}\Big(\frac{
-\lceil\boldsymbol{P}^+\boldsymbol{4}^+\rfloor-\lceil\boldsymbol{P}^+\boldsymbol{4}^-\rfloor-\lceil\boldsymbol{P}^-\boldsymbol{4}^+\rfloor-\lceil\boldsymbol{P}^-\boldsymbol{4}^-\rfloor}{m}\Big)^{2S}.\nonumber\\
\end{eqnarray}
Using the EOM in Eq.~\eqref{eq:integratekminus}, the small component $|\boldsymbol{k}^-\rfloor$ is expanded, and we reproduce the three-point amplitudes in Eq.~\eqref{eq:alar3pt}.

To derive the four-point amplitudes, we need to
contract the spin index $I$ of the internal leg in the three-point amplitudes. 
The all-order s-channel on-shell constraint for the internal momentum gives
\begin{eqnarray}
    &&|\boldsymbol{P}^{+I}\rfloor\lceil\boldsymbol{P}^{+}_I|+|\boldsymbol{P}^{+I}\rfloor\lceil\boldsymbol{P}^{-}_I|+|\boldsymbol{P}^{+I}\rfloor\lceil\boldsymbol{P}^{+}_I|-|\boldsymbol{P}^{-I}\rfloor\lceil\boldsymbol{P}^{-}_I|\nonumber\\
    &=&\frac{m^2\Big(-|3\rfloor\lceil\tilde2|+|\tilde 2\rfloor\lceil3|
    (1+\frac{v\cdot k_1}{m}) (1+\frac{v\cdot k_4}{m})-\frac{|\tilde2\rfloor\lceil3|k_{4\perp}}{m}(1+\frac{v\cdot k_1}{m})-\frac{k_{1\perp}|\tilde2\rfloor\lceil3|}{m}(1+\frac{v\cdot k_4}{m})+\frac{k_{1\perp}|\tilde 2\rfloor\,\lceil3|k_{4\perp}}{m^2}\Big)}{-\left(m+v\cdot k_1\right)\lceil3\tilde2\rfloor+\lceil3|k_{1\perp}|\tilde2\rfloor}.\nonumber\\
\end{eqnarray}
This constrain reduces to Eq.~\eqref{eq:onshellPconstrain} when expanding to the order $\mathcal{O}(k/m)$.

Using the above constraint, then we obtain the s-channel residue 
\begin{eqnarray}
   && \lim_{s\rightarrow m^2}\text{Res}(s)= \mathcal{A}_L(-\textbf{1}^{S},2^{+h},\textbf{P}^{S})\,\mathcal{A}_R(-\textbf{P}^{S},3^{-h},\textbf{4}^{S})\nonumber\\
    &=&m^2\left(\frac{x_{12}}{x_{34}}\right)^h \Big(\frac{\lceil\boldsymbol{1}^+\boldsymbol{4}^+\rfloor\lceil\tilde 23\rfloor-\lceil\boldsymbol{1}^-\boldsymbol{4}^-\rfloor\lceil\tilde23\rfloor+\lceil\boldsymbol{1}^+|(|3\rfloor\lceil\tilde2|+|\tilde2\rfloor\lceil3|)|\boldsymbol{4}^-\rfloor-\lceil\boldsymbol{1}^-|(|3\rfloor\lceil\tilde2|+|\tilde2\rfloor\lceil3|)|\boldsymbol{4}^+\rfloor}{-\left(m+v\cdot k_1\right)\lceil3\tilde2\rfloor+\lceil3|k_{1\perp}|\tilde2\rfloor}\Big)^{2S}.\nonumber\\
\end{eqnarray}

\paragraph{Photon} For photon minimal coupling $h=\pm1$, the multiplication of $x$-factors reads
\begin{eqnarray}
     m^2 \frac{x_{12}}{x_{34}}=-\frac{\Big[\left(m+v\cdot k_1\right)\lceil3\tilde2\rfloor-\lceil3|k_{1\perp}|\tilde2\rfloor\Big]^2}{2mv\cdot k_u+k_u^2},
\end{eqnarray}
thus we derive the velocity-projected photon Compton amplitude 
\begin{eqnarray}\label{eq:vampphotoncompton}
    &&\mathcal{A}_4(-\textbf{1}^S,2^{+1},3^{-1},\textbf{4}^S)\nonumber\\
    &=&\frac{\Big[\left(1+\frac{v\cdot k_1}{m}\right)\lceil3\tilde2\rfloor-\frac{\lceil3|k_{1\perp}|\tilde2\rfloor}{m}\Big]^2}{-4(v\cdot k_s+\frac{k_s^2}{2m})(v\cdot k_u+\frac{k_u^2}{2m})}\Big(\frac{\lceil\boldsymbol{1}^+\boldsymbol{4}^+\rfloor-\lceil\boldsymbol{1}^-\boldsymbol{4}^-\rfloor+\lceil\boldsymbol{1}^+|\frac{|3\rfloor\lceil\tilde2|+|\tilde2\rfloor\lceil3|}{\lceil\tilde23\rfloor}|\boldsymbol{4}^-\rfloor-\lceil\boldsymbol{1}^-|\frac{|3\rfloor\lceil\tilde2|+|\tilde2\rfloor\lceil3|}{\lceil\tilde23\rfloor}|\boldsymbol{4}^+\rfloor}{\left(m+v\cdot k_1\right)+\frac{\lceil3|k_{1\perp}|\tilde2\rfloor}{\lceil\tilde23\rfloor}}\Big)^{2S}.\nonumber\\
\end{eqnarray}
Comparing to the sub-leading order result in Eq.~\eqref{eq:bottomupcompton1}, we see that at the spurious pole $\lceil3\tilde2\rfloor$ exists at order $k^3/m^3$ and higher in the HPET expansion.

\paragraph{Graviton}
Similarly, for graviton minimal coupling $h=\pm2$, the multiplication of $x$-factors reads
\begin{eqnarray}
     m^2 \left(\frac{x_{12}}{x_{34}}\right)^2=-\frac{\Big[\left(m+v\cdot k_1\right)\lceil3\tilde2\rfloor-\lceil3|k_{1\perp}|\tilde2\rfloor\Big]^4}{(2mv\cdot k_u+k_u^2)t},
\end{eqnarray}
and then we derive the velocity-projected Gravitational Compton amplitude 
\begin{eqnarray}\label{eq:vampgravitonncompton}
    &&\mathcal{A}_4(-\textbf{1}^S,2^{+2},3^{-2},\textbf{4}^S)\nonumber\\
    &=&\frac{\Big[\left(1+\frac{v\cdot k_1}{m}\right)\lceil3\tilde2\rfloor-\frac{\lceil3|k_{1\perp}|\tilde2\rfloor}{m}\Big]^4}{-4(v\cdot k_s+\frac{k_s^2}{2m})(v\cdot k_u+\frac{k_u^2}{2m})t}\Big(\frac{\lceil\boldsymbol{1}^+\boldsymbol{4}^+\rfloor-\lceil\boldsymbol{1}^-\boldsymbol{4}^-\rfloor+\lceil\boldsymbol{1}^+|\frac{|3\rfloor\lceil\tilde2|+|\tilde2\rfloor\lceil3|}{\lceil\tilde23\rfloor}|\boldsymbol{4}^-\rfloor-\lceil\boldsymbol{1}^-|\frac{|3\rfloor\lceil\tilde2|+|\tilde2\rfloor\lceil3|}{\lceil\tilde23\rfloor}|\boldsymbol{4}^+\rfloor}{\left(m+v\cdot k_1\right)+\frac{\lceil3|k_{1\perp}|\tilde2\rfloor}{\lceil\tilde23\rfloor}}\Big)^{2S}.\nonumber\\
\end{eqnarray}
In the HPET expansion, the spurious pole exists at order $k^5/m^5$ and higher comparing to Eq.~\eqref{eq:boup4ptgravitoncomp1}.
Specifically, when we set $k_1^{\mu}=0$ such that $p_1^{\mu}=mv^{\mu}$, this amplitude reduces to
\begin{eqnarray}\label{eq:k10gravitatiaolcomp}
    &&\mathcal{A}_4(-\textbf{1}^S,2^{+2},3^{-2},\textbf{4}^S)=\frac{\lceil3\tilde2\rfloor^4}{-4(v\cdot k_s+\frac{k_s^2}{2m})(v\cdot k_u+\frac{k_u^2}{2m})t}\Big(\frac{\lceil\boldsymbol{1}^+\boldsymbol{4}^+\rfloor+\lceil\boldsymbol{1}^+|\frac{|3\rfloor\lceil\tilde2|+|\tilde2\rfloor\lceil3|}{\lceil\tilde23\rfloor}|\boldsymbol{4}^-\rfloor}{m}\Big)^{2S},\nonumber\\
\end{eqnarray}
which agrees with the quantum gravitational Compton amplitudes in Refs.~\cite{Aoude:2020onz,Bautista:2022wjf} after expansion $|\boldsymbol{4}^-\rfloor=\frac{k_{4\perp}}{2m_{k_4}}|\boldsymbol{4}^+\rfloor$. The scalar amplitude is the spinless Schwarzschild amplitude~\cite{Bjerrum-Bohr:2013bxa,Brandhuber:2023hhy,Cangemi:2023bpe}.

\subsection{Projection and Expansion of 4-point Relativistic Amplitudes}\label{subsec:top4pt}

In the relativistic case, the minimal couplings of a photon and two identical massive spin-$S$ particles are given in Refs.~\cite{Arkani-Hamed:2017jhn,Chung:2018kqs} by $M_3^{S,S,+1}=x\frac{\langle \textbf{1}\textbf{2}\rangle^{2S}}{m^{2S-1}}$ and $M_3^{S,S,-1}=\frac{1}{x}\frac{\lbrack \textbf{1}\textbf{2}\rbrack^{2S}}{m^{2S-1}}$. The four-particle bootstrap amplitudes are obtained from them. In the s-channel, the residue follows by gluing the two three-particle amplitudes $M_3(-\textbf{1},2^{+1},\textbf{P})\times M_3(-\textbf{P},3^{-1},\textbf{4})$. For our purpose, the heavy-particle leg $1$ is incoming, contrary to the conventional all-outgoing assignment in the relativistic case, so there is an extra minus sign for $p_1\rightarrow-p_1$, $\tilde\lambda_{1}^{\sigma}\rightarrow-\tilde\lambda_{1}^{\sigma}$, and $x_{12}\rightarrow-x_{12}$. This residue turns out to be properly factorized in the u-channel. Consequently, the Compton amplitude for spin-$S$ heavy particles is
\begin{equation}\label{eq:4ptcompton}
    M_4(-\textbf{1}^S,2^{+1},3^{-1},\textbf{4}^S)=-\frac{\langle3|(-p_1)|2\rbrack^{2-2S}}{(s-m^2)(u-m^2)}(\langle\textbf{1}3\rangle\lbrack\textbf{4}2\rbrack-\lbrack\textbf{1}2\rbrack\langle\textbf{4}3\rangle)^{2S}.
\end{equation}
Similarly, the minimal couplings with the graviton are $M_3^{S,S,+2}=\frac{m}{M_{pl}}x^2\frac{\langle \textbf{1}\textbf{2}\rangle^{2S}}{m^{2S-1}}$ and $M_3^{S,S,-2}=\frac{m}{M_{pl}}\frac{1}{x^2}\frac{\lbrack \textbf{1}\textbf{2}\rbrack^{2S}}{m^{2S-1}}$. Note that the coupling constant for gravity is $\frac{m}{M_{pl}}$, where $M_{pl}$ is the Planck mass. The graviton Compton amplitude is given by
\begin{equation}\label{eq:4ptgravitoncompton}
   M_4(-\textbf{1}^S,2^{+2},3^{-2},\textbf{4}^S)= -\frac{1}{M_{pl}^2}\frac{\langle 3|(-p_1)|2\rbrack^{4-2S}}{(s-m^2)(u-m^2)t}\left(\langle\textbf{1}3\rangle\lbrack\textbf{4}2\rbrack-\lbrack\textbf{1}2\rbrack\langle\textbf{4}3\rangle\right)^{2S}.
\end{equation}
We now project the amplitudes of Eqs.~\eqref{eq:4ptcompton} and \eqref{eq:4ptgravitoncompton} into the $SU(2)$ velocity-projected amplitudes, and then expand them into the HPET amplitudes.

\paragraph{Projection}
We first reduce the photon Compton amplitude $M_4(\textbf{1}^S,2^{+1},3^{-1},\textbf{4}^S)$ with the $v\!\!\!/$-spinors. The heavy-particle propagator reads
\begin{equation}
      \frac{1}{(s-m^2)(u-m^2)}=\frac{1}{4m^2(v\cdot k_s+\frac{k_{s}^2}{2m})(v\cdot k_u+\frac{k_{u}^2}{2m})}.
\end{equation}
The factor $\langle3|(-p_1)|2\rbrack$, which carries the spurious pole for high spin, is expanded into the large mass and the small residual momentum as
\begin{equation}\label{eq:sppoleexpand}
     \la3|(-p_1)|2\rb=-\left(m+v\cdot k_1\right)\lceil3\tilde2\rfloor+\lceil3|k_{1\perp}|\tilde2\rfloor=-m\lceil3\tilde2\rfloor-\frac{1}{2}v\cdot(k_1+k_4)\lceil3\tilde2\rfloor+\frac{1}{2}\lceil3|k_{1\perp}+k_{4\perp}|\tilde2\rfloor,
\end{equation}
where the second equality follows from momentum conservation $p_1=q_2+q_3+p_4$.

Second, the spin dependent part can be projected as
\begin{eqnarray}\label{eq:spincomppart}
    \langle\textbf{1}3\rangle\lbrack\textbf{4}2\rbrack&=&-\lceil\boldsymbol{1}^+3\rfloor\lceil\boldsymbol{4}^+\tilde2\rfloor-\lceil\boldsymbol{1}^+3\rfloor\lceil4^-\tilde2\rfloor+\lceil\boldsymbol{1}^-3\rfloor\lceil\boldsymbol{4}^+\tilde2\rfloor+\lceil\boldsymbol{1}^-3\rfloor\lceil4^-\tilde2\rfloor,\nonumber\\
    \langle\textbf{4}3\rangle\lbrack\textbf{1}2\rbrack&=&-\lceil\boldsymbol{4}^+3\rfloor\lceil\boldsymbol{1}^+\tilde2\rfloor-\lceil\boldsymbol{4}^+3\rfloor\lceil1^-\tilde2\rfloor+\lceil\boldsymbol{4}^-3\rfloor\lceil\boldsymbol{1}^+\tilde2\rfloor+\lceil\boldsymbol{4}^-3\rfloor\lceil1^-\tilde2\rfloor.
\end{eqnarray}
Assembling the propagator expansion, the spurious-pole expansion of Eq.~\eqref{eq:sppoleexpand}, and the spin-structures of Eq.~\eqref{eq:spincomppart}, the velocity-projected Compton amplitude from the top-down projection of $M_4(-\textbf{1}^S,2^{+1},3^{-1},\textbf{4}^S)$ in Eq.~\eqref{eq:4ptcompton} coincides with the bottom-up result of $\mathcal{A}_4(-\textbf{1}^S,2^{+1},3^{-1},\textbf{4}^S)$ in Eq.~\eqref{eq:vampphotoncompton}.

In the graviton Compton amplitude of Eq.~\eqref{eq:4ptgravitoncompton}, all the heavy-particle poles are replaced by the heavy-particle propagators, with $\frac{1}{(s-m^2)(u-m^2)t}=\frac{1}{4m^2(v\cdot k_s+\frac{k_s^2}{2m})(v\cdot k_u+\frac{k_u^2}{2m})t}$. The remaining part of the graviton Compton amplitude is treated in the same way as above, and the velocity-projected graviton Compton amplitude $\mathcal{A}_4(-\textbf{1}^S,2^{+2},3^{-2},\textbf{4}^S)$ matches the bottom-up result of Eqs~\eqref{eq:vampgravitonncompton}, up to the overall gravitational coupling $m^2/M_{pl}^2$.

\paragraph{Expansion}
The massive spinors are decomposed into the large component $|\boldsymbol{k}^+\rfloor\sim\sqrt{m}$ and the small component $|\boldsymbol{k}^-\rfloor\sim\sqrt{m}\times k/m$, and the spin structures are expanded in these components through $k/m$.
The expansion result that only contains the $|\boldsymbol{k}^+\rfloor$ is
\begin{eqnarray}\label{eq:fraction2S4pt}
      \frac{\langle\textbf{1}3\rangle\lbrack\textbf{4}2\rbrack-\lbrack\textbf{1}2\rbrack\langle\textbf{4}3\rangle}{\langle3|-p_1|2\rbrack}
      &=&\frac{\lceil\boldsymbol{1}^+\boldsymbol{4}^+\rfloor}{m}+\frac{1}{2m}\frac{\lceil3|k_{1\perp}+k_{4\perp}|\tilde2\rfloor}{\lceil3\tilde2\rfloor}\frac{\lceil\boldsymbol{1}^+\boldsymbol{4}^+\rfloor}{m}\nonumber\\
    &-&\frac{1}{2m}\frac{\lceil\textbf{1}^+3\rfloor\lceil\textbf{4}^+|k_{4\perp}|\tilde 2\rfloor-\lceil\textbf{4}^+3\rfloor\lceil\textbf{1}^+|k_{1\perp}|\tilde 2\rfloor}{m\lceil3\tilde2\rfloor}\nonumber\\
      &-&\frac{1}{2m}\frac{\lceil\textbf{1}^+\tilde2\rfloor\lceil\textbf{4}^+|k_{4\perp}|3\rfloor-\lceil\textbf{4}^+\tilde 2\rfloor\lceil\textbf{1}^+|k_{1\perp}|3\rfloor}{m\lceil3\tilde2\rfloor}+\mathcal{O}(\frac{k^2}{m^2}).
\end{eqnarray}
where we have used the Schouten identity $-\lceil\boldsymbol{1}^+3\rfloor\lceil\boldsymbol{4}^+\tilde2\rfloor+\lceil\boldsymbol{4}^+3\rfloor\lceil\boldsymbol{1}^+\tilde2\rfloor=-\lceil\textbf{1}^+\textbf{4}^+\rfloor\lceil3\tilde2\rfloor$.
Since $\lceil\textbf{1}^+\textbf{4}^+\rfloor\sim m$, the first term in the above equation is $\mathcal{O}(1)$ and the remaining terms are $\mathcal{O}(k/m)$. The top-down expansion coincides with the bottom-up result of Eq.~\eqref{eq:bottomupcompton1} term by term.

In fact, for $S>1$ in the photon case and $S>2$ in the graviton case, the factor $\langle 3|-p_1|2\rbrack$ appears with a negative power in the relativistic amplitudes of Eqs.~\eqref{eq:4ptcompton} and \eqref{eq:4ptgravitoncompton}, leading to a spurious pole at $\langle3|-p_1|2\rbrack=0$. Rewriting $M_4(-\textbf{1}^S,2^{+1},3^{-1},\textbf{4}^S)$ and $M_4(-\textbf{1}^S,2^{+2},3^{-2},\textbf{4}^S)$ as a positive power of $\langle3|-p_1|2\rbrack$ times the $2S$-th power of the ratio in Eq.~\eqref{eq:fraction2S4pt}, and expanding this ratio in $1/m$, we find that only finitely many terms of the binomial expansion are kept at each order. 

At leading order the ratio reduces to $\lceil\textbf{1}^+\textbf{4}^+\rfloor/m$, so the dependence on the soft brackets $\lceil3\tilde2\rfloor$ is eliminated. Through the orders displayed, each kept term $\sim\mathcal{O}(k/m)$ carries at most one inverse power $\lceil3\tilde2\rfloor^{-1}$, while the prefactor supplies $\lceil3\tilde2\rfloor^2$ for photon and $\lceil3\tilde2\rfloor^4$ for gravitational amplitudes, so no negative power of $\lceil3\tilde2\rfloor$ survives in the leading terms or in the $1/m$ corrections of Eqs.~\eqref{eq:bottomupcompton1} and \eqref{eq:boup4ptgravitoncomp1}. The terms of the binomial expansion with further inverse powers of $\lceil3\tilde2\rfloor$ are precisely the higher-order spin effects, suppressed by additional powers of $1/m$ and discarded by the expansion, so they cannot reintroduce the pole.

To summarize, the Compton amplitudes are obtained both bottom-up, by gluing the HPET amplitudes through factorization consistency, and top-down, by velocity-projecting the relativistic amplitudes; the two routes agree order by order in the $1/m$ expansion, with the sub-leading corrections worked out explicitly in Eqs.~\eqref{eq:bottomupcompton1} and \eqref{eq:boup4ptgravitoncomp1}. These results provide the starting point for the gravitational applications developed in Section~\ref{sec:spmultipole}.

\section{Applications to Kerr Black Hole}\label{sec:spmultipole}
Having constructed the velocity-projected Compton amplitude in Eq.~\eqref{eq:k10gravitatiaolcomp} in the previous sections, we now apply them to the gravitational coupling of the Kerr black hole. In this section, the amplitudes are organized in the post-Minkowskian (PM) expansion of gravitational scattering, which counts powers of Newton's constant $G_{\text{N}}$. We work only at tree level, including the 1PM elastic scattering and the gravitational Compton amplitudes. In Section~\ref{sec:kerr} we match the HPET minimal coupling of Section~\ref{sec:threep} to the multipole Lagrangian of the worldline effective theory, fixing all the Wilson coefficients $C_{S^{n}}=1$, exactly up to order $2S$ for finite spin. In Section~\ref{sec:gfa}, the classical limits of the 4-point amplitudes are discussed, which are consistent with the known results, up to additional contact terms.

We need to confirm the $\hbar$ counting of amplitudes before we derive the classical limit. Comparing to the relativistic amplitudes, one feature of the velocity-projected amplitudes is that, the power counting in $k/m$ also counts the power of $\hbar$.

First, every momentum in the velocity-projected amplitudes has definite $\hbar$ counting:
\begin{itemize}
    \item the residual momentum $k^{\mu}$ of the heavy particle, $p^{\mu}=mv^{\mu}+k^{\mu}$, $k^{\mu}\rightarrow \hbar \, \bar k^{\mu}$;
    \item the soft massless boson momentum, $q^{\mu}\rightarrow \hbar \,\bar q^{\mu}$.
\end{itemize}
In the classical limit, $\hbar\rightarrow 0$,  the residual momenta and the soft momenta vanish $k^{\mu}\rightarrow0, q^{\mu}\rightarrow0$ while the wave numbers $\bar k$ and $\bar q$ remain constant. Consequently, the $\hbar$ counting and the $k/m$ counting of momenta are related by $  k/m \sim q/m \sim \hbar$.

Second, the spin of the heavy particle also has its $\hbar$ counting. In previous
sections, we have treated $S\ll m$ as in the usual HPET. In the classical limit,
however, the classical spin $S_{\text{cl}}=\hbar\, S$ must remain finite as
$\hbar\rightarrow0$, which forces $S\rightarrow\infty$. In this case, the $\hbar$
counting includes:
\begin{itemize}
    \item the spin operator introduced in Eq.~\eqref{eq:Smu}, $S^{\mu}\rightarrow\frac{S_{\text{cl}}}{\hbar }$;
    \item the contraction  $q\cdot S/m\rightarrow \bar q\cdot S_{\text{cl}}/m$.
\end{itemize}
Therefore, the contraction is $q\cdot S/m \sim k\cdot S/m \sim\mathcal{O}(1)$~\cite{Bern:2022kto},
and should be kept for all orders in the $k/m$ expansion. We follow the above power
counting in this section.

Keeping $q\cdot S/m$ of order one makes the velocity-projected
amplitude essential for the classical limit. Every power of $q\cdot S/m$ is then
equally important, so the spin-induced multipoles do not organize into a truncated
$q/m\sim k/m$ expansion. The velocity-projected amplitude retains both the large and
the small components, including $|\boldsymbol{k}^-\rfloor$, so that it keeps the
full $1/m$ expansion and hence the complete multipole tower. An HPET amplitude,
obtained by integrating out the small component, usually contains only part of the tower.
The classical limit below is therefore taken on the complete amplitude.

\subsection{Three-point Amplitudes as Spin-induced Multipole Expansions}\label{sec:kerr}

In the worldline EFT, an action describing the Kerr black hole contains a three-particle interaction, which can be matched to the three-particle amplitude. To be specific,
the one-body effective action of a heavy particle coupled to the graviton reads~\cite{Porto:2005ac,Porto:2008tb,Porto:2008jj,Levi:2015msa,Levi:2018nxp}
\begin{eqnarray}\label{eq:Lsi}
    S_{\text{WL}} = \int d\sigma \left\{ -m\sqrt{u^2} - \frac{1}{2} S_{\mu\nu}\Omega^{\mu\nu} + L_{\text{SI}} \left[ u^\mu, S_{\mu\nu}, g_{\mu\nu}(y^\mu) \right]  \right\},
\end{eqnarray}
where $u^{\mu}$ is the four-velocity, $S_{\mu\nu}$ the spin tensor, and $\Omega^{\mu\nu}$ the angular velocity. The spin interaction of the heavy particle is parameterized as
\begin{eqnarray}\label{eq:LSIlag}
L_{\text{SI}} &=& \sum_{n=1}^{\infty} \frac{(-1)^n}{(2n)!} \frac{C_{\mathrm{S}}^{2n}}{m^{2n-1}} D_{\mu_{2n}} \cdots D_{\mu_3} \frac{E_{\mu_1 \mu_2}}{\sqrt{u^2}} S^{\mu_1} S^{\mu_2} \cdots S^{\mu_{2n-1}} S^{\mu_{2n}} \nonumber\\
&&+ \sum_{n=1}^{\infty} \frac{(-1)^n}{(2n+1)!} \frac{C_{\mathrm{S}}^{2n+1}}{m^{2n}} D_{\mu_{2n+1}} \cdots D_{\mu_3} \frac{B_{\mu_1 \mu_2}}{\sqrt{u^2}} S^{\mu_1} S^{\mu_2} \cdots S^{\mu_{2n}} S^{\mu_{2n+1}}.
\end{eqnarray}
where $D_{\mu}$ is the covariant derivative and the Wilson coefficients $C_{S^{n}}$ are all equal to one for the Kerr black hole. The even orders couple to the mass multipoles through the electric-type field $E_{\mu\nu}$, and the odd orders to the current multipoles through the magnetic-type field $B_{\mu\nu}$; for the Kerr black hole the two towers are not independent, as they share the single rotation parameter.

To specify how the minimal coupling $\mathcal{A}^{(S,S,h)}$ that we derive in Eq.~\eqref{eq:alloderintegrateput3pt} can be applied to the Kerr black hole, we need 
to match $\mathcal{A}^{(S,S,h)}$ with the action $S_{\text{WL}}$. Therefore, we first discuss the spin operator $\hat S^{\mu}$ in the on-shell amplitude, in comparison with $S^{\mu}$ in the spin-dependent interaction $L_{\text{SI}}$.

\paragraph{The Spin Operator}
The $v\!\!\!/$-spinor realizes the little group rotation of the heavy-particle state, and the generator of this rotation is the spin operator. The same operator appears as the spin vector $S^{\mu}$ of the worldline effective theory below, so this identification is the bridge from the velocity-projected amplitudes to the multipole Lagrangian. In the spin-dependent interaction $L_{\text{SI}}$ Eq.~\eqref{eq:LSIlag}, the spin operator is the Pauli-Lubanski vector for the initial state particle 1:
\begin{equation}\label{eq:Smu}
    S^{\mu}=-\frac{1}{2}\epsilon^{\mu\nu\rho\sigma}p_{1\nu}J_{\rho\sigma}.
\end{equation}
To isolate the spin effect,
in this section we choose the reference velocity to coincide with that of one of the heavy particles,
\begin{eqnarray}\label{eq:zerok1}
    k_{1}=0,\quad p_1^{\mu}=mv^{\mu}.
\end{eqnarray}
The only residual momentum available for the orbital insertion is $k_{2\perp}=-q_{\perp}$. However, the relativistic amplitude itself is the same for a general $k_1^{\mu}$ due to the reparameterization invariance of $v^{\mu}$ (Section~\ref{sec:threep} and Appendix~\ref{apprpi}).
With the antisymmetrized products $\sigma^{\mu\nu}\equiv\frac{\mathrm{i}}{4}(\sigma^{\mu}\bar\sigma^{\nu}-\sigma^{\nu}\bar\sigma^{\mu})$ and $\bar\sigma^{\mu\nu}\equiv\frac{\mathrm{i}}{4}(\bar\sigma^{\mu}\sigma^{\nu}-\bar\sigma^{\nu}\sigma^{\mu})$, and using the identities $\sigma^{\mu\nu}=-\frac{1}{2}\mathrm{i}\epsilon^{\mu\nu\rho\kappa}\sigma_{\rho\kappa}$ and $\bar\sigma^{\mu\nu}=\frac{1}{2}\mathrm{i}\epsilon^{\mu\nu\rho\kappa}\bar\sigma_{\rho\kappa}$, the Lorentz generator in the chiral representation and the spin operator from Eqs.~\eqref{eq:Smu} and \eqref{eq:zerok1} are
\begin{eqnarray}
    J^{\mu\nu}=\left[\begin{array}{cc}
        (\sigma^{\mu\nu})_{\alpha}^{~\beta} &0  \\
        0 & (\bar{\sigma}^{\mu\nu})^{\dot\alpha}_{~\dot\beta}
    \end{array}\right],\qquad S^{\mu}=v_{\nu}\left[\begin{array}{cc}
        -\mathrm{i}(\sigma^{\mu\nu})_{\alpha}^{~\beta} &0  \\
        0 & \mathrm{i}(\bar{\sigma}^{\mu\nu})^{\dot\alpha}_{~\dot\beta}
    \end{array}\right].\nonumber
\end{eqnarray}

In the rest frame $v^{\mu}=(1,0,0,0)$, where $\sigma^{i0}=\frac{\mathrm{i}}{2}\sigma^i$ and $\bar\sigma^{i0}=-\frac{\mathrm{i}}{2}\sigma^i$, the non-vanishing spin operator is $\frac{\sigma^i}{2}$ for the left- and right-handed spinors, $S^{\mu}=(0,S^i)$ with $S^{i}=\mathrm{diag}(\frac{\sigma^i}{2},\frac{\sigma^i}{2})$, and the inner product reduces to $(q\cdot S)_{LL'}=-\vec q\cdot(\frac{\vec\sigma}{2})_{LL'}$. The Pauli matrix $(\vec\sigma/2)_L^{~L'}$ is independent of the momentum $\vec q$, and it is exactly the spin operator used in quantum mechanics. Consequently, for general $v^{\mu}$ we can denote the $SU(2)$ little group covariant operator as
\begin{equation}\label{eq:spinopqs}
    (q\cdot \hat S)_{LL'}\equiv-\frac{1}{2}(q_{\perp})_{LL'}.
\end{equation}
Since the reference velocity $v^{\mu}$ is kept arbitrary, the spin operator is defined in a general frame. Its matrix elements only involve the $SU(2)$ little group indices and the perpendicular transfer momentum $q_{\perp}$, with no reference to the rest frame of the particle.
The $n$th power of the spin operator, sandwiched between the velocity-projected states, acts on the symmetrized product of $2S$ heavy-particle spinors as
\begin{eqnarray}\label{eq:spinpower}
    \left[q\cdot \hat S\right]^n|\textbf{1}^+\rfloor^{2S}=\frac{(2S)!}{(2S-n)!}(-\frac{1}{2}q_{\perp})^n|\textbf{1}^+\rfloor^{2S},
\end{eqnarray}
which can be derived from the definition of the Lie algebra after symmetrization~\cite{Chung:2018kqs}. Divided by $n!$, the coefficient $(2S)!/(2S-n)!$ of the above equation becomes the binomial $\binom{2S}{n}$, and the matrix element of the $n$th power of $q\cdot \hat S/m$ between the symmetrized products of the two heavy-particle spinors reads
\begin{eqnarray}\label{eq:spinmatnn}
    \lceil \boldsymbol{2}^+|^{2S}\,\left[\frac{1}{n!}\left(\frac{q\cdot\hat S}{m}\right)^n\,\right]\,|\boldsymbol{1}^+\rfloor^{2S}&=&\binom{2S}{n}\lceil\boldsymbol{1}^+\boldsymbol{2}^+\rfloor^{2S-n}\left(-\frac{1}{2m}\lceil\boldsymbol{1}^+|q_{\perp}|\boldsymbol{2}^+\rfloor\right)^{n}.
\end{eqnarray}
Each term is thus realized by $n$ powers of the spin-operator insertion $q\cdot \hat S/m$ acting on the symmetrized product of the $2S$ heavy-particle spinors.

\paragraph{The Kerr Matching}
 With $k_1=0$, the covariant derivatives are replaced in momentum space by $D^{\mu}\rightarrow q^{\mu}$, and the spin operators act on the symmetrized products of the heavy-particle spinors. In the spin-dependent interaction $L_{\text{SI}}$, the fields $E_{\mu\nu}$ and $B_{\mu\nu}$ are translated into the $x^{2}$ factor, and the insertions of $S^{\mu}$ with their covariant derivatives $D_{\mu}$ correspond to the spin part and the orbital part. The conversion of the multipole insertions in Eq.~\eqref{eq:LSIlag} thus reads
\begin{eqnarray}\label{eq:translation}
    \frac{1}{n!}\frac{1}{m^n}D_{\mu_1}\cdots D_{\mu_n} S^{\mu_1}\cdots S^{\mu_n}&\longrightarrow&  \lceil \boldsymbol{2}^+|^{2S}\,\left[\frac{1}{n!}\left(\frac{q\cdot\hat S}{m}\right)^n\,\right]\,|\boldsymbol{1}^+\rfloor^{2S},
\end{eqnarray}
where this step is the matrix element of the spin operator acting on the heavy particle state; in the non-relativistic limit this is simply $q\cdot S\xrightarrow{\text{NR}}-\vec q\cdot\frac{\vec\sigma}{2}$.
With the above conversion, the worldline action of Eq.~\eqref{eq:Lsi} produces the three-particle amplitude
\begin{eqnarray}\label{eq:Msi}
    \mathcal{A}^{(S,S,+2)}_{\text{WL}} 
    &=&\frac{x^2}{m^{2S}} \frac{\kappa m^2}{2} \sum_{n=0}^{2S}C_{S^{n}}\binom{2S}{n}(\lceil\textbf{2}^+\textbf{1}^+\rfloor)^{2S-n}(-\frac{1}{2m}\lceil\textbf{2}^+|q_{\perp}|\textbf{1}^+\rfloor)^{n}.
\end{eqnarray}
which is consistent with the results in Refs.~\cite{Aoude:2020onz,Chung:2018kqs}. The factor $\lceil\textbf{2}^+\textbf{1}^+\rfloor$ is the spin-singlet contraction of the two heavy-particle spinors. Each further insertion of $-\frac{1}{2m}\lceil\textbf{2}^+|q_{\perp}|\textbf{1}^+\rfloor$ supplies one power of the orbital part coupled to the spin part as the scalar combination. The worldline amplitude is thus the binomial sum over the multipole orders, each weighted by the Wilson coefficient $C_{S^{n}}$.

On the other hand, the minimal coupling of Eq.~\eqref{eq:alloderintegrateput3pt} is already expressed as the HPET amplitude. In the $k_1=0$ frame of Eq.~\eqref{eq:zerok1}, the on-shell condition of the heavy leg, $(mv-q)^2=m^2$, gives $v\cdot q=0$ as well as $k_{2\perp}=-q_{\perp}$, and the small component of the heavy-particle bilinear reduces to
\begin{eqnarray}\label{eq:eomreduce}
    &&\lceil\boldsymbol{1}^+\boldsymbol{2}^-\rfloor=-\frac{1}{2m}\lceil\boldsymbol{1}^+|q_{\perp}|\boldsymbol{2}^+\rfloor,\quad\quad \lceil\boldsymbol{1}^-\boldsymbol{2}^+\rfloor=\lceil\boldsymbol{1}^-\boldsymbol{2}^-\rfloor=0,
\end{eqnarray}
where the contractions involving the small component of leg 1 vanish because they carry one power of the vanishing $k_{1\perp}$. The two structures entering the minimal coupling of Eq.~\eqref{eq:alloderintegrateput3pt}, the singlet $\lceil\textbf{2}^+\textbf{1}^+\rfloor$ and the contraction $\lceil\textbf{1}^+\tilde3\rfloor\lceil\textbf{2}^+\tilde3\rfloor$, are the ones fixed by gauge invariance, and they reduce in this frame to the exact identities
\begin{eqnarray}\label{eq:zerokindition}
    &&\langle\textbf{21}\rangle=\lceil \textbf{2}^+\textbf{1}^+\rfloor-\frac{1}{2m}\lceil\textbf{2}^+|q_{\perp}|\textbf{1}^+\rfloor,\quad\quad
    \langle\textbf{1}3\rangle\langle\textbf{2}3\rangle=\lceil\textbf{1}^+3\rfloor\lceil\textbf{2}^+3\rfloor,
    \nonumber\\
    &&\lbrack\textbf{21}\rbrack=-\lceil \textbf{2}^+\textbf{1}^+\rfloor-\frac{1}{2m}\lceil\textbf{2}^+|q_{\perp}|\textbf{1}^+\rfloor,\quad\quad
    \lbrack\textbf{1}3\rbrack\lbrack\textbf{2}3\rbrack=\lceil\textbf{1}^+\tilde3\rfloor\lceil\textbf{2}^+\tilde3\rfloor,
\end{eqnarray}
where for zero initial residual momentum the terms of order $1/m^2$ and higher all vanish. The choice $k_1=0$ thus merely trades the contraction $\lceil\textbf{1}^+\tilde3\rfloor\lceil\textbf{2}^+\tilde3\rfloor$ for the perpendicular insertion $\lceil\textbf{2}^+|q_{\perp}|\textbf{1}^+\rfloor$, since with the definition of $q_{\perp}$ in Eq.~\eqref{eq:qperp} the identity $\lceil\textbf{1}^+|q_{\perp}|\textbf{2}^+\rfloor=-\frac1{x}\lceil\textbf{1}^+\tilde 3\rfloor\lceil\textbf{2}^+\tilde 3\rfloor=-x\lceil\textbf{1}^+3\rfloor\lceil\textbf{2}^+3\rfloor$ holds, without changing the amplitude.
For the minimal coupling, the non-minimal couplings $g_{i>0}$ of Eq.~\eqref{eq:alloderintegrateput3pt} vanish, and, using the above identities, the minimal gravitational three-particle amplitude reduces to:
\begin{eqnarray}\label{eq:KKtower}
\mathcal{A}_{\text{min}}^{(S,S,+2)}&=&x^2g_0\sum_{n=0}^{2S}\binom{2S}{n}\lceil\boldsymbol{1}^+\boldsymbol{2}^+\rfloor^{2S-n}\left(-\frac{1}{2m}\lceil\boldsymbol{1}^+|q_{\perp}|\boldsymbol{2}^+\rfloor\right)^{n}\nonumber\\
        &=&x^2g_0\sum_{n=0}^{2S}  \lceil \boldsymbol{2}^+|^{2S}\,\left[\frac{1}{n!}\left(\frac{q\cdot\hat S}{m}\right)^n\,\right]\,|\boldsymbol{1}^+\rfloor^{2S},
\end{eqnarray}
Each insertion of $-\frac{1}{2m}\lceil\textbf{2}^+|q_{\perp}|\textbf{1}^+\rfloor$ acts as one power of $q\cdot S/m$, so the expansion in the number of insertions is at the same time the $1/m$ expansion of the heavy-particle amplitude and the spin-induced multipole expansion. In other words, integrating out the small component through Eq.~\eqref{eq:integratekminus} and organizing the multipole moments are one and the same operation in this framework.

The matching between $\mathcal{A}_{\text{WL}}$ and $\mathcal{A}_{\text{min}}$ proceeds order by order in the number of spin-operator insertions, and
the comparison of Eq.~\eqref{eq:KKtower} with the worldline amplitude of Eq.~\eqref{eq:Msi} fixes the Wilson coefficients:
\begin{eqnarray}\label{eq:Cone}
 \mathcal{A}_{\text{WL}}^{(S,S,+2)}= \mathcal{A}_{\text{min}}^{(S,S,+2)} \quad\Longrightarrow C_{S^{n}}=1,
\end{eqnarray}
with $g_0=\frac{1}{m^{2S}}\frac{\kappa m^2}{2}$, for all orders $n\leq 2S$ in the spin-induced multipoles. 
With the truncated exponential $\exp_{2S}(X)\equiv\sum_{n=0}^{2S}\frac{X^n}{n!}$, the finite-spin minimal coupling in Eq.~\eqref{eq:KKtower} can then be written compactly as
\begin{eqnarray}\label{eq:exp2S}
    \mathcal{A}^{(S,S,+2)}_{\text{min}}=\frac{\kappa m^2}{2}x^2\, \frac{\lceil\textbf{2}^+|^{2S}\exp_{2S}\left(\frac{q\cdot \hat S}{m}\right)|\textbf{1}^+\rfloor^{2S}}{m^{2S}},
\end{eqnarray}
which reproduces the multipoles of the Kerr black hole for finite spin with $C_{S^n}=1$. Here the coupling constant is $g_{\rm grav}\equiv\kappa m^2/2$ with $\kappa^2=32\pi G_{\text{N}}$, so that $g_0=g_{\rm grav}/m^{2S}$. The truncation of the exponential at order $2S$ is not an approximation. It is forced by the finite-dimensional representation, since $(q\cdot\hat S)^{n}$ with $n>2S$ vanishes on the spin-$S$ multiplet. The finite-spin quantum amplitude therefore carries only the first $2S$ multipole moments, while the full classical multipole tower of the Kerr black hole is restored by the coherent-state sum over all spins performed below.

In the $k_1=0$ frame, the $2S+1$ independent tensor structures of the three-particle amplitude are in one-to-one correspondence with the $2S+1$ independent spin-induced multipole moments. Matching all the coefficients in this frame is therefore exhaustive, and, since the multipole coefficients are frame independent, reparameterization invariance of $v^{\mu}$ extends the matched amplitude to a general frame. In the spin-exponentiated minimal coupling in the relativistic spinor-helicity formalism~\cite{Guevara:2019fsj}, an additional Lorentz boost is needed to describe the Kerr black hole~\cite{Arkani-Hamed:2019ymq}. Amplitudes can be recognized as operators acting on the Hilbert space of the particle states, and the relativistic spin operator, defined from the Pauli-Lubanski pseudovector, gives an alternative way to express the three-particle amplitudes. In contrast, the velocity-projected spin operator of Eq.~\eqref{eq:spinopqs} carries $SU(2)$ little group indices with respect to the arbitrary reference velocity $v^{\mu}$, so it is defined in a general frame and requires no additional boost.

\subsection{Gravitational 4-point Amplitudes}\label{sec:gfa}

\subsubsection{Quantum Amplitudes}\label{sec:quantumA4pt}

In Section~\ref{sec:fourp} we have derived the higher-spin Compton amplitudes for the general momenta.  To isolate the spin effect of the heavy particle, we set $k_{1}=0$ and $p_1^{\mu}=mv^{\mu}$ as in Eq.~\eqref{eq:zerok1}. In this frame the spurious-pole term reduces exactly to $-m\lceil3\tilde2\rfloor$, and the gravitational Compton amplitudes reduces to the velocity-projected amplitudes in Eq.~\eqref{eq:k10gravitatiaolcomp}.
Denoting the scalar part of Eq.~\eqref{eq:k10gravitatiaolcomp} as  
\begin{eqnarray}\label{eq:scalar4pt}
  \mathcal{A}_{4,\text{Compton}}^0\equiv-\frac{\kappa^2m^2}{4}\frac{\lceil3\tilde 2\rfloor^4}{4\left(v\cdot k_s+\frac{k_{s}^2}{2m}\right)\left(v\cdot k_u+\frac{k_{u}^2}{2m}\right)t},
\end{eqnarray}
the full gravitational Compton amplitude from the expansion of Eq.~\eqref{eq:k10gravitatiaolcomp} reads
\begin{eqnarray}\label{eq:4ptspf}
    \mathcal{A}_4(-\textbf{1}^S,2^{+2},3^{-2},\textbf{4}^S)=\mathcal{A}_{4,\text{Compton}}^0
   \left(\frac{1}{m}\lceil\textbf{1}^+|\mathbb{I}-\frac{(q_3-q_2)\cdot \hat S}{m_{k_4}}+\frac{w\cdot \hat S}{m_{k_4}} -\frac{q_2\cdot q_3}{m_{k_4}}\frac{|\tilde2\rfloor\lceil3|}{m\lceil\tilde23\rfloor}|\textbf{4}^+\rfloor\right)^{2S},
\end{eqnarray}
where we define $ (w\cdot \hat S)_L^{~L'}\equiv v\cdot q_2\, \frac{|3\rfloor_L\lceil\tilde2|^{L'}+|\tilde2\rfloor_L\lceil3|^{L'}}{\lceil\tilde23\rfloor}$. The vector $w$ carries the spurious pole of the Compton amplitude. It originates from the spurious pole  $\la3|-p_1|2\rb=-m\lceil3\tilde2\rfloor$ of the relativistic amplitude.  The shifted mass $m_{k_4}$ is defined in Eq.~\eqref{eq:integratekminus}, which absorbs the residual-momentum corrections of the outgoing heavy leg, $\frac{1}{m_{k_4}}\equiv\frac{1}{m}\left(1+\sum_{n=1}^{\infty}(-\frac{v\cdot k_4}{2m})^n\right)$.
The mass $m_{k_4}$ absorbs the high-order contributions from integrating out the small component $|\textbf{4}^-\rfloor$; it reduces to $m$ in the classical limit or the heavy mass limit, where these contributions vanish.
 In the above equation, the identity term describes the spinless Schwarzschild contribution, the $(q_3-q_2)\cdot \hat S$ term carries the Kerr multipoles inherited from the three-point minimal coupling, the $w\cdot \hat S$ term is the spurious pole structure, and the last term is a higher-order $1/m$ correction that drops out in the classical and heavy mass limits.

With the three-point minimal coupling of Eq.~\eqref{eq:exp2S} describing the Kerr multipole moments in hand, we now compute the elastic gravitational scattering of two heavy particles of arbitrary spins $S_a$ and $S_b$ at 1PM, where only single-graviton exchange contributes.
For the two-particle scattering, the reference velocity of each heavy particle is identified with the four-velocity of its incoming external leg, $p_1^{\mu}=m_a v_a^{\mu}$ and $p_3^{\mu}=m_b v_b^{\mu}$, where $v_a^{\mu}$ and $v_b^{\mu}$ are the four-velocities of the external legs 1 and 3, respectively. The outgoing heavy legs then carry the residual momenta $-q$ and $+q$; in the following, the spinor-helicity variables of the heavy legs 1, 2, 3 and 4, as well as the perpendicular momenta $q_{\perp}$, are defined with respect to the corresponding reference velocity, $v_a^{\mu}$ or $v_b^{\mu}$.

At 1PM, the elastic amplitude is given by the $t$-channel graviton exchange, where the residue of the four-particle amplitude at $q^2\rightarrow0$ factorizes into the product of two three-particle minimal couplings, summed over the graviton helicities,
\begin{eqnarray}\label{eq:1pmres}
    &&\lim_{q^2\rightarrow0}\,\text{Res}(t)=\lim_{q^2\rightarrow0}q^2\,\mathcal{A}_4(-\textbf{1}^{S_a},\textbf{2}^{S_a},-\textbf{3}^{S_b},\textbf{4}^{S_b})=\sum_{h=\pm2}\mathcal{A}_{\text{min}}(-\textbf{1}^S,\textbf{2}^S, q^{h})\mathcal{A}_{\text{min}}(-\textbf{3}^S,\textbf{4}^S, -q^{-h})\nonumber\\
    &=&\frac{\kappa^2m_a^2m_b^2}{4m_a^{2S_a}m_b^{2S_b}}\sum_{\pm}\left[(\frac{x_{12}}{x_{34}})^{\pm2}\left(\lceil \textbf{2}^+\textbf{1}^+\rfloor\mp\frac{1}{2m_a}\lceil\textbf{2}^+|q_{\perp}|\textbf{1}^+\rfloor\right)^{2S_a}\left(\lceil \textbf{4}^+\textbf{3}^+\rfloor\mp\frac{1}{2m_b}\lceil\textbf{4}^+|q_{\perp}|\textbf{3}^+\rfloor\right)^{2S_b}\right],\nonumber\\
\end{eqnarray}
where the heavy legs 1, 2 and 3, 4 belong to the particles $a$ and $b$, respectively, and the minus signs on the legs 1 and 3 denote that they are the incoming particles in the all-outgoing convention. The $x$ factors associated with the two graviton helicities combine into the boost factors of the relative motion, $(\frac{x_{12}}{x_{34}})^{\pm 2}=\gamma^2\,(1\pm v)^2,$ $\gamma=v_a\cdot v_b=\frac{1}{\sqrt{1-v^2}}$,
where $v$ is the relative velocity between particles 1 and 3, and $\gamma$ is the corresponding Lorentz factor. The four-particle amplitude reads
\begin{eqnarray}\label{eq:1pm4pt}
    \mathcal{A}_4(-\textbf{1}^{S_a},\textbf{2}^{S_a},-\textbf{3}^{S_b},\textbf{4}^{S_b})&=&16\pi G_{N}\frac{m_a^2m_b^2}{m_a^{2S_a}m_b^{2S_b}\,t}\,\sum_{\pm}\frac{1}{2}\gamma^2(1\pm v)^2\nonumber\\
    &&\times\left(\lceil \textbf{2}^+\textbf{1}^+\rfloor\mp\frac{1}{2m_a}\lceil\textbf{2}^+|q_{\perp}|\textbf{1}^+\rfloor\right)^{2S_a}\left(\lceil \textbf{4}^+\textbf{3}^+\rfloor\mp\frac{1}{2m_b}\lceil\textbf{4}^+|q_{\perp}|\textbf{3}^+\rfloor\right)^{2S_b}.\nonumber\\
\end{eqnarray}

\subsubsection{Classical Amplitudes}
\paragraph{Classical Kerr Amplitude}
We now derive the classical three-particle amplitude from the quantum on-shell amplitudes, rewriting the result in terms of classical variables. To absorb the little group indices of the particles in the expectation value, we take the states as the coherent-spin states~\cite{Aoude:2021oqj},
\begin{eqnarray}\label{eq:costate}
    |\textbf{1}^+\rfloor\rightarrow|\textbf{1}^+_{\text{coherent}}\rfloor\equiv e^{-|z|^2/2}\sum_{2S=0}^{\infty}\frac{(z_I)^{2S}}{\sqrt{(2S)!}}\left(\frac{|\textbf{1}^{+I}\rfloor}{\sqrt{m}}\right)^{2S},
\end{eqnarray}
where an auxiliary $SU(2)$ wave function $z_I$ is introduced.
Sandwiched between the initial state $|\textbf{1}^+_{\text{coherent}}\rfloor$ and the final state $|\textbf{2}^+_{\text{coherent}}\rfloor$, the expectation value of the spin operator reads $\langle \hat S^{\mu}\rangle= \frac{1}{m}\lceil\textbf{2}^+_{\text{coherent}}|\hat S^{\mu}|\textbf{1}^+_{\text{coherent}}\rfloor$. We can further define a classical vector, the ring radius vector $a^{\mu}$,
\begin{eqnarray}\label{eq:ringradius}
    a^{\mu}\equiv\frac{1}{m}\lceil\textbf{2}^+_{\text{coherent}}|\,\hat S^{\mu}\,|\textbf{1}^+_{\text{coherent}}\rfloor,
\end{eqnarray}
such that the corresponding classical three-particle amplitude is
\begin{eqnarray}\label{eq:classical3pt}
    \mathcal{A}_{\text{min,cl}}\equiv\lim_{\hbar\rightarrow0}\sum_{S}\mathcal{A}_{\text{min,coherent}}^{(S,S,+2)}=\lim_{\hbar\rightarrow0}e^{-|z|^2}\sum_{2S=0}^{\infty}\frac{(z^*_Iz_J)^{2S}}{(2S)!}\,\mathcal{A}_{\text{min}}^{(S,S,+2),IJ}=\frac{\kappa m^2}{2}x^2\,e^{q\cdot a}.
\end{eqnarray}
which is consistent with the results in Refs.~\cite{Guevara:2018wpp,Chung:2018kqs,Guevara:2019fsj,Arkani-Hamed:2019ymq,Guevara:2020xjx}. The classical exponent thus resums the multipole tower beyond the truncation order of any individual finite-spin amplitude, which is the mechanism of spin universality discussed in Refs.~\cite{Guevara:2019fsj,Cangemi:2023bpe}.
The vector $a^{\mu}$ has the dimension of length. In the classical limit, the truncated exponential $\exp_{2S}(X)$ up to order $2S$ is replaced by $\exp(X)$, and the ring radius is $a^{\mu}\rightarrow S^{\mu}_{\text{cl}}/m$.

\paragraph{1PM Elastic Gravitational Scattering}\label{sec:onepm}

In the classical limit, we absorb the little group indices of both heavy particles into the coherent spin states of Eq.~\eqref{eq:costate}, and send $\hbar\rightarrow0$, such that the truncated exponentials $\exp_{2S}$ of the finite-spin amplitudes are promoted to the classical exponentials. The resulting classical 1PM amplitude of Eq.~\eqref{eq:1pm4pt} is
\begin{eqnarray}\label{eq:1pmclassical}
    \mathcal{A}_{4,\text{Elastic,cl}}&=&\lim_{\hbar \rightarrow0}\sum_S \mathcal{A}_4(-\textbf{1}^{S_a},\textbf{2}^{S_a},-\textbf{3}^{S_b},\textbf{4}^{S_b})_{\text{coherent}}\nonumber\\
    &=&16\pi G_{N}\frac{m_a^2m_b^2}{t}\,\sum_{\pm}\frac{1}{2}\gamma^2(1\pm v)^2\exp\left[\pm q\cdot (a_a+a_b)\right],
\end{eqnarray}
which is consistent with the results in Refs.~\cite{Guevara:2018wpp,Guevara:2019fsj,Arkani-Hamed:2019ymq,Chung:2019duq,Parra-Martinez:2020dzs}. The two exponentials carry the complete spin-induced multipole moments of the two objects, resummed into the classical exponent. Note that the $SU(2)$ vector variables are written in the general frame, with the two arbitrary reference velocities $v^{\mu}_a$ and $v^{\mu}_b$; no specific frame has been chosen. In the general frame, the classical vectors $a_a$ and $a_b$ are contracted with the transfer momentum through the $SU(2)$ little group indices,
\begin{eqnarray}\label{eq:qaIJ}
    q\cdot a_a\equiv (q_{\perp})_{L}^{~L'} (a_a)_{L'}^{~L},\quad  q\cdot a_b\equiv (q_{\perp})_{J}^{~J'} (a_b)_{J'}^{~J}.
\end{eqnarray}
To make contact with the standard non-relativistic expressions, we reduce the above result to the leading order of the non-relativistic expansion, where $v_a=(1,0,0,0)$, $v_b=(1,\boldsymbol{v})$ with the spatial component $|\boldsymbol{v}|\ll1$, $p_1=(m_a,\boldsymbol{0})$, $p_3=(m_b,\boldsymbol{0})$, $q_{\perp}=(0,\boldsymbol{q})$ and $a_a=(0,\frac{\boldsymbol{\sigma}_a}{2m_a})$, $a_b=(0,\frac{\boldsymbol{\sigma}_b}{2m_b})$. Here $v_b^{\mu}=(1,\boldsymbol{v})$ is the non-relativistic limit ($\gamma\rightarrow1$, $v\rightarrow0$) of the boosted four-velocity $v_b^{\mu}=\gamma(1,0,0,v)$ used in the example below. Note that the classical information in the literature is usually evaluated in the holomorphic classical limit (HCL)~\cite{Guevara:2017csg}, which requires a specific kinematic configuration. In contrast, the $SU(2)$ contractions of Eq.~\eqref{eq:qaIJ} are defined for arbitrary $v_a^{\mu}$ and $v_b^{\mu}$, so that the classical amplitude can in principle be evaluated in any frame; the non-relativistic limit is taken here only for comparison. In this case the scalar amplitude contributes to the Newton potential
\begin{eqnarray}\label{eq:newton}
 \mathcal{A}^0_{4,\text{Elastic,cl}}\rightarrow \frac{ 16\pi G_{N}m_a^2m_b^2}{-\boldsymbol{ q}^2},\quad V(\boldsymbol{r})=\int \frac{d^3\boldsymbol{q}}{(2\pi)^3}\frac{1}{4m_am_b} \mathcal{A}^0_{4,\text{Elastic,cl}}\,e^{\mathrm{i}\boldsymbol{q}\cdot \boldsymbol{r}}=- \frac{G_{N}m_am_b}{|\boldsymbol{r}|},
\end{eqnarray}
while the exponential factor carries the spin effects from the spin-induced multipole moments. In the non-relativistic expansion, the covariant form Eq.~\eqref{eq:qaIJ} reduces to the $SO(3)$ rotation vectors as~\cite{Chung:2018kqs,Chung:2019duq,Chung:2020rrz}
\begin{eqnarray}\label{eq:nrspin}
    (q_{\perp})_{L}^{~L'}(a_a)_{L'}^{~L}\rightarrow\frac{\boldsymbol{S}_a}{m_a}\cdot\boldsymbol{q},\qquad (q_{\perp})_{J}^{~J'}(a_b)_{J'}^{~J}\rightarrow\frac{\boldsymbol{S}_b}{m_b}\cdot\boldsymbol{q},
\end{eqnarray}
while the Levi-Civita structure introduced by the on-shell equality $\mathrm{i}\epsilon_{\mu\nu\rho\sigma}v_a^{\mu}v_b^{\nu}q^{\rho}a^{\sigma}=\sqrt{\gamma^2-1}(q\cdot a)$ of Refs.~\cite{Guevara:2019fsj,Aoude:2021oqj} gives the spin-orbit part
\begin{eqnarray}\label{eq:nrspinorbit}
    \sqrt{\gamma^2-1}(q_{\perp})_{L}^{~L'}(a_a)_{L'}^{~L}\rightarrow\frac{\boldsymbol{S}_a}{m_a}\cdot\boldsymbol{q}\times\boldsymbol{v},\qquad \sqrt{\gamma^2-1}(q_{\perp})_{J}^{~J'}(a_b)_{J'}^{~J}\rightarrow\frac{\boldsymbol{S}_b}{m_b}\cdot\boldsymbol{q}\times\boldsymbol{v}.
\end{eqnarray}
In fact, the factors of $\gamma$ and $v$ enter the orbital part, giving the spin-orbit couplings, and the above two expressions are related in the HCL limit~\cite{Chung:2019duq}. In the frame $v_a=(1,0,0,0)$ and $v_b=\gamma(1,0,0,v)$, $\boldsymbol{q}\times\boldsymbol{v}$ denotes the orbital angular momentum in momentum space.

\paragraph{Gravitational Compton Amplitudes}\label{sec:gravcompton}

Rewriting this Compton amplitude of Eq.~\eqref{eq:4ptspf} in terms of the classical ring radius vector defined in Eq.~\eqref{eq:ringradius}, we obtain the following classical amplitude:
\begin{eqnarray}\label{eq:classical4pt}
      \mathcal{A}_{4,\text{cl}}(-\textbf{1},2^{+2},3^{-2},\textbf{4})&=&\lim_{\hbar\rightarrow0} \sum_S \mathcal{A}_{4}(-\textbf{1}^S,2^{+2},3^{-2},\textbf{4}^S)_{\text{coherent}}=\mathcal{A}_{4,\text{Compton,cl}}^0 \,e^{(q_3-q_2+w)\cdot a},
\end{eqnarray}
where $\mathcal{A}_{4,\text{Compton,cl}}^0=-\frac{\kappa^2}{4}\frac{\left(p_1\cdot F_2\cdot F_3\cdot p_1\right)^2}{\left(2p_1\cdot q_2\right)\left(2p_1\cdot q_3\right)t}$ is the classical scalar Compton amplitude in Eq.~\eqref{eq:clascalar4pt}.

At the quantum amplitude level, the original exponentiated gravitational Compton amplitude $\mathcal{A}_4$ derived in Eq.~\eqref{eq:4ptspf} has the same factorization channels as the amplitude in Ref.~\cite{Arkani-Hamed:2017jhn}. At the classical level, this amplitude $\mathcal{A}_{4,\text{cl}}$ in Eq.~\eqref{eq:classical4pt} also factorizes the same way as the higher-spin results in Refs.~\cite{Cangemi:2023ysz,Cangemi:2023bpe}. In the special case of polar scattering, where the pole in $w$ is directly canceled by the numerator, this classical Compton amplitude $\mathcal{A}_{4,\text{cl}}$ becomes a simple spurious-pole-free exponential~\cite{Bautista:2022wjf,Cangemi:2023bpe}. Our classical Compton amplitude differs in two respects. First, the coherent-spin state of Eq.~\eqref{eq:costate} is defined from the heavy-particle state instead of the Poincar\'{e} state, with the small component of the heavy particle integrated out. Second, since Eq.~\eqref{eq:classical4pt} is derived purely from the factorization, the potential additional contact terms are undetermined for the Kerr black hole, in contrast to Ref.~\cite{Cangemi:2023bpe}, and fixing them requires further physical input beyond the scope of this work.

To close this section, let us summarize the relation of our construction to the existing literature. First, the multipole matching is performed at an arbitrary reference velocity $v^{\mu}$. Since the $SU(2)$ covariant spin operator of Eq.~\eqref{eq:spinopqs} carries the spin degrees of freedom in a general frame, the matching performed once in the $k_1=0$ frame is extended to any frame by reparameterization invariance. The Kerr coefficients are thereby reproduced without invoking the boost part of the Hilbert space matching through which the reference-momentum formalism of Ref.~\cite{Chung:2019duq} achieves the same finite-spin result, or the additional Lorentz boost introduced for the spin-exponentiated form~\cite{Guevara:2019fsj,Arkani-Hamed:2019ymq}; the classical quantities are likewise defined without restricting to the HCL kinematics~\cite{Guevara:2017csg}. Second, integrating out the small component via Eq.~\eqref{eq:integratekminus} and organizing the multipole moments are one and the same operation. Each spin-operator insertion in the minimal coupling is simultaneously one power of $1/m$ and one multipole order, so that the Kerr condition $C_{S^{n}}=1$ is reproduced exactly up to order $2S$ for finite spin. Third, the truncation of the spin exponential at order $2S$ is forced by the finite-dimensional representation rather than imposed by hand, and the coherent-state sum over all spins restores the full classical multipole tower, giving a transparent derivation of the spin universality of Refs.~\cite{Chung:2019duq,Guevara:2019fsj,Cangemi:2023bpe}. With these inputs, the 1PM elastic amplitude and the classical gravitational Compton amplitude agree with the known results.

\section{Conclusion}

In this work we have developed a new on-shell spinor-helicity formalism for the scattering of heavy particles of arbitrary spin $S$. Projecting both the heavy-particle and soft-boson spinors replaces their Lorentz spinor indices with the common little group indices of a time-like reference velocity $v^{\mu}$.  We have shown that this formalism provides the on-shell counterpart of HPET, with each amplitude corresponding one-to-one to an HPET operator, in close connection with the constructive heavy particle effective theory and the effective operator construction for general non-relativistic theories.

The central step is an intermediate velocity-projected amplitude that retains both
the large and the small components, including $|\boldsymbol{k}^-\rfloor$, connecting
the relativistic and the HPET amplitudes. This amplitude can be constructed directly
from little group symmetry and kinematics, similar to the relativistic three-particle
amplitudes. The reparameterization condition takes a much simpler form on it, greatly
simplifying the calculation and making the HPET operator matching manifest. Compared with the HPET amplitudes, the velocity-projected amplitude keeps the complete multipole tower for the
classical limit of spinning amplitudes, where keeping the classical spin
$S_{\text{cl}}=\hbar S$ fixed as $\hbar\to0$ implies $S\to\infty$, so that
$k\cdot S/m\sim\mathcal{O}(1)$ must be kept to all orders in the $1/m$ expansion.


Our main results are the following. First, we have determined the most general $SU(2)$ invariant three-point velocity-projected amplitudes with two equal-mass particles of arbitrary finite spin $S$ and a soft massless boson of helicity $h$, given the $SU(2)$ little group symmetry and the degrees of freedom of the heavy particle. The construction is based on the velocity-projected spinors of the large and small components, so that the expansion in $k/m$ is explicit for every degree of freedom and the reparameterization invariance is substantially simplified. Compared with directly integrating out the AHH amplitudes, this construction yields more general Wilson coefficients, and while the  AHH amplitudes  are matched as a special case, our results are not limited to them. Second, we have obtained the velocity-projected quantum Compton amplitudes for arbitrary finite spin $S$, which is free of spurious poles up to the sub-leading order in $1/m$, reproduced as well by expanding the relativistic amplitude, the two routes agreeing order by order. In the heavy mass limit $m\rightarrow\infty$ the spin effects decouple, in direct analogy with the heavy quark spin symmetry.
    
    As a consistency check, we have applied our results to the gravitational scattering related to the Kerr black hole, and the conclusions agree with the literature. The three-particle minimal coupling with a graviton matches the multipole Lagrangian of the worldline effective theory on the HPET amplitudes, with all the Wilson coefficients $C_{S^{n}}=1$ of the Kerr black hole, exactly up to the maximal order $2S$ for finite spin, and the matching holds at an arbitrary reference velocity $v^{\mu}$, hence in a general frame, without the boost part of the Hilbert-space matching of the reference-momentum formalism, the HCL kinematics, or an additional Lorentz boost. The 1PM elastic scattering and the gravitational Compton amplitudes agree with the literature, up to additional contact terms.
    
Further developments include a systematic treatment of contact terms and interaction-dependent small-component relations, extensions to higher-point amplitudes through on-shell recursion, and applications at higher post-Minkowskian orders. The connection to HPET operator structures also motivates applications to non-relativistic QCD and potential non-relativistic QCD, as well as effective descriptions of dark matter with arbitrary spin.

\section*{Acknowledgements}
We would like to thank Fei Teng for valuable discussions and comments on the draft. This work is supported by the
National Science Foundation of China under Grants No. 12375099, No. 12347105  and No. 12047503, and the National Key Research and Development Program of China Grant No. 2020YFC2201501,
No. 2021YFA0718304.

\appendix

\section{Heavy Particle States and Wave Function}\label{app:HPET}
In an effective theory containing a heavy particle, the heavy mass $m$
naturally separates the dynamics into two distinct scales. This separation
is encoded by a timelike reference vector $v^{\mu}$ with
$v^2=1$, the velocity that labels the heavy particle, through the
momentum decomposition
\begin{equation}\label{eq:peqvk}
    p^{\mu}=mv^{\mu}+k^{\mu},\quad  m\gg  k.
\end{equation}
Here $v^{\mu}$ carries the large momentum, while $k^{\mu}$ is
the small residual momentum that fluctuates during interactions. For
example, in heavy quark effective theory the residual momentum is
of order $\Lambda_{\text{QCD}}$, much smaller than the heavy
quark mass $m_{Q}$. This decomposition reflects the superselection
rule~\cite{Georgi:1990um} for heavy-particle
propagation: the velocity sector is labeled by $v^\mu$, while the quantum fluctuation information is
encoded in $k^\mu$.

Below the heavy-particle production threshold, heavy-particle
number is conserved, so the ground state of the effective theory is labeled by the velocity sector and differs
from the conventional relativistic vacuum $|0\rangle$.
We denote this heavy-particle vacuum by $|\Omega\rangle$, which satisfies
\begin{equation}
    \langle \Omega|\hat P^{\mu}|\Omega\rangle \equiv mv^{\mu},
\end{equation}
where $v^\mu$ is the velocity introduced in Eq.~\eqref{eq:peqvk}.
Physically, $|\Omega\rangle$ is the ground state of the
sector containing a single heavy particle moving with velocity $v^\mu$.
In contrast, the relativistic vacuum $|0\rangle$ has vanishing
momentum expectation value and describes the absence of any particle
excitations.

In the rest frame $v^{\mu}=(1,0,0,0)$, the symmetry preserved
by $|\Omega\rangle$ reduces to spatial rotations and translations. The
full Poincar\'{e} symmetry is therefore spontaneously broken according to
\begin{equation}
    \mathbb{R}^{3,1}\rtimes SO(3,1)\longrightarrow  \mathbb{R}^{3,1}\rtimes SO(3)
    \simeq\mathbb{R}^1\times \text{E}(3),
\end{equation}
with $\text{E}(3)$ the three-dimensional Euclidean group. In this
symmetry-breaking pattern, the unbroken $SO(3)$ (and its double cover $SU(2)$) is the little
group that preserves the velocity $v^\mu$, under which the spin indices
of the heavy particle transform linearly. On the other hand, the
Lorentz boosts are spontaneously broken and nonlinearly
realized, manifesting as Wigner rotations in the transformation of the
heavy-particle states.

As discussed in Ref.~\cite{Li:2025ejk}, for a rotation $R\in SU(2)$
the spin-$S$ heavy-particle state transforms linearly under the unbroken $SU(2)$
symmetry as
\begin{equation}\label{eq:staterotation}
    U(R)|v,\vec k, I \rangle=\sum_{ I ^{\prime}}D^{(s)}_{ I ^{\prime} I }(R)|v,R\vec k, I ^{\prime}\rangle,
\end{equation}
where $ I $ denotes the spin index of the particle. Under the broken
Lorentz boost $\Lambda$, however, the state transforms nonlinearly as
\begin{equation}\label{eq:stateboost}
    U(\Lambda)|v,\vec k, I \rangle=\sum_{ I ^{\prime}}D^{(s)}_{ I ^{\prime} I }[W(\Lambda p,v)]|v,\vec k^{\prime}, I ^{\prime}\rangle,
\end{equation}
 where the Wigner rotation $W(\Lambda p,v)$ keeps $v^{\mu}$ invariant, and
$k^{\prime\mu}=\Lambda^{\mu}_{~\nu}p^{\nu}-mv^{\mu}$. The linear transformation
in Eq.~\eqref{eq:staterotation} preserves $|\vec k|$, while the boost
in Eq.~\eqref{eq:stateboost} mixes heavy-particle states with different $|\vec k|$.

The transformations above act on an infinite-dimensional unitary
representation of the heavy-particle state. To extract a finite-dimensional representation,  we introduce the  spin-1/2 heavy-particle wave
function $u_{ L}^{ I }(k)$ which corresponds to the $v\!\!\!/$-spinor $|\boldsymbol{k}^+\rfloor$, 
\begin{eqnarray}
    u_{ L}^{ I }(k)\longleftrightarrow |\boldsymbol{k}^+\rfloor_L^I.
\end{eqnarray}
Here $ L$ denotes the fundamental representation index of the
unbroken $SU(2)$ group.

In the rest frame $v^{\mu}=(1,0,0,0)$, rotations keep
$v^{\mu}$ invariant, and the rest-frame wave function $u_{ L}^{ I }(v)$
satisfies the constraint: $ \sum_{ I '}D^{(s)}_{ I  I ^{\prime}}(R)u_{ L}^{ I '}(v)
    =\sum_{\bar  L}D_{ L\bar  L}(R)u_{\bar  L}^{ I }(v)$.
The wave function at general momentum is then obtained from the rest-frame
wave function by the standard
boost $L_{ L\bar L}$:
$u_{ L}^{ I }(k)=\sum_{\bar L}L_{ L\bar  L}(\vec k)~u_{\bar L}^{ I }(v)$.
For $S=1/2$, $ L=1,2$, the standard boosts $S_L$ and
$S_R$ of the Weyl spinors $u_L(k)$ and $u_R(k)$ satisfy
$u_L(k)=S_L u(v)$ and $u_R(k)=S_R u(v)$. The standard boost matrix
$L_{ L\bar  L}(\vec k)=\frac{1}{2}\left(S_L + S_R \right)
= \sqrt{\frac{m+E}{2m}}\delta_{ L\bar  L}$ connects the rest-frame
wave function to the general-momentum wave function. 

Combining the constraint and the boost relation above, one finds that $u_{ L}^{ I }(k)$ transforms linearly under the spatial rotation
\begin{equation}
    \sum_{ I '}D^{(s)}_{ I  I ^{\prime}}(R)u_{ L}^{ I '}(Rk)
    =\sum_{\bar  L}D_{ L\bar  L}(R)u_{\bar  L}^{ I }(k),
\end{equation}
and transforms under the Lorentz boost as
\begin{equation}
    \sum_{ I '}D^{(s)}_{ I  I ^{\prime}}[W(\Lambda p,p_0)]u_{ L}^{ I '}(k')
    =\sum_{\bar  L}\left[L(k')D[W(\Lambda p,v)]L(k)^{-1}\right]_{ L\bar L}u_{\bar  L}^{ I }(k).
\end{equation}
These two relations are the wave-function counterparts of Eqs.~\eqref{eq:staterotation} and \eqref{eq:stateboost}.

For a heavy particle of general spin $S$, the wave function $u_{ L }^{ I }$ is obtained by the symmetric direct product of $2S$ spin-$1/2$ representations:
\begin{equation}
    \text{spin-}S: u_{\{ L_1}^{ \{I _1}(k) \cdots u_{ L_{2S}\}}^{ I _{2S}\}}(k),
\end{equation}
where $ I $ labels the $2S+1$ spin states and the $2S$ little group indices
$\{ L_1,\ldots, L_{2S}\}$ are totally symmetric. This is the wave-function counterpart of the symmetrized $v\!\!\!/$-spinor construction of Section~\ref{sec:spinor}, and it establishes the dictionary between the heavy-particle states and the on-shell variables used in the main text.

\section{Comparison with Literature}\label{subsec:comp}

We compare the $v\!\!\!/$-spinors with the on-shell heavy particle variables of Ref.~\cite{Aoude:2020onz}, emphasizing the different index structures and treatments of the auxiliary small component. Table~\ref{tab:comparisonspinors} summarizes the comparison row by row, and in the following we go through the rows in turn and cross-check the two expansion results.
\begin{table}[H]
    \centering
    \begin{tabular}{|c|c|c|c|}
    \hline
        & $v\!\!\!/$-spinor in this work & Spinors in Ref.~\cite{Aoude:2020onz} \\
         \hline
       Index after projection & $SU(2)_{LG}$&$SL(2,\mathbb{C})$ \\
       \hline
       Large component&$|\boldsymbol{k}^+\rfloor \Leftrightarrow N(x) $&
       $\begin{array}{l}
            |\boldsymbol{p}_v\rangle=v^{L}_{\alpha}|\boldsymbol{k}^+\rfloor_{L}  \\
            |\boldsymbol{p}_v\rbrack=\tilde v^{L\dot\alpha}|\boldsymbol{k}^+\rfloor_{L}\\
            \text{ Eq.~\eqref{eq:relationspinorpvk}}
       \end{array}$
   \\
       \hline
         Small component
    &$|\boldsymbol{k}^-\rfloor\Leftrightarrow\tilde N(x)$ &  \\
       \hline
       Massless spinor&$|q\rfloor$, $|\tilde q\rfloor$&$\begin{array}{l}
           |q\rangle=v_{\alpha}^L|q\rfloor_L   \\
            |\tilde q\rbrack=\tilde v^{L\dot\alpha} |\tilde q\rfloor_L
       \end{array}$
        \\
        &$\begin{array}{rc}
             |q\rfloor|q\rfloor+|\tilde q\rfloor|\tilde q\rfloor&\Leftrightarrow \vec B(x)   \\
        \mathrm{i}\left(|q\rfloor|q\rfloor-|\tilde q\rfloor|\tilde q\rfloor\right)&\Leftrightarrow\vec E(x)
        \end{array}$&\\
       \hline
       Reparameterization& $\begin{array}{l}
            |\boldsymbol{k}^+\rfloor \longrightarrow|\boldsymbol{k}^+\rfloor- \delta v|\boldsymbol{k}^-\rfloor,\\
|\boldsymbol{k}^-\rfloor\longrightarrow |\boldsymbol{k}^-\rfloor- \delta v|\boldsymbol{k}^+\rfloor,
       \end{array}$  &
       \begin{tiny}
           $\begin{array}{c}
               |\mathbf{p}_v\rangle =\frac{1}{1 - \frac{(k-2m\delta v)^2}{4m^2}} \left[ \left(1 - \frac{k^2}{4m^2} + \frac{k\!\!\!/\delta v\!\!\!/}{2m}\right) |\mathbf{p}_w\rangle - \delta v\!\!\!/ |\mathbf{p}_w] \right], \\
|\mathbf{p}_v] = \frac{1}{1 - \frac{(k-2m\delta v)^2}{4m^2}}  \left[ \left(1 - \frac{k^2}{4m^2} + \frac{k\!\!\!/\delta v\!\!\!/}{2m}\right) |\mathbf{p}_w] - \delta v\!\!\!/ |\mathbf{p}_w\rangle \right].
           \end{array}$
       \end{tiny}
       \\
       &Eq.~\eqref{eq:k+-rpi}&Eq.~\eqref{eq:refRPI}\\
      \hline
      Expansion results&\begin{scriptsize}$
     \begin{array}{ll}
           \la\textbf{21}\ra&=\lceil\boldsymbol{2}^+\boldsymbol{1}^+\rfloor-\lceil\boldsymbol{2}^+\boldsymbol{1}^-\rfloor\\
           &-\lceil\boldsymbol{2}^-\boldsymbol{1}^+\rfloor+\lceil\boldsymbol{2}^-\boldsymbol{1}^-\rfloor \\
           \Leftrightarrow&N_2^{\dagger}(x)N_1(x)\\
           &+\frac{1}{2m}N_2^{\dagger}(x)\mathrm{i}\vec\sigma\cdot\vec\nabla N_1(x)\\
           &+\frac{1}{2m}\mathrm{i}\vec\sigma\cdot\vec\nabla N_2^{\dagger}(x)N_1(x)\\
           &-\frac{1}{4m^2}\vec\sigma\cdot \vec\nabla N^{\dagger}_2(x)\vec\sigma\cdot\vec\nabla N_1(x)\\
           &+\cdots
      \end{array}$
      \end{scriptsize}&\begin{scriptsize}
         $ \begin{array}{rl}
             \langle \textbf{21} \rangle &= \frac{m^2}{m_{k_2} m_{k_1}} \left[ \frac{m_{k_1}}{m} \langle \textbf{2}_v \textbf{1}_v \rangle \right.\\
             &\left.+ \frac{1}{4m} [\textbf{2}_v q] \langle q \textbf{1}_v \rangle - \frac{x^{-1}}{4m} [\textbf{2}_v q] [q \textbf{1}_v] \right]
         \end{array}$
      \end{scriptsize}\\
      \hline
    \end{tabular}
    \caption{Comparison of the on-shell variables. The small component is integrated out and is absent in Ref.~\cite{Aoude:2020onz}. In our work, we can keep the small component $|\boldsymbol{k}^-\rfloor$ as the auxiliary variable in determining the amplitudes and the reparameterization. The small component can be integrated out in the final result.}
    \label{tab:comparisonspinors}
\end{table}

In heavy quark (particle) effective theory, the projected wave function $\mathcal{U}^I_v(p)$ is obtained from the Dirac wave function $\mathcal{U}^I(p)$ by $\mathcal{U}^I_v(p)=\frac{1+v\!\!\!/}{2}\mathcal{U}^I(p)$. Analogously, the on-shell heavy particle variables of Ref.~\cite{Aoude:2020onz}, which carry $SL(2,\mathbb{C})$ indices with the little group index $I$ suppressed by the $\textbf{BOLD}$ symbol, are defined as
\begin{eqnarray}\label{eq:onshelhpetdef}
    \left[\begin{array}{c}
         |\boldsymbol{p}_v\rangle  \\
         |\boldsymbol{p}_v]
    \end{array}\right]=\left(\frac{1+v\!\!\!/}{2}\right)\left[\begin{array}{c}
         |\boldsymbol{p}\rangle  \\
         |\boldsymbol{p}]
    \end{array}\right].
\end{eqnarray}
The definition uses the momentum decomposition $p^{\mu}=mv^{\mu}+k^{\mu}$ and the Dirac equation $\frac{p\!\!\!/}{m}\mathcal{U}^I(p)=\mathcal{U}^I(p)$; the residual momentum $k_{\alpha\dot\beta}$ is not explicitly computed. Through the time component decomposition $v\cdot p=m+v\cdot k$ and the on-shell condition, the time component $v\cdot k$ is traded for the scalar $k^2$, so that the projected variables depend only on $m_k\equiv m+\frac{v\cdot k}{2}=m-\frac{k^2}{4m}$.

The first row of Table~\ref{tab:comparisonspinors} concerns the index structure after the projection. The variables of Eq.~\eqref{eq:onshelhpetdef} retain the $SL(2,\mathbb{C})$ spinor indices, whereas in the $v\!\!\!/$-spinors these indices have been contracted with the velocity spinors, so only the little group that stabilizes $v^\mu$ remains manifest. Lorentz invariance itself is retained through reparameterization invariance.

The second row compares the large components, which carry the same physical information. In our construction, the kinematic data are encoded in the total momentum and its on-shell spinors, while the reference velocity specifies the decomposition. Utilizing the momentum decomposition $p^{\mu}=mv^{\mu}+k^{\mu}$ and the Dirac equation, the projected variables defined in Eq.~\eqref{eq:onshelhpetdef} are expressed directly in terms of the total momentum and the velocity,
\begin{eqnarray}\label{eq:refpvtotal}
    |\boldsymbol{p}_v\rangle&=&\frac{1}{2}\left(|\boldsymbol{p}\rangle+v_{\alpha\dot\beta}|\boldsymbol{p}]\right),\nonumber\\
    |\boldsymbol{p}_v]&=&\frac{1}{2}\left(|\boldsymbol{p}]+v^{\dot\alpha\beta}|\boldsymbol{p}\rangle\right),
\end{eqnarray}
so that the explicit residual-momentum dependence is removed from the projection. Applying the expansion Eq.~\eqref{eq:expandAHH} to the right-hand sides and contracting the velocity spinors, the on-shell heavy particle spinor with $SL(2,\mathbb{C})$ indices defined in Eq.~\eqref{eq:onshelhpetdef} is converted into our $v\!\!\!/$-spinor with $SU(2)$ indices by
\begin{eqnarray}\label{eq:relationspinorpvk}
    \left\{
    \begin{array}{lll}
         v^{L}_{\alpha}|\boldsymbol{k}^+\rfloor_{L} & =&|\boldsymbol{p}_v\rangle, \\
        \\
         \tilde v^{L\dot\alpha}|\boldsymbol{k}^+\rfloor_{L}&=&|\boldsymbol{p}_v].
    \end{array}
    \right.
\end{eqnarray}
The large component $|\boldsymbol{k}^+\rfloor$ is thus the spinor-helicity counterpart of the non-relativistic field $N(x)$.

The third row concerns the auxiliary small component. Apart from the large $SU(2)$ heavy-particle spinor, we also keep the small component $|\boldsymbol{k}^-\rfloor_{L}$ as an auxiliary variable. By contrast, the small degree of freedom is directly integrated out in the on-shell heavy particle variables of Ref.~\cite{Aoude:2020onz}. To restore the implicit Lorentz boost invariance, Ref.~\cite{Aoude:2020onz} adopts a top-down matching and uses the tree-level integrate-out equations of motion,
$\mathcal{U}^I(p)=\left(1+\frac{1}{2m}\left(1+\frac{k\cdot v}{2m}\right)^{-1}(k\!\!\!/-k\cdot v)\right)\mathcal{U}_v^I(p)$,
expanding the relativistic spinors as
\begin{eqnarray}\label{eq:sec:refexpandrelativistic}
    \left\{
    \begin{array}{lll}
         |\boldsymbol{p}\rangle
         & =&\frac{m}{2m_k} |\boldsymbol{p}_v\rangle+\frac{1}{2m_k}p_{\alpha\dot\beta} |\boldsymbol{p}_v]\\
         \\
          |\boldsymbol{p}]&=&\frac{m}{2m_k} |\boldsymbol{p}_v]+\frac{1}{2m_k}p^{\dot\alpha\beta} |\boldsymbol{p}_v\rangle.
    \end{array}
    \right.
\end{eqnarray}
Since the small component has been integrated out, this expansion expresses the relativistic spinors solely in terms of the projected variables $|\boldsymbol{p}_v\rangle$ and $|\boldsymbol{p}_v]$, which retain only the large component. On our side, the large component is dynamical, and the small component, suppressed by $k/m$, contributes only at higher orders; it is eliminated at the final step by the equations of motion Eq.~\eqref{eq:integrateoutk-spinor}, order by order in $k/m$.

The fourth row compares the soft photon spinors. Ref.~\cite{Aoude:2020onz} uses the relativistic spinors $|q\rangle$ and $|\tilde q]$ directly, without projecting them onto the velocity. These relativistic spinors are converted into our $v\!\!\!/$-spinors of Eq.~\eqref{eq:qfloordef} by projecting them onto the velocity spinors, and the conversion can be inverted as $|q\rangle=v_{\alpha}^{L}|q\rfloor_{L}$ and $|\tilde q\rbrack=\tilde v^{L\dot\alpha}|\tilde q\rfloor_{L}$. In the $v^{\mu}$ frame the photon field strength reduces to the electric and magnetic fields $\vec E$ and $\vec B$, and under the velocity projection these are represented by the symmetric $SU(2)$ tensor combinations $|q\rfloor|q\rfloor+|\tilde q\rfloor|\tilde q\rfloor$ and $\mathrm{i}\left(|q\rfloor|q\rfloor-|\tilde q\rfloor|\tilde q\rfloor\right)$, which correspond to $\vec B$ and $\vec E$, respectively, as listed in Table~\ref{tab:comparisonspinors}. These are precisely the structures that couple the soft photon to the heavy-particle operators.

The fifth row compares the two implementations of reparameterization invariance. Under the reparameterization $w\rightarrow v$, the variables of Ref.~\cite{Aoude:2020onz} transform as
\begin{eqnarray}\label{eq:refRPI}
|\mathbf{p}_v\rangle &= \left(1 - \frac{(k-2m\delta v)^2}{4m^2}\right)^{-1} \left[ \left(1 - \frac{k^2}{4m^2} + \frac{k\!\!\!/\delta v\!\!\!/}{2m}\right) |\mathbf{p}_w\rangle - \delta v\!\!\!/ |\mathbf{p}_w] \right], \nonumber\\
|\mathbf{p}_v] &= \left(1 - \frac{(k-2m\delta v)^2}{4m^2}\right)^{-1} \left[ \left(1 - \frac{k^2}{4m^2} + \frac{k\!\!\!/\delta v\!\!\!/}{2m}\right) |\mathbf{p}_w] - \delta v\!\!\!/ |\mathbf{p}_w\rangle \right].
\end{eqnarray}
In the bottom-up construction, by contrast, the projected heavy spinors transform linearly under reparameterization,
\begin{eqnarray}\label{eq:k+-rpi}
    |\boldsymbol{k}^+\rfloor &\longrightarrow&|\boldsymbol{k}^+\rfloor- \delta v|\boldsymbol{k}^-\rfloor,\nonumber\\
        |\boldsymbol{k}^-\rfloor&\longrightarrow& |\boldsymbol{k}^-\rfloor- \delta v|\boldsymbol{k}^+\rfloor.
\end{eqnarray}
Together with the transformation of the projected soft boson discussed in the next section, Eq.~\eqref{eq:k+-rpi} provides a compact implementation of reparameterization invariance that retains the auxiliary variable. This use of an auxiliary variable can also be found in Ref.~\cite{Sundrum:1997ut}.

Finally, the last row cross-checks the two expansion results. The original relativistic massive spinor product, expanded with Eq.~\eqref{eq:sec:refexpandrelativistic}, reads
\begin{eqnarray}\label{eq:refexp21}
    \langle \textbf{21} \rangle = \frac{m^2}{m_{k_2} m_{k_1}} \left[ \frac{m_{k_1}}{m} \langle \textbf{2}_v \textbf{1}_v \rangle + \frac{1}{4m} [\textbf{2}_v q] \langle q \textbf{1}_v \rangle - \frac{x^{-1}}{4m} [\textbf{2}_v q] [q \textbf{1}_v] \right].
\end{eqnarray}
An additional minus sign in the last term comes from the different definitions of the $x$ factor. This integrated-out expansion depends explicitly on the photon spinors $|q\rangle$ and $|q]$, whereas the original relativistic product does not. On our side, utilizing the expansion Eq.~\eqref{eq:expandAHH} for the two heavy particles and the velocity normalization $\langle v_{I}v_{J}\rangle=\epsilon_{IJ}$, the same product is expanded in the $v\!\!\!/$-spinor basis as
\begin{equation}\label{eq:vspinexp21}
    \langle \textbf{21}\rangle=\lceil\boldsymbol{2}^+\boldsymbol{1}^+\rfloor-\lceil\boldsymbol{2}^+\boldsymbol{1}^-\rfloor-\lceil\boldsymbol{2}^-\boldsymbol{1}^+\rfloor+\lceil\boldsymbol{2}^-\boldsymbol{1}^-\rfloor .
\end{equation}
The four terms are ordered by the number of small components, and each small component carries one power of $k/m$, so the expansion has definite power counting with the first term dominant at leading order. Eliminating the small components with Eq.~\eqref{eq:integrateoutk-spinor} order by order in $k/m$, the successive terms reproduce the operator expansion of the non-relativistic fields listed in the last row of Table~\ref{tab:comparisonspinors}, with the leading term corresponding to $N_2^{\dagger}(x)N_1(x)$. Since both Eq.~\eqref{eq:refexp21} and Eq.~\eqref{eq:vspinexp21} are exact rewritings of the same relativistic spinor product, the two expansions agree with each other order by order in $k/m$. In our expansion, the heavy-particle $v\!\!\!/$-spinors and the residual momenta are the only ingredients, so no photon spinors appear and the power counting remains manifest at every order.

An additional consistency check of the conversion Eq.~\eqref{eq:relationspinorpvk} concerns the minimal coupling of Ref.~\cite{Aoude:2020onz}, where a special case is adopted when $k_1^{\mu}=0$, leading to a redundancy in the on-shell heavy particle variables, $\la\textbf{1}_v\textbf{2}_v\ra=-\lb\textbf{1}_v\textbf{2}_v\rb$. However, this redundancy actually always holds, due to the on-shellness of $v^{\mu}$ and the constraint $v\!\!\!/\mathcal{U}_v^s(p)=\mathcal{U}_v^s(p)$. This can be seen directly in terms of our $v\!\!\!/$-spinor, since
$\la\textbf{1}_v\textbf{2}_v\ra= v^{\alpha L}|\boldsymbol{1}^+\rfloor_L v_{\alpha}^{L'}|\boldsymbol{2}^+\rfloor_{L'}=\lceil\boldsymbol{1}^+\boldsymbol{2}^+\rfloor$
and
$-\lb\textbf{1}_v\textbf{2}_v\rb=-\tilde v^{L}_{\dot\alpha}|\boldsymbol{1}^+\rfloor_{L}\tilde v^{L'\dot\alpha}|\boldsymbol{2}^+\rfloor_{L'}=\lceil\boldsymbol{1}^+\boldsymbol{2}^+\rfloor$.
The $v\!\!\!/$-spinor structures can be matched systematically to heavy-particle operators. In the present variables, the relativistic spinor product admits the direct expansion in Eq.~\eqref{eq:expandAHH}, where the small components are integrated out by Eq.~\eqref{eq:integrateoutk-spinor}, order by order in $k/m$. The following section develops the corresponding amplitude basis and its relation to heavy-particle operators.

\section{Amplitude-operator Correspondence}\label{app:aoc}

In this appendix we make explicit the correspondence between the integrated-out on-shell amplitudes $\mathcal{A}$ and the non-relativistic operators, following the construction of Ref.~\cite{Li:2026eym}, which relates the operators to $SU(2)$ semi-standard Young tableaux. Each box of a tableau carries a little group index $L$; columns are antisymmetrized with the invariant tensor $\epsilon_{LL'}$, and rows are symmetrized. Contracting the column indices with the external $v\!\!\!/$-spinors and field strengths then produces the corresponding amplitude. In the present notation this correspondence reads:
\begin{eqnarray}
\begin{Young}
	$ L_1$\cr
 $ L_{2}$\cr
\end{Young}&\sim& \epsilon_{ L_1 L_2}\times|\boldsymbol{1}^+\rfloor^{ L_1}|\boldsymbol{2}^+\rfloor^{ L_2}\nonumber\\
\nonumber\\
&=&\lceil\boldsymbol{1}^+\boldsymbol{2}^+\rfloor,\nonumber\\
\nonumber\\
     \begin{Young}
	$ L_1$&$ L_{2}$\cr
 $ L_{3}$&$ L_3$\cr
\end{Young} &\sim& 
\epsilon_{ L_1 L_3}\epsilon_{ L_2 L_3}\times |\boldsymbol{1}^+\rfloor^{ L_1}|\boldsymbol{2}^+\rfloor^{ L_2}|F_3\rfloor^{ L_3}|F_3\rfloor^{ L_3}\nonumber\\
\nonumber\\
&=&\lceil\boldsymbol{1}^+F_3\rfloor\lceil\boldsymbol{2}^+ F_3\rfloor,
\nonumber\\
\nonumber\\
      \begin{Young}
	$ L_1$&$ L_2$&$ L_2'$\cr
 $ L_2'$&$ L_3$&$ L_3$\cr
\end{Young}&\sim& \epsilon_{ L_1 L_2'}\epsilon_{ L_2 L_3}\epsilon_{ L_2' L_3}\times|\boldsymbol{1}^+\rfloor^{ L_1}|\boldsymbol{2}^+\rfloor^{ L_2}|F_3\rfloor^{ L_3}|F_3\rfloor^{ L_3}(k_{2\perp})^{ L_2' L_2'}\nonumber\\
\nonumber\\
&=&\lceil\boldsymbol{1}^+|k_{2\perp}|F_3\rfloor\lceil\boldsymbol{2}^+F_3\rfloor.
\end{eqnarray}
Here $L$ denotes the little group index of the reference velocity; for the rest frame $v^{\mu}=(1,0,0,0)$, this group reduces to the spatial rotation group.
The first tableau contracts the two heavy-particle indices into the basic bilinear $\lceil\boldsymbol{1}^+\boldsymbol{2}^+\rfloor$. In the second and third tableaux, the two additional boxes associated with the soft photon are contracted with the on-shell field strength, and in the third tableau a transverse momentum $k_{2\perp}$ is inserted between the heavy leg and the field strength. Here $F$ denotes the on-shell soft photon field strength, representing either the magnetic or the electric field,
\begin{equation}
    |F_3\rfloor|F_3\rfloor=\left\{\begin{array}{ll}
         |B_3\rfloor|B_3\rfloor\equiv |3\rfloor|3\rfloor+|\tilde 3\rfloor|\tilde3\rfloor,  \\
        \\|E_3\rfloor|E_3\rfloor\equiv \mathrm{i}|3\rfloor|3\rfloor-\mathrm{i}|\tilde 3\rfloor|\tilde3\rfloor.
    \end{array}\right.
\end{equation}

Consequently, the complete basis for the integrated-out non-relativistic on-shell amplitudes can be constructed using the method of Ref.~\cite{Li:2026eym}. Each tableau carries a definite heavy-particle power counting. The amplitudes up to dimension 7 are listed in Tab.~\ref{Tab:dim567ampssyt}, while the explicit operators with magnetic or electric fields up to dimension 6 are listed in Tab.~\ref{tab:aoc}.








\begin{table}[H]
    \centering
    \begin{tabular}{|c|c|c|c|}
    \hline
         dim& SSYT& Amplitudes   \\
         \hline
         5& \begin{Young}
	$ L_1$&$ L_{2}$\cr
 $ L_{3}$&$ L_3$\cr
\end{Young}&$\lceil\boldsymbol{1}^+F_3\rfloor\lceil\boldsymbol{2}^+ F_3\rfloor$
         \\
         \hline
         6&
       
      \begin{Young}
	$ L_1$&$ L_2$&$ L_2'$\cr
 $ L_2'$&$ L_3$&$ L_3$\cr
\end{Young}&$\lceil\boldsymbol{1}^+|k_{2\perp}|F_3\rfloor\lceil\boldsymbol{2}^+F_3\rfloor$

\\
\cline{2-3}
 & \begin{Young}
	$ L_1$&$ L_2'$&$ L_2'$\cr
 $ L_2$&$ L_3$&$ L_3$\cr
\end{Young}&$-\lceil\boldsymbol{1}^+\boldsymbol{2}^+\rfloor\lceil F_3|k_{2\perp}|F_3\rfloor$
\\
\hline   

7& \begin{Young}
	$ L_1$&$ L_2$&\small{$ L_2''$}&\small{$ L_2''$}\cr
 $ L_2'$&$ L_2'$&$ L_3$&$ L_3$\cr
\end{Young}& $\lceil\boldsymbol{1}^+|k_{2\perp}|\boldsymbol{2}^+\rfloor\lceil F_3|k_{2\perp}|F_3\rfloor$

\\
\cline{2-3}

&\begin{Young}
	$ L_1$&$ L_2$&$ L_2'$&\small{$ L_2''$}\cr
 $ L_2'$&\small{$ L_2''$}&$ L_3$&$ L_3$\cr
\end{Young}& $\lceil\boldsymbol{1}^+|k_{2\perp}|F_3\rfloor\lceil \boldsymbol{2}^+|k_{2\perp}|F_3\rfloor$

\\
\cline{2-3}

&\begin{Young}
	$ L_1$&$ L_2$&$ L_2'$&$ L_2'$\cr
 $ L_3$&$ L_3$&$ L_3$&$ L_3$\cr
\end{Young}& $-\lceil\boldsymbol{1}^+F_3\rfloor\lceil \boldsymbol{2}^+F_3\rfloor\text{tr}[k_{2\perp}q_{\perp}]$
\\
\hline

    \end{tabular}
\caption{Correspondence between the semi-standard Young tableaux and the non-relativistic on-shell amplitudes. Here $F_3$ denotes either the electric field $E$ or the magnetic field $B$. With the heavy-particle power counting, this table lists the complete operator/amplitude basis at dimensions five, six and seven for interactions containing one heavy-fermion bilinear and one on-shell soft photon.}\label{Tab:dim567ampssyt}
\end{table}

In Tab.~\ref{tab:aoc}, the same dictionary is applied to the electric and magnetic field strengths, where $F_3$ in Tab.~\ref{Tab:dim567ampssyt} is replaced by either $\boldsymbol{B}$ or $\boldsymbol{E}$ according to the identification above. The resulting operators are displayed up to dimension six.
\begin{table}[H]\small
    \centering
    \begin{tabular}{|c|c|c|}
    \hline
        Young tableau& Amplitudes&Operators\\
         \hline
         
    \begin{Young}
	$ L_1$&$ L_3$\cr
 $ L_3$&$ L_2$\cr
\end{Young}& $\lceil\boldsymbol{1}^+3\rfloor\lceil3\boldsymbol{2}^+\rfloor+\lceil\boldsymbol{1}^+\tilde3\rfloor\lceil\tilde3\boldsymbol{2}^+\rfloor$&$N_1^{\dagger}\boldsymbol{\sigma}\cdot\boldsymbol{B}_3N_2$\\
\hline

\begin{Young}
$ L_1$&$  L_3$&$  L_2'$\cr
 $  L_2$&$ L_2'$&$ L_3$\cr
\end{Young}
        & $\lceil\boldsymbol{1}^+\boldsymbol{2}^+\rfloor\lceil3|k_{2\perp}|3\rfloor+\lceil\boldsymbol{1}^+\boldsymbol{2}^+\rfloor\lceil\tilde3|k_{2\perp}|\tilde3\rfloor$
        &$2N_1^{\dagger}\boldsymbol{B}_3\cdot\boldsymbol{D}N_2$
        \\
        \hline

\begin{Young}
$ L_1$&$  L_3$&$  L_2'$\cr
 $  L_3$&$ L_2'$&$ L_2$\cr
\end{Young}$-$\begin{Young}
$ L_1$&$  L_2'$&$  L_3$\cr
 $  L_2'$&$ L_3$&$ L_2$\cr
\end{Young}
&
$\begin{array}{c}
     \left(\lceil\boldsymbol{1}^+3\rfloor\lceil3|k_{2\perp}|\boldsymbol{2}^+\rfloor+\lceil\boldsymbol{1}^+\tilde3\rfloor\lceil\tilde3|k_{2\perp}|\boldsymbol{2}^+\rfloor\right)  \\
     -\left(\lceil\boldsymbol{1}^+|k_{2\perp}|3\rfloor\lceil3\boldsymbol{2}^+\rfloor+\lceil\boldsymbol{1}^+|k_{2\perp}|\tilde3\rfloor\lceil\tilde3\boldsymbol{2}^+\rfloor\right) 
\end{array}$&$2iN_1^{\dagger}\boldsymbol{\sigma}\cdot\boldsymbol{B}_3\times\boldsymbol{D}N_2$

\\
\hline

 \begin{Young}
	$ L_1$&$ L_3$\cr
 $ L_3$&$ L_2$\cr
\end{Young}& $i\lceil\boldsymbol{1}^+3\rfloor\lceil3\boldsymbol{2}^+\rfloor-i\lceil\boldsymbol{1}^+\tilde3\rfloor\lceil\tilde3\boldsymbol{2}^+\rfloor$&$N_1^{\dagger}\boldsymbol{\sigma}\cdot\boldsymbol{E}_3N_2$\\
\hline

\begin{Young}
$ L_1$&$  L_3$&$  L_2'$\cr
 $  L_2$&$ L_2'$&$ L_3$\cr
\end{Young}
        & $i\lceil\boldsymbol{1}^+\boldsymbol{2}^+\rfloor\lceil3|k_{2\perp}|3\rfloor-i\lceil\boldsymbol{1}^+\boldsymbol{2}^+\rfloor\lceil\tilde3|k_{2\perp}|\tilde3\rfloor$
        &$2N_1^{\dagger}\boldsymbol{E}_3\cdot\boldsymbol{D}N_2$
        \\
        \hline

\begin{Young}
$ L_1$&$  L_3$&$  L_2'$\cr
 $  L_3$&$ L_2'$&$ L_2$\cr
\end{Young}$-$\begin{Young}
$ L_1$&$  L_2'$&$  L_3$\cr
 $  L_2'$&$ L_3$&$ L_2$\cr
\end{Young}
&
$\begin{array}{c}
     \left(i\lceil\boldsymbol{1}^+3\rfloor\lceil3|k_{2\perp}|\boldsymbol{2}^+\rfloor-i\lceil\boldsymbol{1}^+\tilde3\rfloor\lceil\tilde3|k_{2\perp}|\boldsymbol{2}^+\rfloor\right)  \\
     -\left(i\lceil\boldsymbol{1}^+|k_{2\perp}|3\rfloor\lceil3\boldsymbol{2}^+\rfloor-i\lceil\boldsymbol{1}^+|k_{2\perp}|\tilde3\rfloor\lceil\tilde3\boldsymbol{2}^+\rfloor\right) 
\end{array}$&$2iN_1^{\dagger}\boldsymbol{\sigma}\cdot\boldsymbol{E}_3\times\boldsymbol{D}N_2$

\\
\hline

    \end{tabular}
\caption{Amplitude-operator correspondence up to dimension six for effective interactions involving the magnetic field $\boldsymbol{B}$ and the electric field $\boldsymbol{E}$. The tableaux are read in the same way as in Tab.~\ref{Tab:dim567ampssyt}.} \label{tab:aoc}
\end{table}

The dictionary displayed in these tables is used in Sections~\ref{sec:spinor} and \ref{sec:threep} to translate the integrated-out amplitudes into heavy-particle operators.

  

\section{Reparameterization of Three-Point Amplitudes}\label{apprpi}

In this appendix we spell out the reparameterization calculation quoted in subsection~\ref{sec:bottomup}. We start from the most general ansatz for the dimension-four $(1/2,1/2,+1)$ three-particle amplitude built out of the four little group structures of Eq.~\eqref{eq:b3pt1}, with arbitrary coefficients $c_i$ to be fixed by reparameterization invariance. Factoring out the common dimensionful coefficient $1/(v\cdot q)$, we parametrize the amplitude as
\begin{equation}
    \frac{A}{v\cdot q}\equiv\frac{1}{v\cdot q}\left(c_1\lceil\boldsymbol{1}^-\tilde3\rfloor\lceil\boldsymbol{2}^+\tilde3\rfloor+c_2\lceil\boldsymbol{1}^+\tilde3\rfloor\lceil\boldsymbol{2}^-\tilde3\rfloor+
    c_3\lceil\boldsymbol{1}^+\tilde3\rfloor\lceil\boldsymbol{2}^+\tilde3\rfloor+
    c_4\lceil\boldsymbol{1}^-\tilde3\rfloor\lceil\boldsymbol{2}^-\tilde3\rfloor\right).
\end{equation}
Under an infinitesimal reparameterization, the building blocks transform according to the rules of Eq.~\eqref{eq:rpiall}; in particular the numerator and the scalar factor $v\cdot q$ transform as
\begin{equation}
    \left\{\begin{array}{ll}
         A&\longrightarrow A+\delta A  \\
         v\cdot q& \longrightarrow v\cdot q+2\delta v\cdot q,
    \end{array}\right.
\end{equation}
and the requirement that the full amplitude be invariant, $\frac{A}{v\cdot q}=\frac{A+\delta A}{v\cdot q
+2\delta v\cdot q}$, leads, to first order in $\delta v$, to
\begin{equation}\label{deltavA}
    2\delta v\cdot q A=v\cdot q\delta A.
\end{equation}
Notice that $v\cdot q=\frac{1}{2}\lceil3\tilde3\rfloor$ and $2\delta v\cdot q=-\lceil3|\delta v|\tilde3\rfloor$, so that the LHS of Eq.~\eqref{deltavA} reads
\begin{equation}
    2\delta v \cdot q A=-\lceil3|\delta v|\tilde3\rfloor\left(c_1\lceil\boldsymbol{1}^-\tilde3\rfloor\lceil\boldsymbol{2}^+\tilde3\rfloor+c_2\lceil\boldsymbol{1}^+\tilde3\rfloor\lceil\boldsymbol{2}^-\tilde3\rfloor+
    c_3\lceil\boldsymbol{1}^+\tilde3\rfloor\lceil\boldsymbol{2}^+\tilde3\rfloor+
    c_4\lceil\boldsymbol{1}^-\tilde3\rfloor\lceil\boldsymbol{2}^-\tilde3\rfloor\right).
\end{equation}
On the other hand, the RHS is obtained by applying the transformation rules to each of the four structures in $A$. For a generic structure, each heavy-particle leg contributes one variation in which $|\boldsymbol{k}^{\pm}\rfloor$ is replaced by the opposite component contracted with $\delta v$, and each photon leg contributes one variation in which $|\tilde3\rfloor$ is replaced by $-|\delta v|\tilde3\rfloor$. Collecting these variations, the RHS is
\begin{equation}
\begin{array}{lll}
        v\cdot q\delta A &=&\frac{1}{2}\lceil3\tilde3\rfloor
        \\
       & \times(&c_1\lceil\boldsymbol{1}^+|\delta v|\tilde3\rfloor\lceil\boldsymbol{2}^+\tilde3\rfloor
        -c_1\lceil\boldsymbol{1}^-|\delta v|\tilde3\rfloor\lceil\boldsymbol{2}^+\tilde3\rfloor
        +c_1\lceil\boldsymbol{1}^-\tilde3\rfloor\lceil\boldsymbol{2}^-|\delta v|\tilde3\rfloor
        -c_1\lceil\boldsymbol{1}^-\tilde3\rfloor\lceil\boldsymbol{2}^+|\delta v|\tilde3\rfloor  \\
     & +&c_2\lceil\boldsymbol{1}^-|\delta v|\tilde3\rfloor\lceil\boldsymbol{2}^-\tilde3\rfloor
        -c_2\lceil\boldsymbol{1}^+|\delta v|\tilde3\rfloor\lceil\boldsymbol{2}^-\tilde3\rfloor
        +c_2\lceil\boldsymbol{1}^+\tilde3\rfloor\lceil\boldsymbol{2}^+|\delta v|\tilde3\rfloor
        -c_2\lceil\boldsymbol{1}^+\tilde3\rfloor\lceil\boldsymbol{2}^-|\delta v|\tilde3\rfloor \\
        &+&c_3\lceil\boldsymbol{1}^-|\delta v|\tilde3\rfloor\lceil\boldsymbol{2}^+\tilde3\rfloor
        -c_3\lceil\boldsymbol{1}^+|\delta v|\tilde3\rfloor\lceil\boldsymbol{2}^+\tilde3\rfloor
        +c_3\lceil\boldsymbol{1}^+\tilde3\rfloor\lceil\boldsymbol{2}^-|\delta v|\tilde3\rfloor
        -c_3\lceil\boldsymbol{1}^+\tilde3\rfloor\lceil\boldsymbol{2}^+|\delta v|\tilde3\rfloor\\
        &+&c_4\lceil\boldsymbol{1}^+|\delta v|\tilde3\rfloor\lceil\boldsymbol{2}^-\tilde3\rfloor
        -c_4\lceil\boldsymbol{1}^-|\delta v|\tilde3\rfloor\lceil\boldsymbol{2}^-\tilde3\rfloor
        +c_4\lceil\boldsymbol{1}^-\tilde3\rfloor\lceil\boldsymbol{2}^+|\delta v|\tilde3\rfloor
        -c_4\lceil\boldsymbol{1}^-\tilde3\rfloor\lceil\boldsymbol{2}^-|\delta v|\tilde3\rfloor),
\end{array}
\end{equation}
after collecting the like terms,
\begin{equation}\label{LHS}
    \begin{array}{lll}
         v\cdot q\delta A&=\frac{1}{2}\lceil3\tilde3\rfloor
         \\
         &\times\left((c_1-c_3)\lceil\boldsymbol{1}^+|\delta v|\tilde3\rfloor\lceil\boldsymbol{2}^+\tilde3\rfloor
       -(c_1-c_3)\lceil\boldsymbol{1}^-|\delta v|\tilde3\rfloor\lceil\boldsymbol{2}^+\tilde3\rfloor\right.
        \\
        &+(c_1-c_4)\lceil\boldsymbol{1}^-\tilde3\rfloor\lceil\boldsymbol{2}^-|\delta v|\tilde3\rfloor
        -(c_1-c_4)\lceil\boldsymbol{1}^-\tilde3\rfloor\lceil\boldsymbol{2}^+|\delta v|\tilde3\rfloor  \\
     &+(c_2-c_4)\lceil\boldsymbol{1}^-|\delta v|\tilde3\rfloor\lceil\boldsymbol{2}^-\tilde3\rfloor
       -(c_2-c_4)\lceil\boldsymbol{1}^+|\delta v|\tilde3\rfloor\lceil\boldsymbol{2}^-\tilde3\rfloor
        \\
        &\left.+(c_2-c_3)\lceil\boldsymbol{1}^+\tilde3\rfloor\lceil\boldsymbol{2}^+|\delta v|\tilde3\rfloor
        -(c_2-c_3)\lceil\boldsymbol{1}^+\tilde3\rfloor\lceil\boldsymbol{2}^-|\delta v|\tilde3\rfloor\right).
    \end{array}
\end{equation}
To compare the two sides of Eq.~\eqref{deltavA}, we rewrite the left-hand side in the spinor basis of the right-hand side. The two sides are written in different spinor bases, so we first reexpress the combinations $\lceil3|\delta v|\tilde3\rfloor\lceil\boldsymbol{1}^{\pm}\tilde3\rfloor$ appearing in the left-hand side in terms of the structures appearing in the right-hand side. Due to the Schouten identity,
\begin{equation}
\begin{array}{ll}
       \left(\lceil3|\delta v|\tilde3\rfloor\lceil\boldsymbol{1}^-\tilde3\rfloor\right)\lceil\boldsymbol{2}^+\tilde3\rfloor&=\left(-\lceil3\boldsymbol{1}^-\rfloor\lceil\tilde3|\delta v|\tilde3\rfloor+\lceil3\tilde3\rfloor\lceil\boldsymbol{1}^-|\delta v|\tilde3\rfloor\right)\lceil\boldsymbol{2}^+\tilde3\rfloor  \\
     &=\left(v\cdot q\lceil\boldsymbol{1}^-\boldsymbol{2}^+\rfloor-\lceil\boldsymbol{1}^-|q_{\perp}|\boldsymbol{2}^+\rfloor\right)\lceil\tilde3|\delta v|\tilde3\rfloor+\lceil3\tilde3\rfloor\lceil\boldsymbol{1}^-|\delta v|\tilde3\rfloor\lceil\boldsymbol{2}^+\tilde3\rfloor\\
     &=-\lceil\boldsymbol{1}^-|q_{\perp}|\boldsymbol{2}^+\rfloor\lceil\tilde3|\delta v|\tilde3\rfloor+v\cdot q\lceil\boldsymbol{1}^-\boldsymbol{2}^+\rfloor\lceil\tilde3|\delta v|\tilde3\rfloor+2v\cdot q\lceil\boldsymbol{1}^-|\delta v|\tilde3\rfloor\lceil\boldsymbol{2}^+\tilde3\rfloor
     \\
     &=(v\cdot k_2+v\cdot q)\lceil\boldsymbol{1}^+\boldsymbol{2}^+\rfloor\lceil\tilde3|\delta v|\tilde3\rfloor+(2m+v\cdot k_2)\lceil\boldsymbol{1}^-\boldsymbol{2}^-\rfloor\lceil\tilde3|\delta v|\tilde3\rfloor
     \\
     &+v\cdot q\lceil\boldsymbol{1}^-\boldsymbol{2}^+\rfloor\lceil\tilde3|\delta v|\tilde3\rfloor+2v\cdot q\lceil\boldsymbol{1}^-|\delta v|\tilde3\rfloor\lceil\boldsymbol{2}^+\tilde3\rfloor,
\end{array}
\end{equation}
where by definition
\begin{equation}
    \lceil\boldsymbol{1}^-3\rfloor\lceil\boldsymbol{2}^+\tilde3\rfloor=v\cdot q\lceil\boldsymbol{1}^-\boldsymbol{2}^+\rfloor-\lceil\boldsymbol{1}^-|q_{\perp}|\boldsymbol{2}^+\rfloor,
\end{equation}
and we have used momentum conservation $k_{1\perp}=k_{2\perp}+q_{\perp}$ and the heavy-particle equations of motion Eq.~\eqref{eq:EOMfermion}, such that
\begin{equation}
\begin{array}{lll}
     \lceil\boldsymbol{1}^-|q_{\perp}|\boldsymbol{2}^+\rfloor&=&\lceil\boldsymbol{1}^-|k_{1\perp}|\boldsymbol{2}^+\rfloor-\lceil\boldsymbol{1}^-|k_{2\perp}|\boldsymbol{2}^+\rfloor  \\
     &=&-(v\cdot k_2+v\cdot q)\lceil\boldsymbol{1}^+\boldsymbol{2}^+\rfloor -(2m+v\cdot k_2)\lceil\boldsymbol{1}^-\boldsymbol{2}^-\rfloor.
\end{array}
\end{equation}
For the same reason, we have
\begin{equation}
\begin{array}{ll}
       \left(\lceil3|\delta v|\tilde3\rfloor\lceil\boldsymbol{1}^+\tilde3\rfloor\right)\lceil\boldsymbol{2}^-\tilde3\rfloor
     &=-\lceil\boldsymbol{1}^+|q_{\perp}|\boldsymbol{2}^-\rfloor\lceil\tilde3|\delta v|\tilde3\rfloor+v\cdot q\lceil\boldsymbol{1}^+\boldsymbol{2}^-\rfloor\lceil\tilde3|\delta v|\tilde3\rfloor+2v\cdot q\lceil\boldsymbol{1}^+|\delta v|\tilde3\rfloor\lceil\boldsymbol{2}^-\tilde3\rfloor
     \\
     &=(2m+v\cdot k_2+v\cdot q)\lceil\boldsymbol{1}^-\boldsymbol{2}^-\rfloor\lceil\tilde3|\delta v|\tilde3\rfloor+(v\cdot k_2)\lceil\boldsymbol{1}^+\boldsymbol{2}^+\rfloor\lceil\tilde3|\delta v|\tilde3\rfloor
     \\
     &+v\cdot q\lceil\boldsymbol{1}^+\boldsymbol{2}^-\rfloor\lceil\tilde3|\delta v|\tilde3\rfloor+2v\cdot q\lceil\boldsymbol{1}^+|\delta v|\tilde3\rfloor\lceil\boldsymbol{2}^-\tilde3\rfloor,
\end{array}
\end{equation}

\begin{equation}
\begin{array}{ll}
       \left(\lceil3|\delta v|\tilde3\rfloor\lceil\boldsymbol{1}^+\tilde3\rfloor\right)\lceil\boldsymbol{2}^+\tilde3\rfloor
     &=-\lceil\boldsymbol{1}^+|q_{\perp}|\boldsymbol{2}^+\rfloor\lceil\tilde3|\delta v|\tilde3\rfloor+v\cdot q\lceil\boldsymbol{1}^+\boldsymbol{2}^+\rfloor\lceil\tilde3|\delta v|\tilde3\rfloor+2v\cdot q\lceil\boldsymbol{1}^+|\delta v|\tilde3\rfloor\lceil\boldsymbol{2}^+\tilde3\rfloor
     \\
     &=(2m+v\cdot k_2+v\cdot q)\lceil\boldsymbol{1}^-\boldsymbol{2}^+\rfloor\lceil\tilde3|\delta v|\tilde3\rfloor+(2m+v\cdot k_2)\lceil\boldsymbol{1}^+\boldsymbol{2}^-\rfloor\lceil\tilde3|\delta v|\tilde3\rfloor
     \\
     &+v\cdot q\lceil\boldsymbol{1}^+\boldsymbol{2}^+\rfloor\lceil\tilde3|\delta v|\tilde3\rfloor+2v\cdot q\lceil\boldsymbol{1}^+|\delta v|\tilde3\rfloor\lceil\boldsymbol{2}^+\tilde3\rfloor,
\end{array}
\end{equation}

\begin{equation}
\begin{array}{ll}
       \left(\lceil3|\delta v|\tilde3\rfloor\lceil\boldsymbol{1}^-\tilde3\rfloor\right)\lceil\boldsymbol{2}^-\tilde3\rfloor
     &=-\lceil\boldsymbol{1}^-|q_{\perp}|\boldsymbol{2}^-\rfloor\lceil\tilde3|\delta v|\tilde3\rfloor+v\cdot q\lceil\boldsymbol{1}^-\boldsymbol{2}^-\rfloor\lceil\tilde3|\delta v|\tilde3\rfloor+2v\cdot q\lceil\boldsymbol{1}^-|\delta v|\tilde3\rfloor\lceil\boldsymbol{2}^-\tilde3\rfloor
     \\
     &=(v\cdot k_2+v\cdot q)\lceil\boldsymbol{1}^+\boldsymbol{2}^-\rfloor\lceil\tilde3|\delta v|\tilde3\rfloor+(v\cdot k_2)\lceil\boldsymbol{1}^-\boldsymbol{2}^+\rfloor\lceil\tilde3|\delta v|\tilde3\rfloor
     \\
     &+v\cdot q\lceil\boldsymbol{1}^-\boldsymbol{2}^-\rfloor\lceil\tilde3|\delta v|\tilde3\rfloor+2v\cdot q\lceil\boldsymbol{1}^-|\delta v|\tilde3\rfloor\lceil\boldsymbol{2}^-\tilde3\rfloor.
\end{array}
\end{equation}
The left-hand side is then written as
\begin{equation}
    \begin{array}{lll}
       2\delta v\cdot q A  &=& -
       \lceil\boldsymbol{1}^+\boldsymbol{2}^+\rfloor\lceil\tilde3|\delta v|\tilde3\rfloor\left((c_1+c_2)v\cdot k_2+(c_1+c_3)v\cdot q\right)
       \\
         &&-\lceil\boldsymbol{1}^-\boldsymbol{2}^-\rfloor \lceil\tilde3|\delta v|\tilde3\rfloor\left((c_1+c_2)2m+(c_1+c_2)v\cdot k_2+(c_2+c_4)v\cdot q\right)
         \\
         &&-\lceil\boldsymbol{1}^-\boldsymbol{2}^+\rfloor \lceil\tilde3|\delta v|\tilde3\rfloor\left((c_1+c_3)v\cdot q+(c_3+c_4)v\cdot k_2+2c_3m\right)
         \\
         &&-\lceil\boldsymbol{1}^+\boldsymbol{2}^-\rfloor \lceil\tilde3|\delta v|\tilde3\rfloor\left((c_2+c_4)v\cdot q+(c_3+c_4)v\cdot k_2+2c_3m\right)
         \\
         &&-2c_1v \cdot q\lceil\boldsymbol{1}^-|\delta v|\tilde 3\rfloor\lceil\boldsymbol{2}^+\tilde3\rfloor
         \\
         &&-2c_2v \cdot q\lceil\boldsymbol{1}^+|\delta v|\tilde 3\rfloor\lceil\boldsymbol{2}^-\tilde3\rfloor
         \\
         &&
         -2c_3v \cdot q\lceil\boldsymbol{1}^+|\delta v|\tilde 3\rfloor\lceil\boldsymbol{2}^+\tilde3\rfloor\\
         &&
         -2c_4v \cdot q\lceil\boldsymbol{1}^-|\delta v|\tilde 3\rfloor\lceil\boldsymbol{2}^-\tilde3\rfloor.
    \end{array}
\end{equation}
In the first four lines of the above equation, $\delta v$ is carried by the factor $\lceil\tilde3|\delta v|\tilde3\rfloor$; to compare with Eq.~\eqref{LHS} we redistribute it onto either of the two $\tilde3$ spinors. Once again, due to the Schouten identity,
\begin{equation}
    \lceil\boldsymbol{1}^+\boldsymbol{2}^+\rfloor\lceil\tilde3|\delta v|\tilde3\rfloor=-\lceil\boldsymbol{1}^+|\delta v|\tilde3\rfloor\lceil\boldsymbol{2}^+\tilde3\rfloor+\lceil\boldsymbol{1}^+\tilde3\rfloor\lceil\boldsymbol{2}^+|\delta v|\tilde3\rfloor,
\end{equation}
so that the LHS becomes
\begin{equation}\label{RHS}
    \begin{array}{lll}
       2\delta v\cdot q A  &=& -
       \lceil\boldsymbol{1}^+\tilde3\rfloor\lceil\boldsymbol{2}^+|\delta v|\tilde3\rfloor\left((c_1+c_2)v\cdot k_2+(c_1+c_3)v\cdot q\right)
       \\
         &&-\lceil\boldsymbol{1}^-\tilde3\rfloor\lceil\boldsymbol{2}^-|\delta v|\tilde3\rfloor\left((c_1+c_2)2m+(c_1+c_2)v\cdot k_2+(c_2+c_4)v\cdot q\right)
         \\
         &&-\lceil\boldsymbol{1}^-\tilde3\rfloor\lceil\boldsymbol{2}^+|\delta v|\tilde3\rfloor\left((c_1+c_3)v\cdot q+(c_3+c_4)v\cdot k_2+2c_3m\right)
         \\
         &&-\lceil\boldsymbol{1}^+\tilde3\rfloor\lceil\boldsymbol{2}^-|\delta v|\tilde3\rfloor\left((c_2+c_4)v\cdot q+(c_3+c_4)v\cdot k_2+2c_3m\right)
         \\
         &&-\lceil\boldsymbol{1}^-|\delta v|\tilde 3\rfloor\lceil\boldsymbol{2}^+\tilde3\rfloor((c_1-c_3)v \cdot q-(c_3+c_4)v\cdot k_2-2c_3m)
         \\
         &&-\lceil\boldsymbol{1}^+|\delta v|\tilde 3\rfloor\lceil\boldsymbol{2}^-\tilde3\rfloor((c_2-c_4)v \cdot q-(c_3+c_4)v\cdot k_2-2c_3m)
         \\
         &&
         -\lceil\boldsymbol{1}^+|\delta v|\tilde 3\rfloor\lceil\boldsymbol{2}^+\tilde3\rfloor(-(c_1+c_2)v\cdot k_2 -(c_1-c_3)v\cdot q)\\
         &&
         -\lceil\boldsymbol{1}^-|\delta v|\tilde 3\rfloor\lceil\boldsymbol{2}^-\tilde3\rfloor(-(c_1+c_2)2m-(c_1+c_2)v\cdot k_2-(c_2-c_4)v\cdot q).
    \end{array}
\end{equation}
Comparing Eq.~\eqref{LHS} and Eq.~\eqref{RHS} term by term, since the spinor structures on the two sides are linearly independent, the amplitude is RPI when
\begin{equation}
    c_2=-c_1, \quad c_4=-c_3=0.
\end{equation}
Thus the invariant amplitude is unique up to an overall factor, and we could set $c_1=1$. The RPI dimension-four three-particle $(1/2,1/2, +1)$ amplitude is then
\begin{equation}
    M_3^+=\frac{1}{v\cdot q}\left(\lceil\boldsymbol{1}^-\tilde3\rfloor\lceil\boldsymbol{2}^+\tilde3\rfloor-\lceil\boldsymbol{1}^+\tilde3\rfloor\lceil\boldsymbol{2}^-\tilde3\rfloor\right).
\end{equation}
This is precisely the coefficient of $g_0$ in Eq.~\eqref{eq:spins3pt1}. Similarly, the $(1/2,1/2, -1)$ amplitude is
\begin{equation}
    M_3^-=\frac{1}{v\cdot q}\left(\lceil\boldsymbol{1}^-3\rfloor\lceil\boldsymbol{2}^+3\rfloor-\lceil\boldsymbol{1}^+3\rfloor\lceil\boldsymbol{2}^-3\rfloor\right).
\end{equation}

Note that we have the following local form of the three-particle amplitudes:
\begin{eqnarray}\label{eq:local3ptform}
    x\left(\lceil\boldsymbol{1}^+\boldsymbol{2}^+\rfloor-\lceil\boldsymbol{1}^+\boldsymbol{2}^-\rfloor-\lceil\boldsymbol{1}^-\boldsymbol{2}^+\rfloor+\lceil\boldsymbol{1}^-\boldsymbol{2}^-\rfloor\right)&=&\frac{1}{v\cdot q}\left(\lceil\boldsymbol{1}^-\tilde 3\rfloor\lceil\boldsymbol{2}^+\tilde 3\rfloor-\lceil\boldsymbol{1}^+\tilde3\rfloor\lceil\boldsymbol{2}^-\tilde3\rfloor\right),\nonumber\\
    -x^{-1}\left(\lceil\boldsymbol{1}^+\boldsymbol{2}^+\rfloor+\lceil\boldsymbol{1}^+\boldsymbol{2}^-\rfloor+\lceil\boldsymbol{1}^-\boldsymbol{2}^+\rfloor+\lceil\boldsymbol{1}^-\boldsymbol{2}^-\rfloor\right)&=&\frac{1}{v\cdot q}\left(\lceil\boldsymbol{1}^-3\rfloor\lceil\boldsymbol{2}^+3\rfloor-\lceil\boldsymbol{1}^+3\rfloor\lceil\boldsymbol{2}^-3\rfloor\right).
\end{eqnarray}
To derive these expressions, we use momentum conservation $k_{1\perp}=k_{2\perp}+q_{\perp}$, and the Schouten identity to decompose the $x$ factor:
\begin{eqnarray}
\lceil\tilde3|k_{2\perp}|\tilde{3}\rfloor\lceil\boldsymbol{1}^{\pm}\boldsymbol{2}^{\pm}\rfloor&=&\lceil\boldsymbol{1}^{\pm}\tilde{3}\rfloor\lceil\boldsymbol{2}^{\pm}|k_{2\perp}|\tilde{3}\rfloor-\lceil\boldsymbol{1}^{\pm}|k_{2\perp}|\tilde{3}\rfloor\lceil\boldsymbol{2}^{\pm}\tilde{3}\rfloor,\nonumber\\
    &=&\lceil\boldsymbol{1}^{\pm}\tilde{3}\rfloor\lceil\boldsymbol{2}^{\pm}|k_{2\perp}|\tilde{3}\rfloor-\lceil\boldsymbol{1}^{\pm}|k_{1\perp}|\tilde{3}\rfloor\lceil\boldsymbol{2}^{\pm}\tilde{3}\rfloor+\lceil\boldsymbol{1}^{\pm}|q_{\perp}|\tilde{3}\rfloor\lceil\boldsymbol{2}^{\pm}\tilde{3}\rfloor.\nonumber
\end{eqnarray}
Applying the equations of motion to remove the three-momenta and summing over the four sign combinations of the heavy-particle legs, these identities collect into
\begin{equation}
\nonumber
    \lceil\tilde{3}|k_{2\perp}|\tilde{3}\rfloor(\lceil\boldsymbol{1}^+\boldsymbol{2}^+\rfloor-\lceil\boldsymbol{1}^+\boldsymbol{2}^-\rfloor-\lceil\boldsymbol{1}^-\boldsymbol{2}^+\rfloor+\lceil\boldsymbol{1}^-\boldsymbol{2}^-\rfloor)=2m(\lceil\boldsymbol{1}^-\tilde{3}\rfloor\lceil\boldsymbol{2}^+\tilde{3}\rfloor-\lceil\boldsymbol{1}^+\tilde{3}\rfloor\lceil\boldsymbol{2}^-\tilde{3}\rfloor).
\end{equation}
A similar calculation applies to the $h=-1$ helicity amplitude. Therefore we obtain the result Eq.~\eqref{eq:local3ptform}.

The second independent RPI structure in that equation, the non-minimal coupling proportional to $g_1$, is determined by solving the analogous condition $\delta A=0$ for the four structures without the overall factor $1/(v\cdot q)$; that condition forces the four coefficients to be equal. Together with the $g_1$ terms, this completes the two invariant amplitudes used in subsection~\ref{sec:bottomup}.

\bibliography{ref}
\end{document}